\documentclass{iopart}
\usepackage{iopams}

\begin{document}
\newcommand{\ExpandDerivations}{1}
\title{$\Lambda$-renormalized Einstein-Schr\"{o}dinger theory with spin-0 and spin-1/2 sources}
\author{J. A. Shifflett}
\address{Department of Physics,
Washington University, St.~Louis, Missouri 63130
}

%\date{October 29, 2004} %\today or an explicit date

\begin{abstract}
The Einstein-Schr\"{o}dinger theory is extended to include spin-0 and spin-1/2
sources, and the theory is derived from a Lagrangian density which allows
other fields to be easily added. The original
theory is also modified by including a cosmological constant caused by zero-point
fluctuations. This cosmological constant which multiplies the symmetric
metric is assumed to be nearly cancelled
by Schr\"{o}dinger's ``bare'' cosmological constant which multiplies the
nonsymmetric fundamental tensor, such that the total ``physical'' cosmological
matches measurement. We show that the resulting
$\Lambda$-renormalized Einstein-Schr\"{o}dinger theory closely
approximates ordinary Einstein-Maxwell theory and one-particle quantum mechanics.
In particular, the field equations match the ordinary Einstein and Maxwell
equations except for additional terms which are $<\!10^{-16}$ of the usual terms for
worst-case field strengths and rates-of-change accessible to measurement.
We also show that the theory predicts the exact Lorentz force equation and
the exact Klein-Gordon and Dirac equations.
And the theory becomes exactly Einstein-Maxwell theory
and one-particle quantum mechanics
in the limit as the cosmological constant from zero-point fluctuations goes to infinity.
Lastly, we discuss the merits of our Lagrangian density
compared to the Einstein-Maxwell Lagrangian density.
\end{abstract}

\newcommand{\rmt}{\sqrt{2}\,i}
\newcommand{\ca}{c_1}
\newcommand{\cb}{c_2}
\newcommand{\cc}{c_3}
\newcommand{\sR}{{{^*}R}}
\newcommand{\Nbar}{W}
\newcommand{\hR}{\mathcal R}
\newcommand{\tR}{\tilde\hR}
\newcommand{\hG}{\mathcal G}
\newcommand{\tG}{\tilde G}
\newcommand{\tB}{R}
\newcommand{\cUps}{\check\Upsilon}
\newcommand{\bUps}{\bar\Upsilon}
\newcommand{\tGam}{\tilde\Gamma}
\newcommand{\nGam}{\widehat\Gamma}
\newcommand{\sGam}{{^*\Gamma}}
\newcommand{\ftilde}{\tilde f}
\newcommand{\dual}{\vartheta}
\newcommand{\dualtilde}{\tilde\vartheta}
\newcommand{\Fdash}{\raise1pt\hbox{\rlap\textendash} F}
\newcommand{\uacute}{\acute{u}}
\newcommand{\hf}{\hat f}
\newcommand{\hj}{\hat j}
\newcommand{\hbj}{\hat\mathbf{j}}
\newcommand{\rmg}{\sqrt{-g}}
\newcommand{\rmN}{\sqrt{\!-N}}
\newcommand{\ehat}{Q}
\newcommand{\ff}{\,\ell}
\newcommand{\Aphi}{A}
\newcommand{\Gk}{G}
\newcommand{\Fdashoverf}{\raise1pt\hbox{\rlap\textendash} I}
\newcommand{\mum}{\mu}
\newcommand{\Ddag}{\overleftarrow D}
\newcommand{\ord}{{\mathcal O}}
\newcommand{\HR}{\Re}
\def\Stacksymbols #1#2#3#4{\def\theguybelow{#2}
   \def\verticalposition{\lower#3pt}
   \def\spacingwithinsymbol{\baselineskip0pt\lineskip#4pt}
   \mathrel{\mathpalette\intermediary#1}}
\def\intermediary#1#2{\verticalposition\vbox{\spacingwithinsymbol
     \everycr={}\tabskip0pt
     \halign{$\mathsurround0pt#1\hfil##\hfil$\crcr#2\crcr
            \theguybelow\crcr}}}

% Einstein Maxwell spacetimes (04.40.Nr)
% cosmological constant (98.80.Es)
% dark matter (95.35.+d)
% exact solutions of general relativity (04.20.Jb)
% unified field theories and models (12.10.-g)
% relativity and gravitation (95.30.Sf) - probably covered by (04.40.Nr)
% alternative theories of gravity (04.50.+h)
% No more than 4 of these are supposed to be used
\pacs{04.40.Nr,98.80.Es,12.10.-g,04.50.+h}% PACS codes
%\keywords{Einstein-Schrodinger Theory, Hermitian Theory of Relativity,
%Schrodinger Affine Field Theory, Einstein-Straus Theory,
%Cosmological Constant, Zero-Point Fluctuations}
%Use showkeys class option

\ead{shifflet@hbar.wustl.edu}

\section{\label{Introduction}Introduction}
The Einstein-Schr\"{o}dinger theory has had a long and distinguished history.
Two full length books and over 100 research articles have been published on it
(see the bibliographies of \cite{Hlavaty,Shifflett}). Both Einstein and Schr\"{o}dinger
devoted a section of one of their books to it. All of this was done without anyone being
able to definitely connect the theory to reality, and interest in the theory has faded as a result.
%and it has fallen out of favor for this reason. Today, the theory is thought by many scientists
%to be a dead theory. We will show that in a slightly modified form it is very much alive.

%It was shown early on that the original theory apparently does not predict the
%Lorentz force\cite{Callaway,Infeld}. This was rather cruelly pointed out in a
%popular general relativity text\cite{Misner} which notes that the theory ``has been shown to
%lead to the wrong equation of motion for a charged particle. It moves as if uncharged no matter
%how much charge is piled on its back. If that theory were correct, no cyclotron could operate,
%no atom could exist, and life itself would be impossible''. This and other derogatory statements
%have caused a very negative perception of the theory by those who know little about it.

In this paper and in two previous papers\cite{Shifflett,Shifflett2} it is shown
that the Einstein-Schr\"{o}dinger theory can be made to closely approximate
ordinary Einstein-Maxwell theory by simply including a large cosmological term
$\Lambda_z g_{\nu\mu}$ in the field equations, where $g_{\nu\mu}$ is the symmetric metric.
It is well known that zero-point fluctuations should be expected to cause
just such a cosmological term\cite{Zeldovich,Sahni,Peskin}, so this
can be viewed as a kind of zeroth order quantization effect. We assume the
cosmological term $\Lambda_z g_{\nu\mu}$ is nearly
cancelled by Schr\"{o}dinger's ``bare'' cosmological term $\Lambda_b N_{\nu\mu}$,
where $N_{\nu\mu}$ is the nonsymmetric fundamental tensor and
$g_{\nu\mu}\!\ne\!N_{(\nu\mu)}$. The resulting ``physical'' cosmological constant
$\Lambda=\Lambda_b+\Lambda_z$ can then be consistent with measurement,
hence the name $\Lambda$-renormalized Einstein-Schr\"{o}dinger theory (LRES theory).

In \cite{Shifflett}, the electro-vac version of this theory was
described in detail, and close approximations of Maxwell's equations and the Einstein
equations were derived. An exact electric monopole solution was derived which closely approximates
the Reissner-Nordstr\"{o}m solution, and the equations-of-motion for such solutions found using
the Einstein-Infeld-Hoffman method\cite{EinsteinInfeld,Wallace} were shown to definitely exhibit the
Lorentz force. In \cite{Shifflett2}, the theory was expressed in Newman-Penrose tetrad
form, some critical calculations were confirmed using tetrad methods,
and the electric monopole solution was shown to be of Petrov-type D.
A classical hydrodynamics extension of the theory was also developed, and the exact Lorentz force
equation was derived. In the present paper,
the theory is further extended to include spin-0 and spin-1/2 sources.
It is also derived from a Palatini type of Lagrangian density which allows other fields
to be easily added, including the additional fields of the Standard Model.

%There is a common opinion that Einstein produced little of scientific value during
%the last 30 years of his life.
%Supposedly he lost the creativity of his youth and became out of touch with modern physics.
%People even go so far as to pronounce why he failed in his search for a unified field theory,
%often with little knowledge of his work. It is said that he failed because he was unaware of
%the additional fields of the Standard Model, or because he strayed too much into pure
%mathematics and neglected physical intuition, or because he rejected quantum mechanics.
Einstein's work\cite{EinsteinStraus,Einstein3,EinsteinBianchi,EinsteinKaufman,EinsteinMOR}
on this theory began in 1946, and he held to the same theory until his death in 1955.
In his papers he never calls the theory a unified field theory, being careful
not to claim anything he could not prove. He derived the theory using arguments
similar to those he used to derive ordinary general relativity\cite{Schilpp,EinsteinBianchi},
and he used a Palatini type of Lagrangian density\cite{EinsteinBianchi}.

Schr\"{o}dinger's contributions\cite{SchrodingerI,SchrodingerIII,SchrodingerSTS}
to the theory began shortly after Einstein's. However, he came upon the theory
independently, by considering the simplest Lagrangian density he could think of, namely the
square-root of the determinant of the Ricci tensor composed of only an affinity.
%When he began his investigation of this Lagrangian density, he had no idea that it had any
%relation to Einstein's theory.
He found that the resulting field
equations were identical to Einstein's theory, but with a cosmological constant.
In fact, Schr\"{o}dinger's simple derivation only works if this cosmological constant
is not zero. This is somewhat important for the present theory in the sense that one can
claim that this ``bare'' cosmological constant is not simply being appended onto the theory,
but is instead an inherent part of it.
%In this paper we will show another simple derivation which also leads to the same theory,
%namely the derivation by requiring the Lagrangian density transplant into itself.

Many others have made important contributions to the theory, most of which are listed
in the bibliographies of \cite{Hlavaty,Shifflett}. Of particular significance
to our modified theory are contributions related to the choice of
metric\cite{Hely,Treder57,JohnsonI,Antoci3}, the generalized contracted Bianchi
identity\cite{Hely,JohnsonI,Antoci3}, the inclusion of charge
currents\cite{Antoci3,Borchsenius}, the inclusion of an energy-momentum tensor\cite{Treder57},
and exact solutions with a cosmological constant\cite{Papapetrou,Takeno}.

In our modified theory, Schr\"{o}dinger's bare cosmological constant nearly
cancels a large cosmological constant caused by zero-point
fluctuations\cite{Zeldovich}. This is not a radical proposal!
In fact, the apparent absence of a cosmological constant from zero-point fluctuations
has been a longstanding problem in conventional Standard-Model physics\cite{Sahni}.
Zero-point fluctuations are essential to quantum electrodynamics, and are the
cause of the Casimir force\cite{Sahni} and other effects.
%and spontaneous emission\cite{Griffiths}.
%Even if zero-point fluctuations do not contribute to the cosmological constant,
%there are other elements of the Standard-Model which should be expected to cause
%a large cosmological constant, and LRES theory would be very similar with
%such a substitution.
To quote from the quantum field theory text \emph{Peskin and Schroeder}\cite{Peskin}
``We have no understanding of why $\lambda$ is so much smaller than the vacuum energy
shifts generated in the known phase transitions of particle physics, and so much again
smaller than the underlying field zero-point energies.''
In the present theory, the fine-tuning of cosmological constants is not so objectionable
when one considers that it is similar to mass/charge/field-strength renormalization
in quantum electrodynamics. For example, to cancel electron self-energy in quantum electrodynamics,
the ``bare'' electron mass becomes large for a cutoff frequency
$\omega_c\!\sim\!1/({\rm Planck~length})$, and infinite if $\omega_c\!\rightarrow\!\infty$,
but the total ``physical'' mass remains small. In a similar manner, to cancel zero-point energy
in the present theory, the ``bare'' cosmological constant
$\Lambda_b\!\sim\!\omega_c^4({\rm Planck~length})^2$ becomes large
if $\omega_c\!\sim\!1/({\rm Planck~length})$, and infinite if
$\omega_c\!\rightarrow\!\infty$, but the total ``physical'' $\Lambda$ remains small.
This can be viewed as a kind of vacuum energy renormalization of the original
Einstein-Schr\"{o}dinger theory to account for zero-point fluctuations.
With this quantum mechanical effect included, the theory closely
approximates Einstein-Maxwell theory when $\omega_c\!\sim\!1/({\rm Planck~length})$,
and it becomes exactly Einstein-Maxwell theory in the limit as $\omega_c\!\rightarrow\!\infty$.

%This theory does not appear
%to predict any other fields of the Standard Model besides the electromagnetic field,
%at least not in an obvious way.
There are other reasons for pursuing this theory besides the unification of gravitation and electromagnetism.
Firstly, the theory suggests untried approaches to
a complete unified field theory. For example, one could consider the theory with higher
dimensions as in the Kaluza-Klein approach.
%and this is why the dimension ``n'' is left unspecified in this paper.
Another possibility is to consider the theory with non-Abelian fields as in the Standard Model.
This theory also offers an untried approach to the quantization of gravity.
While the theory is almost identical to Einstein-Maxwell theory at ordinary frequencies,
it becomes different near the Planck frequency, and it is possible that this could
fix some of the infinities that occur when attempting to quantize ordinary general relativity.
Since quantization of gravity is such an important topic, and previous approaches have been
worked over exhaustively, any new and well motivated approach should be welcome.
%The theory also needs to be quantized, which is likely to be challenging.
%It may seem like a step
%backwards to unify gravitation with electromagnetism, because we know how to quantize electromagnetism
%by itself, and this is lost when the two are unified. But from the opposite perspective,
In the present paper we are considering a quantization effect, namely a cosmological constant
$\Lambda_z$ caused by zero-point fluctuations, but the second quantization or multiparticle
quantization of the theory is otherwise outside the scope of the paper.
The hypothesis behind this paper is that an unquantized unified field theory might
come in the form of a purely classical theory where fermions are represented not by fields,
but by singular solutions of the field equations. We assume that the quantization
process must also include a first quantization step in which wave-functions are
substituted for particles to create a one-particle quantum mechanical theory.
With this assumption, a charged spin\nobreak-\nobreak1/2 field must be added
to the Lagrangian density, but this could perhaps be viewed as the first quantization of an
elementary Kerr-Newman-like solution. While this idea is speculative, the theory presented
here is rather conventional. We show that spin-0 and spin-1/2 fields can be added to
the theory in the same way they are added to ordinary Einstein-Maxwell theory.

This paper is organized as follows. In \S\ref{LagrangianDensity} we discuss the Lagrangian
density, and its extensions for the electro-vac, classical hydrodynamics, spin-0 and spin-1/2
cases. In \S\ref{EinsteinEquations}-\S\ref{Connections} we show that all of the extra terms
in our Einstein and Maxwell equations have a $\Lambda_b$ in the denominator so that they
go to zero as $\Lambda_b\!\rightarrow\!\infty$. And with a large $\Lambda_b$
caused by zero-point fluctuations, we show that these extra terms are all
$<\!10^{-16}$ of the usual terms for worst-case field strengths and rates-of-change
accessible to measurement.
In \S\ref{LorentzForce}-\S\ref{KleinGordonDirac} we show that the theory predicts the ordinary
Lorentz force equation and the ordinary Klein-Gordon and Dirac equations.
In \S\ref{Discussion} we discuss the merits of our Lagrangian density
compared to the Einstein-Maxwell Lagrangian density.

\section{\label{LagrangianDensity}The Lagrangian Density}
Ordinary vacuum general relativity
%with cosmological constant $\Lambda_b$ and dimension ``n''
can be derived from a Palatini Lagrangian density,
\begin{eqnarray}
\label{GR}
\fl~~~~{\mathcal L}(\Gamma^{\lambda}_{\!\rho\tau},g_{\rho\tau})
=-\frac{\lower1pt\hbox{$1$}}{16\pi}\rmg\left[\,g^{\mu\nu}
R_{\nu\mu}({\Gamma})+(n\!-\!2)\Lambda_b\,\right].
\end{eqnarray}
Here and throughout this paper we are assuming that $n\!=\!4$, but the dimension ``n''
will be included in the equations to show how easily the results can be generalized
to arbitrary dimension.
The original unmodified Einstein-Schr\"{o}dinger
theory
%\cite{SchrodingerI,SchrodingerIII,SchrodingerSTS,EinsteinStraus,Einstein3,EinsteinBianchi,EinsteinKaufman,EinsteinMOR}
can be derived from a generalization of (\ref{GR}) formed from a connection
$\nGam^\alpha_{\nu\mu}$ and a fundamental tensor $N_{\nu\mu}$
with no symmetry properties,
\begin{eqnarray}
\label{Palatini}
\fl~~~~{\mathcal L}(\nGam^{\lambda}_{\!\rho\tau},N_{\rho\tau})
=-\frac{\lower1pt\hbox{$1$}}{16\pi}\rmN\left[N^{\dashv\mu\nu}
\hR_{\nu\mu}({\nGam})+(n\!-\!2)\Lambda_b\,\right].
\end{eqnarray}
%\bigbreak

Our $\Lambda$-renormalized Einstein-Schr\"{o}dinger (LRES) theory includes a cosmological
constant $\Lambda_z$ caused by zero-point fluctuations, and allows other fields,
\begin{eqnarray}
\label{JSlag1}
\fl~~~~{\mathcal L}(\nGam^{\lambda}_{\!\rho\tau},N_{\rho\tau})
&=&-\frac{\lower1pt\hbox{$1$}}{16\pi}\rmN\left[N^{\dashv\mu\nu}
\hR_{\nu\mu}({\nGam})+(n\!-\!2)\Lambda_b\,\right]\nonumber\\
\fl&&\!-\frac{\lower1pt\hbox{$1$}}{16\pi}\rmg\,(n\!-\!2)\Lambda_z+{\mathcal L}_m(\psi_e,g_{\mu\nu},A_\sigma\dots),
\end{eqnarray}
where the ``bare'' $\Lambda_b$ obeys $\Lambda_b\!\approx\!-\Lambda_z$ so that the ``physical'' $\Lambda$ matches measurement,
\begin{eqnarray}
\label{Lambdadef}
\Lambda=\Lambda_b+\Lambda_z,
\end{eqnarray}
and the metric and electromagnetic potential are defined as
\begin{eqnarray}
\label{gdef}
\rmg\,g^{\mu\nu}=\rmN N^{\dashv(\mu\nu)},\\
\label{A}
\Aphi_\nu=\frac{\lower1pt\hbox{$1$}}{2(n\!-\!1)}\,{\nGam}{^\sigma_{\![\sigma\nu]}}\rmt\Lambda_b^{\!-1/2}.
\end{eqnarray}
Here and throughout this paper we use geometrized units with $c\!=\!G\!=\!1$,
the symbols
$\raise2pt\hbox{$_{(~)}$}$ and $\raise2pt\hbox{$_{[~]}$}$ around indices
indicate symmetrization and antisymmetrization, ``n'' is the dimension,
$N\!=\!det(N_{\mu\nu})$, and $N^{\dashv\sigma\nu}$ is the inverse of $N_{\nu\mu}$ so that
$N^{\dashv\sigma\nu}N_{\nu\mu}=\delta^\sigma_\mu$.
The ${\mathcal L}_m$ term is not to include a $\rmg\,F^{\mu\nu}\!F_{\mu\nu}$ part
but may contain the rest of the Standard Model.
In (\ref{JSlag1}), $\hR_{\nu\mu}(\nGam)$
%is not the usual Ricci tensor but
is a form of the so-called Hermitianized Ricci tensor\cite{EinsteinStraus},
%\begin{eqnarray}
%\label{HermitianizedRicci0}
%\hR_{\nu\mu}(\nGam)
%&=&\nGam^\alpha_{\nu\mu,\alpha}
%-\nGam^\alpha_{\nu\alpha,\mu}
%+\nGam^\sigma_{\nu\mu}\nGam^\alpha_{\sigma\alpha}
%-\nGam^\sigma_{\nu\alpha}\nGam^\alpha_{\sigma\mu}
%+\nGam^\alpha_{[\nu|\alpha,|\mu]}\\
%\label{HermitianizedRicci}
%&=&\nGam^\alpha_{\nu\mu,\alpha}
%-\nGam^\alpha_{(\nu|\alpha,|\mu)}
%+\nGam^\sigma_{\nu\mu}\nGam^\alpha_{\sigma\alpha}
%-\nGam^\sigma_{\nu\alpha}\nGam^\alpha_{\sigma\mu}.
%\end{eqnarray}
\begin{eqnarray}
\label{HermitianizedRicci0}
\fl~~~~~~~~\hR_{\nu\mu}(\nGam)
%&=&\nGam^\alpha_{\nu\mu,\alpha}
%-~\nGam^\alpha_{\!(\alpha\nu),\mu}
%~+\nGam^\sigma_{\nu\mu}\nGam^\alpha_{\!(\alpha\sigma)}
%-\nGam^\sigma_{\nu\alpha}\nGam^\alpha_{\sigma\mu}
%\!-\nGam^\tau_{\![\tau\nu]}\nGam^\alpha_{\![\alpha\mu]}/(n\!-\!1)
%+\nGam^\alpha_{\!(\alpha[\nu),\mu]}\\
\label{HermitianizedRicci}
=\nGam^\alpha_{\nu\mu,\alpha}
-\nGam^\alpha_{\!(\alpha(\nu),\mu)}
+\nGam^\sigma_{\nu\mu}\nGam^\alpha_{\!(\alpha\sigma)}
-\nGam^\sigma_{\nu\alpha}\nGam^\alpha_{\sigma\mu}
\!-\nGam^\tau_{\![\tau\nu]}\nGam^\alpha_{\![\alpha\mu]}/(n\!-\!1).
\end{eqnarray}
This tensor reduces to the ordinary Ricci tensor for symmetric fields, where we have
$\Gamma^\alpha_{\![\nu\mu]}\!=\!0$ and
$\Gamma^\alpha_{\alpha[\nu,\mu]}\!=\!R^\alpha_{~\alpha\mu\nu}/2\!=\!0$.

It is helpful to decompose ${\nGam}{^\alpha_{\nu\mu}}$
into another connection ${\tGam}{^\alpha_{\nu\mu}}$, and $A_\sigma$ from~(\ref{A}),
\begin{eqnarray}
\label{gamma_natural}
\fl~~~~~~~~~~~~~~~~~~~~~{\nGam}{^\alpha_{\nu\mu}}&=&\tGam{^\alpha_{\nu\mu}}
+(\delta^\alpha_\mu\Aphi_\nu
\!-\delta^\alpha_\nu \Aphi_\mu)\rmt\Lambda_b^{\!1/2},\\
\label{gamma_tilde}
\fl~~~~~~~~~~{\rm where}~~~\tGam{^\alpha_{\nu\mu}}
&=&{\nGam}{^\alpha_{\nu\mu}}\!+
%\frac{\lower1pt\hbox{$1$}}{(n\!-\!1)}
(\delta^\alpha_\mu{\nGam}{^\sigma_{\![\sigma\nu]}}
-\delta^\alpha_\nu{\nGam}{^\sigma_{\![\sigma\mu]}})/(n\!-\!1).
\end{eqnarray}
By contracting (\ref{gamma_tilde}) on the right and left we see that
$\tGam{^\alpha_{\nu\mu}}$ has the symmetry
\begin{eqnarray}
\label{JScontractionsymmetric}
\tGam^\alpha_{\nu\alpha}
%\!=\!(n\nGam^\alpha_{\nu\alpha}\!-\nGam^\alpha_{\alpha\nu})/(n-1)
\!=\!\nGam^\alpha_{\!(\nu\alpha)}
\!=\!\tGam^\alpha_{\alpha\nu},
\end{eqnarray}
so it has only $n^3\!-\!n$ independent components.
%In \ref{ExtractionofConnectionAddition} it is shown that
%\begin{eqnarray}
Using ${\mathcal R}_{\nu\mu}(\nGam)\!=\!{\mathcal R}_{\nu\mu}(\tGam)+2A_{[\nu,\mu]}\rmt\Lambda_b^{\!1/2}$
%\end{eqnarray}
from (\ref{RnGam}), the Lagrangian density (\ref{JSlag1}) can be rewritten in terms of
${\tGam}{^\alpha_{\nu\mu}}$ and $A_\sigma$,
\begin{eqnarray}
\label{JSlag3}
\fl{\mathcal L}(\nGam^{\lambda}_{\!\rho\tau},N_{\rho\tau})
&=&\!-\frac{\lower1pt\hbox{$1$}}{16\pi}\rmN\left[N^{\dashv\mu\nu}(\tR_{\nu\mu}
%+\ca\tGam^\alpha_{\alpha[\nu,\mu]}
+2\Aphi_{[\nu,\mu]}\rmt\Lambda_b^{\!1/2})+(n\!-\!2)\Lambda_b\,\right]\nonumber\\
\fl&&\!-\frac{\lower1pt\hbox{$1$}}{16\pi}\rmg\,(n\!-\!2)\Lambda_z+{\mathcal L}_m(\psi_e,g_{\mu\nu},A_\sigma\dots).
\end{eqnarray}
Here $\tR_{\nu\mu}\!=\!\hR_{\nu\mu}(\tGam)$, and from (\ref{JScontractionsymmetric})
the Hermitianized Ricci tensor (\ref{HermitianizedRicci}) simplifies to
\begin{eqnarray}
\label{HermitianizedRiccit}
\tR_{\nu\mu}
&=&\tGam^\alpha_{\nu\mu,\alpha}
-\tGam^\alpha_{\alpha(\nu,\mu)}
+\tGam^\sigma_{\nu\mu}\tGam^\alpha_{\sigma\alpha}
-\tGam^\sigma_{\nu\alpha}\tGam^\alpha_{\sigma\mu}.
\end{eqnarray}
From (\ref{gamma_natural},\ref{JScontractionsymmetric}), $\tGam^\alpha_{\nu\mu}$ and $A_\nu$ fully
parameterize $\nGam^\alpha_{\nu\mu}$ and can be treated as independent variables.
So when we set $\delta{\mathcal L}/\delta\tGam^\alpha_{\nu\mu}\!=0$
and $\delta{\mathcal L}/\delta A_\nu\!=0$,
the same field equations must result as with
$\delta{\mathcal L}/\delta\nGam^\alpha_{\nu\mu}\!=0$.
It is simpler to calculate the field equations using
$\tGam^\alpha_{\nu\mu}$ and $A_\nu$ instead of
$\nGam^\alpha_{\nu\mu}$, so we will follow this method.

We will usually assume that $\Lambda_z$ is limited by a
cutoff frequency\cite{Sakharov,Padmanabhan,Padmanabhan2,Ashtekar}
\begin{eqnarray}
\label{cutoff}
\omega_c\!\sim\!1/l_P,
\end{eqnarray}
where
%\begin{eqnarray}
%\label{Planck}
$l_P\!=\!({\rm Planck~length})\!=\!\sqrt{\hbar G/c^3}\!=\!1.6\times 10^{-33}cm$.
%\end{eqnarray}
Then from (\ref{Lambdadef},\ref{cutoff}) and assuming all of the known
fundamental particles we have\cite{Sahni},
\begin{eqnarray}
\label{Lambdab}
\Lambda_b&\approx&-\Lambda_z\sim C_z \omega_c^4 l_P^2
\sim 10^{66}\,{\rm cm}^{-2},\\
\label{Cz}
C_z&=&\frac{\lower2pt\hbox{1}}{2\pi}\!\left({{\lower2pt\hbox{fermion}}\atop{\raise2pt\hbox{spin~states}}}
-{{\lower2pt\hbox{boson}}\atop{\raise2pt\hbox{spin~states}}}\right)
\sim \frac{\lower2pt\hbox{60}}{2\pi}
%~~~~ C_z={\rm(fermions\!-\!bosons)}/2\pi\sim 60/2\pi
\end{eqnarray}
and from astronomical measurements
\begin{eqnarray}
\label{Lambda}
\Lambda &\approx&1.4\times 10^{-56}{\rm cm}^{-2},~~~~
\label{LambdaoverLambdab}
\Lambda/\Lambda_b\sim 10^{-122}.
\end{eqnarray}
However, it might be more correct to
fully renormalize with $\omega_c\!\rightarrow\!\infty$,
$|\Lambda_z|\!\rightarrow\!\infty$, $\Lambda_b\!\rightarrow\!\infty$
as in quantum electrodynamics.
To account for this possibility we will prove that
\begin{eqnarray}
\label{LRESlimit}
{{\lower2pt\hbox{lim}}\atop{\raise2pt\hbox{$\Lambda_b\!\rightarrow\!\infty$}}}
\left(\lower2pt\hbox{LRES}\atop{\raise2pt\hbox{theory}}\right)
\!=\!\left(\lower2pt\hbox{Einstein-Maxwell}\atop{\raise2pt\hbox{theory}}\right).
\end{eqnarray}

The Hermitianized Ricci tensor (\ref{HermitianizedRicci}) has the following invariance properties
\begin{eqnarray}
\label{transpositionsymmetric}
\hR_{\nu\mu}(\nGam^T)=\hR_{\mu\nu}(\nGam),~~~~~~~~~~~~~~~~({\rm T=transpose})\\
\label{gaugesymmetric}
\hR_{\nu\mu}(\nGam^\alpha_{\rho\tau}\!+\delta^\alpha_{[\rho}\varphi_{,\tau]})=
\hR_{\nu\mu}(\nGam^\alpha_{\rho\tau})~~~{\rm for~an~arbitrary}~\varphi(x^\sigma).
\end{eqnarray}
From (\ref{transpositionsymmetric},\ref{gaugesymmetric}), the Lagrangian densities (\ref{JSlag1},\ref{JSlag3}) are invariant under charge conjugation,
\begin{eqnarray}
\label{transposition}
\fl~~Q\!\rightarrow\!-Q,~~
A_\sigma\!\rightarrow\!-A_\sigma,~~
\tGam^\alpha_{\nu\mu}\!\rightarrow\!\tGam^\alpha_{\mu\nu},~~
\nGam^\alpha_{\nu\mu}\!\rightarrow\!\nGam^\alpha_{\mu\nu},~~
N_{\nu\mu}\!\!\rightarrow\!N_{\mu\nu},~~
N^{\dashv\nu\mu}\!\rightarrow\!N^{\dashv\mu\nu},
\end{eqnarray}
and also under an electromagnetic gauge transformation
\begin{eqnarray}
\label{gauge}
\fl~~\psi\!\rightarrow\!\psi e^{i\phi},~~
A_\alpha\!\rightarrow\! A_\alpha\!-\!\frac{\hbar}{Q}\phi_{,\alpha},~~
\tGam^\alpha_{\rho\tau}\!\rightarrow\!\tGam^\alpha_{\rho\tau},~~
\nGam^\alpha_{\rho\tau}\!\rightarrow\!\nGam^\alpha_{\rho\tau}\!+\frac{2\hbar}{Q}\delta^\alpha_{[\rho}\phi_{,\tau]}\rmt\Lambda_b^{\!1/2},
\end{eqnarray}
assuming that ${\mathcal L}_m$ is invariant.
If $\Lambda_b\!>0,~\Lambda_z\!<0$ as in (\ref{Lambdab}) then
$\tGam^\alpha_{\nu\mu}$, $\nGam^\alpha_{\nu\mu}$, $N_{\nu\mu}$ and $N^{\dashv\nu\mu}$ are all Hermitian,
$\tR_{\nu\mu}$ and $\hR_{\nu\mu}(\nGam)$ are Hermitian from (\ref{transpositionsymmetric}),
and $g_{\nu\mu}$, $A_\sigma$ and ${\mathcal L}$ are real from (\ref{gdef},\ref{A},\ref{JSlag1},\ref{JSlag3}).
If instead $\Lambda_b\!<0,~\Lambda_z\!>0$, then all of the fields are real.
%The allowance of Hermitian fields is not a peculiarity of the present
%theory but is in fact a basic part of the original Einstein-Schr\"{o}dinger
%theory\cite{EinsteinStraus,EinsteinBianchi}.
%Finally, let us discuss the ``simplicity'' of the Lagrangian density (\ref{JSlag1}),
%which is of some relevance in any attempt to unify the laws of physics.
%As shown in equation (\ref{ordinary}) of \S\ref{Discussion}, the original Einstein-Schr\"{o}dinger theory
%can be derived using a Lagrangian density formed from the ordinary Ricci tensor $R_{\nu\mu}(\nGam)$,
%which may seem more acceptable than he Htermitianized Ricci tensor $\hR_{\nu\mu}(\nGam)$.
%We should point out that unlike the Riemann tensor, which can be derived from parallel transporting
%a vector along separate paths to the same point, the familiar Ricci tensor has no such geometrical
%significance. The importance of the Ricci tensor is that it is part of the Einstein equations
%and the Hilbert Lagrangian in ordinary general relativity with symmetric fields.
%It may seem unappealing that the Hermitianized Ricci tensor (\ref{HermitianizedRicci}) contains
%more terms than the ordinary Ricci tensor $R_{\nu\mu}(\nGam)$.

Note that (\ref{gdef}) defines $g^{\mu\nu}$ unambiguously because $\rmg\!=\![-det(\rmg\,g^{\mu\nu})]^{1/(n-2)}$.
In this theory the metric $g_{\mu\nu}$ is used for measuring space-time intervals, and for calculating geodesics,
and for raising and lowering of indices. The covariant derivative ``;'' is always
done using the Christoffel connection formed from $g_{\mu\nu}$,
\begin{eqnarray}
\label{Christoffel}
\Gamma^\alpha_{\nu\mu}&=&\frac{\lower2pt\hbox{$1$}}{2}\,g^{\alpha\sigma}(g_{\mu\sigma,\nu}
+g_{\sigma\nu,\mu}-g_{\nu\mu,\sigma}).
\end{eqnarray}
%If ${\mathcal L}_m\!=\!0$, the metric (\ref{gdef}) is the unique choice for which the divergence
%of the Einstein equations vanishes when using (\ref{Christoffel}) for the covariant derivative.
%If ${\mathcal L}_m\!\ne\!0$,
With the metric (\ref{gdef}), the divergence of the Einstein equations
vanishes if ${\mathcal L}_m\!=\nobreak\!0$,
and it gives the exact Lorentz force equation if ${\mathcal L}_m\!\ne\!0$.
%where (\ref{Christoffel}) is used for the covariant derivative.
And when $N_{\mu\nu}$ and $\nGam^\alpha_{\mu\nu}$ are symmetric,
the definition (\ref{gdef}) requires $g_{\mu\nu}\!=\!N_{\mu\nu}$,
the definition (\ref{A}) requires $A_\sigma\!=\!0$,
%$\delta{\mathcal L}/\delta\nGam^\alpha_{\nu\mu}\!=\!0$ requires
%$\nGam^\alpha_{\nu\mu}\!=\!\Gamma^\alpha_{\nu\mu}$,
and the theory reduces to ordinary general relativity without electromagnetism.

The electromagnetic field is defined in terms of the potential (\ref{A}) as usual
\begin{eqnarray}
\label{Fdef}
F_{\mu\nu}=A_{\nu,\mu}-A_{\mu,\nu}.
\end{eqnarray}
However, we will also define a lowercase $f_{\mu\nu}$
\begin{eqnarray}
\label{fdef}
\rmg\,f^{\mu\nu}=\rmN N^{\dashv[\nu\mu]}\Lambda_b^{\!1/2}/\rmt.
\end{eqnarray}
Then from (\ref{gdef}), $g^{\mu\nu}$ and $f^{\mu\nu}\rmt\Lambda_b^{\!-1/2}$ are
the symmetric and antisymmetric
parts of a total field,
\begin{eqnarray}
\label{Wdef}
\Nbar^{\mu\nu}\!=\!(\rmN/\rmg\,)N^{\dashv\nu\mu}=g^{\mu\nu}\!+\!f^{\mu\nu}\rmt\Lambda_b^{\!-1/2}.
\end{eqnarray}
We will see that the field equations require $f_{\mu\nu}\!\approx\!F_{\mu\nu}$
to a very high precision, so it is mainly just a matter of terminology which one
is called the electromagnetic field.

%\newpage
Note that there are many possible nonsymmetric generalizations
of the Ricci tensor besides
the Hermitianized Ricci tensor $\hR_{\nu\mu}(\nGam)$
from (\ref{HermitianizedRicci})
and the ordinary Ricci tensor $R_{\nu\mu}(\nGam)$.
For example, we could form any weighted average of
$R_{\nu\mu}(\nGam)$,
$R_{\mu\nu}(\nGam)$,
$R_{\nu\mu}(\nGam^T)$
and $R_{\mu\nu}(\nGam^T)$,
and then add any linear combination of the tensors
$R^\alpha{_{\alpha\nu\mu}}(\nGam)$,
$R^\alpha{_{\alpha\nu\mu}}(\nGam^T)$,
%$\nGam^\alpha_{\alpha[\nu,\mu]}$,
%$\nGam^\alpha_{\![\nu|\alpha,|\mu]}$,
%$\nGam^\alpha_{\![\alpha\nu],\mu}\!\!+\nGam^\alpha_{\mu\sigma}\nGam^\sigma_{\![\alpha\nu]}$,
$\nGam^\alpha_{\![\nu\mu]}\nGam^\sigma_{\![\sigma\alpha]}$,
$\nGam^\alpha_{\![\nu\sigma]}\nGam^\sigma_{\![\mu\alpha]}$,
and $\nGam^\alpha_{\![\alpha\nu]}\nGam^\sigma_{\![\sigma\mu]}$.
All of these generalized Ricci tensors would be linear in $\nGam^\alpha_{\nu\mu,\sigma}$, quadratic in $\nGam^\alpha_{\nu\mu}$,
and would reduce to the ordinary Ricci tensor for symmetric fields.
Even if we limit the tensor to only four terms, there are still eight possibilities.
We assert that invariance properties like (\ref{transpositionsymmetric},\ref{gaugesymmetric})
are the most sensible way to choose among the different alternatives,
not criteria such as the number of terms in the expression.

%\newpage
For the electro-vac case as in \cite{Shifflett}, matter is represented by solutions
to the field equations (which are allowed to have singularities) and we have
\begin{eqnarray}
{\mathcal L}_m=0.
\end{eqnarray}
For the classical hydrodynamics case as in \cite{Shifflett2}, we can form a rather artificial
${\mathcal L}_m$ which depends on a mass scalar density $\boldsymbol\mu$ and a velocity
vector $u^\nu$, neither of which is constrained (that is we will not require
$\delta{\mathcal L}/\delta\boldsymbol\mu\!=0$ or $\delta{\mathcal L}/\delta u^\sigma\!=0$),
\begin{eqnarray}
\label{Lhydrodynamic}
{\mathcal L}_m&=&-\frac{\boldsymbol\mu Q}{m} u^\nu A_{\nu}
-\frac{\boldsymbol\mu}{2} u^\alpha g_{\alpha\sigma}u^\sigma.
\end{eqnarray}
For the spin-0 case as in \cite{Birrell}, matter is represented with a scalar
wave-function $\psi$,
\begin{eqnarray}
\label{LKleinGordon}
{\mathcal L}_m&=&\rmg\,\frac{1}{2}\left(\frac{\hbar^2}{m}\bar\psi \Ddag\!_\mu D^\mu \psi
-m \bar\psi\psi\right),\\
\label{derivKleinGordon}
D_\mu&=&\frac{\partial}{\partial x^\mu}+\frac{iQ}{\hbar}A_\mu~~,~~
\Ddag\!_\mu=\frac{\overleftarrow \partial}{\partial x^\mu}-\frac{iQ}{\hbar}A_\mu.
\end{eqnarray}
For the spin-1/2 case as in \cite{Birrell}, matter is represented by a four-component
wave-function $\psi$, and things are defined using tetrads $e_{(a)}{^\sigma}$,
\begin{eqnarray}
\label{LDirac}
\fl~~~~~~~~~~{\mathcal L}_m&=&\rmg\left(\frac{i\hbar}{2}(\bar\psi\gamma^\sigma D_\sigma\psi
-\bar\psi\overleftarrow D_\sigma\gamma^\sigma\psi)
-m\bar\psi\psi\right),\\
\fl~~~~~~~~~~~\gamma^\sigma&=&\gamma^{(a)}e_{(a)}{^\sigma}~~~,~~~
\gamma^{(0)}=
\pmatrix{
I&~0\cr
0\!&\!-I
}~~,~~
\gamma^{(i)}=
\pmatrix{
0&\sigma_i\cr
-\sigma_i&0
},\\
\fl~~~~~~g^{(a)(b)}
&=&e^{(a)}{_\tau}\,e^{(b)}{_\sigma}\,g^{\tau\sigma}
=\frac{1}{2}(\gamma^{(a)}\gamma^{(b)}+\gamma^{(b)}\gamma^{(a)})
%\tiny{
%=\pmatrix{
%~1&&&~\cr
%&\!\!\!\!\!-1&&~\cr
%&&\!\!\!\!\!-1&~\cr
%&&&\!\!\!\!\!-1~
%}},\\
=\pmatrix{
\!1~~~\atop ~~~-1&\cr
&\!\!\!\!\!\!\!\!\!\!\!-1~~\atop ~~-1\cr
},\\
\label{decomposition}
\fl~~~~~~~~~~g^{\tau\sigma}&=&e_{(a)}{^\tau}e_{(b)}{^\sigma}g^{(a)(b)}
=\frac{1}{2}(\gamma^\tau\gamma^\sigma+\gamma^\sigma\gamma^\tau),\\
\label{inverses}
\fl ~~e^{(a)}{_\tau}e_{(c)}{^\tau}&=&\delta^{(a)}_{(c)}~~~,~~~
e^{(a)}{_\tau}e_{(a)}{^\sigma}=\delta^{\sigma}_{\tau},\\
\label{derivDirac}
\fl~~~~~~~~~~D_\mu&=&\frac{\partial}{\partial x^\mu}+\check\Gamma_\mu+\frac{iQ}{\hbar}A_\mu~~,~~
\Ddag\!_\mu=\frac{\overleftarrow \partial}{\partial x^\mu}+\check\Gamma^\dag_\mu-\frac{iQ}{\hbar}A_\mu,\\
\fl~~~~~~~~~~\check\Gamma_\mu&=&\frac{1}{2}\Sigma^{(a)(b)}e_{(a)}{^\sigma}e_{(b)\sigma;\mu}~~~~,~~~~
\label{sigma}
\Sigma^{(a)(b)}=\frac{i}{2}(\gamma^{(a)}\gamma^{(b)}-\gamma^{(b)}\gamma^{(a)}).
\end{eqnarray}
In the equations above, $m$ is mass, $Q$ is charge
and the $\sigma_i$ are the Pauli spin matrices.
%\begin{eqnarray}
%\label{Planck}
%&&~~~l_P
%=\sqrt{\frac{\hbar G}{c^3}} =1.6\times\!10^{-33}{\rm cm}.
%\end{eqnarray}
In (\ref{derivKleinGordon},\ref{derivDirac}) the conjugate derivative
operator $\Ddag\!_\mu$ is made to operate from right to left to simplify
subsequent calculations.
The spin-0 and spin-1/2 ${\mathcal L}_m{'s}$ are the ordinary expressions
for quantum fields in curved space\cite{Birrell}.

\section{\label{EinsteinEquations}The Einstein Equations}
The Einstein equations are obtained by setting $\delta{\mathcal L}/\delta N_{\nu\mu}\!=0$.
However, the field equations must be the same
if we instead use $g^{\nu\mu}$ and $f^{\nu\mu}$ as the independent variables. The field
equations must also be the same if we use $\rmN N^{\dashv\mu\nu}$ as the independent
variable, and since this is simplest, we will follow this method. Before calculating the
field equations, we need some preliminary results.
From (\ref{gdef}) we get,
\begin{eqnarray}
\label{gcontravariantderiv}
\fl~~~~~\frac{\partial\left(\rmg g^{\rho\tau}\right)}{\partial(\rmN N^{\dashv\mu\nu})}
&=&\delta^{(\rho}_\mu\delta^{\tau)}_\nu,\\
\label{gcovariantderiv}
\fl~~~~~\frac{\partial(g_{\tau\sigma}/\rmg)}{\partial(\rmN N^{\dashv\mu\nu})}
&=&-\frac{g_{\tau(\nu}g_{\mu)\sigma}}{\rmg\rmg}
~~~~~~\left({\rm because}~~
\frac{\partial\left(\rmg g^{\rho\tau}g_{\tau\sigma}/\rmg\right)}
{\partial(\rmN N^{\dashv\mu\nu})}=0\right).
\end{eqnarray}
Using (\ref{gdef})
and the identities $det(sM^{})\!=s^n det(M^{})$,~$det(M^{-1}_{})\!=1/det(M^{})$ gives
\begin{eqnarray}
%\fl~~~~~\rmg\,g^{\rho\tau}&=&\rmN N^{\dashv(\rho\tau)},\\
%\fl~~~~~\rmg\,f^{\rho\tau}&=&-\rmN N^{\dashv[\rho\tau]},\\
\label{rmN2}
\fl~~~~~~\rmN&=&(-det(\rmN N^{\dashv..}))^{1/(n-2)},\\
\label{rmg2}
\fl~~~~~~\rmg&=&(-det(\rmg\,g^{..}))^{1/(n-2)}
=(-det(\rmN N^{\dashv(..)}))^{1/(n-2)}.
\end{eqnarray}
Using (\ref{rmN2},\ref{rmg2}) and the identity
$\partial(det(M^{\cdot\cdot}))/\partial M^{\mu\nu}\!=M^{-1}_{\nu\mu}det(M^{\cdot\cdot})$ gives
\begin{eqnarray}
\label{rmNderiv}
\fl~~~~~\frac{\partial\rmN}{\partial(\!\rmN N^{\dashv\mu\nu})}
\!&=&\!\frac{(-det(\rmN N^{\dashv..}))^{1/(n-2)-1+1}}{(n\!-\!2)}\frac{N_{\nu\mu}}{\rmN}
=\frac{N_{\nu\mu}}{(n\!-\!2)},\\
\label{rmgderiv}
\fl~~~~~\frac{\partial\rmg}{\partial(\!\rmN N^{\dashv\mu\nu})}
\!&=&\!\frac{(-det(\rmg\,g^{..}))^{1/(n-2)-1+1}}{(n\!-\!2)}\frac{g_{\nu\mu}}{\rmg}
=\frac{g_{\nu\mu}}{(n\!-\!2)}.
\end{eqnarray}
Note that from (\ref{gcontravariantderiv},\ref{gcovariantderiv},\ref{rmgderiv}),
if ${\mathcal L}_m$ depends only on $g_{\nu\mu}$ and $\rmg$, and not on
$N_{\nu\mu}$ and $\rmN$, then $\partial{\mathcal L}_m/\partial(\rmN N^{\dashv\mu\nu})$
gives the same result as $\partial{\mathcal L}_m/\partial(\rmg\,g^{\mu\nu})$.

For the spin-1/2 case we will need the derivative
$\partial(\rmg\,e_{(a)}{^\tau})/\partial(\rmN N^{\dashv\mu\nu})$.
Multiplying (\ref{decomposition}) by $\rmg\,e^{(b)}{_\sigma}$ and taking its
derivative with respect to $\rmg\,g^{\mu\nu}$ gives
\begin{eqnarray}
\label{decomposition2}
\fl&&~~~~~\rmg\,g^{\tau\sigma}e^{(b)}{_\sigma}=\rmg\,e_{(a)}{^\tau}g^{(a)(b)},\\
\label{der1}
\fl&&~~~~~\delta^\tau_{(\mu}\delta^\sigma_{\nu)} e^{(b)}{_\sigma}+\rmg\,g^{\tau\sigma}\frac{\partial e^{(b)}{_\sigma}}{\partial(\rmg\,g^{\mu\nu})}
=\frac{\partial(\rmg\,e_{(a)}{^\tau})}{\partial(\rmg\,g^{\mu\nu})} g^{(a)(b)}.
\end{eqnarray}
Taking the derivative of (\ref{inverses}) with respect to
$\rmg\,g^{\mu\nu}$ and using (\ref{rmgderiv}) gives
\begin{eqnarray}
\label{der2}
\fl~~~~~~~0=e^{(a)}{_\tau}\frac{\partial(\rmg\,e_{(c)}{^\tau})}{\partial(\rmg\,g^{\mu\nu})}
-\frac{g_{\nu\mu}}{(n\!-\!2)}\delta^{(a)}_{(c)}
+\rmg\,\frac{\partial e^{(a)}{_\tau}}{\partial(\rmg\,g^{\mu\nu})}e_{(c)}{^\tau}.
\end{eqnarray}
Substituting (\ref{der2}) into (\ref{der1}) we finally get
\begin{eqnarray}
\fl\frac{\partial(\rmg\,e_{(a)}{^\tau})}{\partial(\rmg\,g^{\mu\nu})} g^{(a)(b)}-e^{(b)}{_{(\nu}}\delta^\tau_{\mu)}
&=&\rmg\,g^{\tau\sigma}\frac{\partial e^{(b)}{_\sigma}}{\partial(\rmg\,g^{\mu\nu})}\\
&=&\,g^{\tau\sigma}\!\left(\frac{g_{\nu\mu}}{(n\!-\!2)}e^{(b)}{_\sigma}-e^{(b)}{_\tau}\frac{\partial(\rmg\,e_{(a)}{^\tau})}{\partial(\rmg\,g^{\mu\nu})}\,e^{(a)}{_\sigma}\right)\\
&=&\frac{g_{\nu\mu}}{(n\!-\!2)}e^{(b)\tau}-g^{(a)(b)}\frac{\partial(\rmg\,e_{(a)}{^\tau})}{\partial(\rmg\,g^{\mu\nu})},\\
\fl\frac{\partial(\rmg\,e_{(a)}{^\tau})}{\partial(\rmN N^{\dashv\mu\nu})}
=\frac{\partial(\rmg\,e_{(a)}{^\tau})}{\partial(\rmg g^{\mu\nu})}
&=&\frac{1}{2}e_{(a)}{_{(\nu}}\delta^\tau_{\mu)}+\frac{g_{\nu\mu}}{2(n\!-\!2)}e_{(a)}{^\tau}.
\end{eqnarray}
\bigskip

Now we are ready to calculate the field equations.
Setting $\delta{\mathcal L}/\delta(\rmN N^{\dashv\mu\nu})\!=0$ and using (\ref{rmNderiv},\ref{rmgderiv}) gives,
\begin{eqnarray}
\fl~~~~~~~~0&=&-16\pi\left[\frac{\partial{\mathcal L}}{\partial(\rmN N^{\dashv\mu\nu})}
-\left(\frac{\partial{\mathcal L}}{\partial(\rmN N^{\dashv\mu\nu})_{\!,\,\omega}}\right){_{\!,\,\omega}}\right]\\
\label{para}
\fl~~~~~~~~&=&\tR_{\nu\mu}
\!+2\Aphi_{[\nu,\mu]}\rmt\Lambda_b^{\!1/2}
\!+\Lambda_b N_{\nu\mu}
\!+\Lambda_z g_{\nu\mu}
\!-8\pi S_{\nu\mu},
\end{eqnarray}
where $S_{\nu\mu}$ and the energy-momentum tensor $T_{\nu\mu}$ are defined by
\begin{eqnarray}
\label{Tdef}
\fl~~~~~S_{\nu\mu}\!&=&2\frac{\delta{\mathcal L}_m}{\delta(\rmN N^{\dashv\mu\nu})}\\
\fl &=&T_{\nu\mu}\!-\frac{1}{(n-2)}g_{\nu\mu}T^\alpha_\alpha,~~~~\\
\fl~~~~~T_{\nu\mu}&=&S_{\nu\mu}\!-\frac{1}{2}g_{\nu\mu}S^\alpha_\alpha.
\end{eqnarray}
Taking the symmetric and antisymmetric parts of (\ref{para}) and using (\ref{Fdef}) gives
\begin{eqnarray}
\label{JSsymmetric}
\fl&&~~~~~ \tR_{(\nu\mu)}
+\Lambda_b N_{(\nu\mu)}+\Lambda_z g_{\nu\mu}
=8\pi\!\left(T_{\nu\mu}
-\frac{\lower1pt\hbox{$1$}}{(n-2)}g_{\nu\mu}T^\alpha_\alpha\right),\\
%\label{JScurl}
%\fl~~~~~~\tR_{[\nu\mu,\sigma]}+\Lambda_b N_{[\nu\mu,\sigma]}=0,\\
\label{JSantisymmetric}
\fl&&~~~~~~N_{[\nu\mu]}=
F_{\nu\mu}\rmt\Lambda_b^{\!-1/2}
\!-\tR_{[\nu\mu]}\Lambda_b^{\!-1}.
\end{eqnarray}
Also from the curl of (\ref{JSantisymmetric}) we get
\begin{eqnarray}
\label{JScurl}
\fl~~~~~~\tR_{[\nu\mu,\sigma]}+\Lambda_b N_{[\nu\mu,\sigma]}=0.
\end{eqnarray}
%These equations with $T_{\nu\mu}\!=\!0$ are often used as part of the definition of the original Einstein-Schr\"{o}dinger theory.

To put (\ref{JSsymmetric}) into a form which looks more like the ordinary Einstein equations,
we need some preliminary results.
The definitions (\ref{gdef},\ref{fdef}) of $g_{\nu\mu}$ and $f_{\nu\mu}$
can be inverted to give $N_{\nu\mu}$ in terms of $g_{\nu\mu}$ and $f_{\nu\mu}$.
An expansion in powers of $\Lambda_b^{\!-1}$ is derived in \ref{ApproximateFandg},
and confirmed by tetrad methods in \cite{Shifflett2},
\begin{eqnarray}
\label{approximateNbar}
\fl~~~~~ N_{(\nu\mu)}&\!\!=\!& g_{\nu\mu}-2\!\left({f_\nu}^\sigma f_{\sigma\mu}
-\frac{1}{2(n\!-\!2)}g_{\nu\mu}f^{\rho\sigma}\!f_{\sigma\rho}\right)\!\Lambda_b^{\!-1}
+(f^4)\Lambda_b^{\!-2}\dots\\
\label{approximateNhat}
\fl~~~~~ N_{[\nu\mu]}&\!\!=\!& f_{\nu\mu}\rmt\Lambda_b^{\!-1/2}
+(f^3)\Lambda_b^{\!-3/2}\dots.
\end{eqnarray}
Here the notation $(f^3)$ and $(f^4)$ refers to terms like
$f_{\nu\alpha}f^\alpha{_\sigma}f^\sigma{_\mu}$ and
$f_{\nu\alpha}f^\alpha{_\sigma}f^\sigma{_\rho}f^\rho{_\mu}$.
Let us consider the size of these higher order terms relative to the leading order term
for worst-case fields accessible to measurement.
%The connection equations (\ref{JSconnection}) describe an implicit algebraic dependence
%of $\tGam^\alpha_{\nu\beta}$ on $N_{\nu\mu}$ and $N_{\nu\mu,\beta}$.
%The solution for $\tGam^\alpha_{\nu\beta}(N_{..})$ yields the Christoffel
%connection (\ref{Christoffel}) in the symmetric case.
%From \cite{Shifflett} we know there is an exact electric monopole solution
%for this theory which approximates a $f^1{_0}\!\sim\!Q/r^2$ field.
%For a solar mass extremal charged black hole with $r\!=\!Q\!=\!M_\odot=1.48\times 10^5cm$ we have
%\begin{eqnarray}
%\label{BHskew}
%|f^1{_0}|_{BH}^2/\Lambda_b\sim
%(Q/r^2)^2/\Lambda_b\sim 10^{-76}
%\end{eqnarray}
In geometrized units an elementary charge has
\begin{eqnarray}
\label{redef}
\fl~~~~~~~~~Q_e=e\sqrt{\frac{G}{c^4}}=\sqrt{\frac{e^2}{\hbar c}\frac{G\hbar}{c^3}}=\sqrt{\alpha}\,l_P=1.38\times 10^{-34}cm
\end{eqnarray}
where $\alpha =e^2/\hbar c$ is the fine structure constant
and $l_P\!=\!\sqrt{G\hbar/c^3}$ is Planck's constant.
If we assume that charged particles retain $f^1{_0}\!\sim\!Q/r^2$
down to the smallest radii probed by high energy particle physics
experiments ($10^{-17}{\rm cm}$) we have from (\ref{redef},\ref{Lambdab}),
\begin{eqnarray}
\label{highenergyskew}
|f^1{_0}|^2/\Lambda_b\sim (Q_e/(10^{-17})^2)^2/\Lambda_b\sim 10^{-66}.
\end{eqnarray}
Here $|f^1{_0}|$ is assumed to be in some
standard spherical or cartesian coordinate system. If an equation has a tensor term which can
be neglected in one coordinate system, it can be neglected in any coordinate system,
so it is only necessary to prove it in one coordinate system.
The fields at $10^{-17}{\rm cm}$ from an elementary charge
would be larger than near any macroscopic charged object,
and would also be larger than the strongest plane-wave fields.
Therefore the higher order terms in (\ref{approximateNbar}-\ref{approximateNhat})
must be $<\!10^{-66}$ of the leading order terms, so they will be completely negligible for most purposes.
%The ``smallness'' of $N_{[\mu\nu]}$ is discussed in
%\S\ref{SymmetricPart}, first paragraph, and in \ref{Monopole}, \S 4, last paragraph.

In \S\ref{Connections} we will calculate the connection equations resulting
from $\delta{\mathcal L}/\delta\tGam^\alpha_{\nu\mu}\!=\!0$. Solving these equations gives
(\ref{upsilonsymmetric},\ref{upsilonantisymmetric},\ref{tGminusG},\ref{antisymmetricpreliminary}),
which can be abbreviated as
\begin{eqnarray}
\fl~~~~~\tGam^\alpha_{(\nu\mu)}&=&\Gamma^\alpha_{\nu\mu}+\ord(\Lambda_b^{\!-1}),
%+({\rm terms~like}~
%f^{\alpha\tau}\!f_{\tau(\nu;\mu)}\Lambda_b^{\!-1}~{\rm and}~
%j^\rho f^\alpha{_\rho}\,g_{\nu\mu}\Lambda_b^{\!-1}),\\
\,~~~~~~\tGam^\alpha_{[\nu\mu]}=\ord(\Lambda_b^{\!-1/2}),\\
%({\rm terms~like}~
%f_{\nu\mu;}{^\alpha}i\Lambda_b^{\!-1/2}~{\rm and}~
%j_{[\nu}\delta^\alpha_{\mu]}i\Lambda_b^{\!-1/2}),\\
\label{tGapproxG}
\fl~~~~~\tG_{\nu\mu}&=&G_{\nu\mu}\!+\ord(\Lambda_b^{\!-1}),
%+({\rm terms~like}~
%f^\sigma{_{\nu;\alpha}}f^\alpha{_{\mu;\sigma}}\Lambda_b^{\!-1},~
%f^{\alpha\tau}f_{\tau(\nu;\,\mu)}{_{;\alpha}}\Lambda_b^{\!-1}~{\rm and}~
%j_\nu j_\mu\Lambda_b^{\!-1}),\\
\label{tRapprox0}
~~~~~~\tR_{[\nu\mu]}=\ord(\Lambda_b^{\!-1/2}),
%({\rm terms~like}~
%f_{[\nu\mu,\alpha];}{^\alpha}i\Lambda_b^{\!-1/2},~
%f^\alpha{_{[\nu;[\mu];\alpha]}}i\Lambda_b^{\!-1/2}~{\rm and}~
%j_{[\nu,\mu]}i\Lambda_b^{\!-1/2}),
\end{eqnarray}
where $\Gamma^\alpha_{\nu\mu}$ is the Christoffel connection (\ref{Christoffel}),
$\tR_{\nu\mu}\!=\!{\mathcal R}_{\nu\mu}(\tGam)$, $R_{\nu\mu}\!=\!R_{\nu\mu}(\Gamma)$ and
\begin{eqnarray}
\label{genEinstein}
\fl ~~~~~\tG_{\nu\mu}&=&\tR_{(\nu\mu)}-\frac{\lower1pt\hbox{$1$}}{2}\,g_{\nu\mu}\tR^\rho_\rho,
~~~~~~~G_{\nu\mu}=R_{\nu\mu}-\frac{\lower1pt\hbox{$1$}}{2}\,g_{\nu\mu}R.
\end{eqnarray}
In (\ref{tGapproxG}), $\ord(\Lambda_b^{\!-1})$ and $\ord(\Lambda_b^{\!-1/2})$ indicate terms like
$f^\sigma{_{\nu;\alpha}}f^\alpha{_{\mu;\sigma}}\Lambda_b^{\!-1}$
and $f_{[\nu\mu,\alpha];}{^\alpha}\Lambda_b^{\!-1/2}$.
%Here we are using the notation $(f^{2\prime})\!=\!{\rm terms~like~} f^{\alpha\tau}\!f_{\tau(\nu;\mu)}$,
%$(f^{\prime})\!=\!{\rm terms~like~} f_{\nu\mu;}{^\alpha}$,
%$(f^{2\prime\prime})\!=\!{\rm terms~like~} f^\sigma{_{\nu;\alpha}}f^\alpha{_{\mu;\sigma}}$
%and $(f^{\prime\prime})\!=\!{\rm terms~like~} f^\alpha{_{[\nu;[\mu];\alpha]}}$.

From the antisymmetric part of the field equations (\ref{JSantisymmetric})
and (\ref{approximateNhat},\ref{tRapprox0}) we get
\begin{eqnarray}
\label{fapproxF}
\fl~~~~~~f_{\nu\mu}&=&F_{\nu\mu}+\ord(\Lambda_b^{\!-1}).
\end{eqnarray}
So $f_{\nu\mu}$ and $F_{\nu\mu}$ only differ by terms with $\Lambda_b$ in the denominator,
and the two become identical in the limit as $\Lambda_b\!\rightarrow\!\infty$.
Combining the symmetric part of the field equations (\ref{JSsymmetric}) with its contraction,
and substituting (\ref{genEinstein},\ref{approximateNbar},\ref{Lambdadef})
\ifnum\ExpandDerivations=1
\begin{eqnarray}
\fl N_{(\nu\mu)}
-\frac{\lower0pt\hbox{$1$}}{2}\,g_{\nu\mu}N^\rho_\rho
\!&=&g_{\nu\mu}-2\left({f_\nu}^\sigma f_{\sigma\mu}
-\frac{1}{2(n\!-\!2)}g_{\nu\mu}f^{\rho\sigma}\!f_{\sigma\rho}\right)\!\Lambda_b^{\!-1}\nonumber\\
\fl&&-\frac{1}{2}g_{\nu\mu}n+g_{\nu\mu}\!\left(f^{\rho\sigma}\!f_{\sigma\rho}
-\frac{1}{2(n\!-\!2)}nf^{\rho\sigma}\!f_{\sigma\rho}\right)\!\Lambda_b^{\!-1}
+(f^4)\Lambda_b^{\!-2}\dots\nonumber\\
\!&=&g_{\nu\mu}\left(1-\frac{n}{2}\right)
-2{f_\nu}^\sigma f_{\sigma\mu}\Lambda_b^{\!-1}\nonumber\\
\fl&&+g_{\nu\mu}\!\left(\frac{1}{(n\!-\!2)}+\!1
\!-\frac{n}{2(n\!-\!2)}\right)f^{\rho\sigma}\!f_{\sigma\rho}\Lambda_b^{\!-1}
+(f^4)\Lambda_b^{\!-2}\dots\nonumber\\
\!&=&-2\left({f_\nu}^\sigma f_{\sigma\mu}
-\frac{1}{4}g_{\nu\mu}f^{\rho\sigma}\!f_{\sigma\rho}\right)\!\Lambda_b^{\!-1}
-\left(\frac{n}{2}-1\right)g_{\nu\mu}
+(f^4)\Lambda_b^{\!-2}\dots\nonumber
\end{eqnarray}
\fi
gives the Einstein equations
\begin{eqnarray}
\label{Einstein}
\fl~~~~~~\tG_{\nu\mu}&=&8\pi T_{\nu\mu}
-\Lambda_b\!\left(N_{(\nu\mu)}
\!-\!\frac{\lower0pt\hbox{1}}{2}g_{\nu\mu}N^\rho_\rho\right)
\!+\!\Lambda_z\!\left(\frac{n}{2}-1\right)g_{\nu\mu},\\
%\end{eqnarray}
%\begin{eqnarray}
\label{Einstein2}
%\fl~~~~~~\tG_{\nu\mu}
\fl &=&8\pi T_{\nu\mu}
+2\left({f_\nu}^\sigma\! f_{\sigma\mu}
\!-\!\frac{\lower0pt\hbox{$1$}}{4}g_{\nu\mu}f^{\rho\sigma}\!f_{\sigma\rho}\right)
+\Lambda\left(\frac{n}{2}-1\right)g_{\nu\mu}+(f^4)\Lambda_b^{\!-1}\dots.
\end{eqnarray}
So from (\ref{fapproxF},\ref{tGapproxG}), equation (\ref{Einstein2}) differs from the
ordinary Einstein equations only by terms with $\Lambda_b$ in the denominator,
and it becomes identical to the ordinary Einstein equations in the limit as $\Lambda_b\!\rightarrow\!\infty$.
In \S\ref{Connections} we will examine how close the approximation is for the very large value
$\Lambda_b\!\sim\!10^{66}cm^{-2}$ from (\ref{Lambdab}).

From (\ref{Tdef}) we see that $S_{\nu\mu}$ and $T_{\nu\mu}$ are different for each ${\mathcal L}_m$ case.
For the electro-vac case
\begin{eqnarray}
\fl~~~~~~S_{\nu\mu}=0,\\
\fl~~~~~~T_{\nu\mu}=0.
\end{eqnarray}
For the classical hydrodynamics case (\ref{Lhydrodynamic}),
\begin{eqnarray}
\fl~~~~~~S_{\nu\mu}&=&\frac{\boldsymbol\mu}{\rmg}\left( u_\nu u_\mu-\frac{\lower1pt\hbox{$1$}}{(n-2)}g_{\nu\mu}u^\alpha u_\alpha\right),\\
\label{Thydrodynamic}
\fl~~~~~~T_{\nu\mu}&=&\frac{\boldsymbol\mu}{\rmg}\,u_\nu u_\mu.
\end{eqnarray}
For the spin-0 case (\ref{LKleinGordon}) as in \cite{Birrell},
\begin{eqnarray}
\fl~~~~~~S_{\nu\mu}
\!&=&\!\frac{1}{m}\left(\hbar^2\bar\psi\Ddag\!_{(\nu} D_{\mu)}\psi-\frac{\lower1pt\hbox{$1$}}{(n\!-\!2)}g_{\nu\mu}m^2\bar\psi\psi\right),\\
\label{TKleinGordon}
\fl~~~~~~T_{\nu\mu}
\!&=&\!\frac{1}{m}\left(\hbar^2\bar\psi\Ddag\!_{(\nu} D_{\mu)}\psi-\frac{\lower2pt\hbox{$1$}}{2}g_{\nu\mu}(\hbar^2\bar\psi\Ddag\!_\sigma D^\sigma\psi\!-\!m^2\bar\psi\psi)\right).
\end{eqnarray}
For the spin-1/2 case (\ref{LDirac}) as in \cite{Birrell},
\begin{eqnarray}
\fl~~~~~~S_{\nu\mu}
&=&\frac{i\hbar}{2}\left(\bar\psi\gamma_{(\nu} D_{\mu)}\psi
-\bar\psi\Ddag\!_{(\mu} \gamma_{\nu)}\psi
-\frac{\lower1pt\hbox{$1$}}{(n\!-\!2)}g_{\nu\mu}(\bar\psi\gamma^{\sigma} D_{\sigma}\psi
-\bar\psi\Ddag\!_{\sigma} \gamma^{\sigma}\psi)\right),\\
\label{TDirac}
\fl~~~~~~T_{\nu\mu}
&=&\frac{i\hbar}{2}\left(\bar\psi\gamma_{(\nu} D_{\mu)}\psi
-\bar\psi\Ddag\!_{(\mu} \gamma_{\nu)}\psi\right).
\end{eqnarray}
Note that in the purely classical limit as $i\hbar D_\sigma\psi\!\rightarrow\!p_\sigma\psi$~,
$-i\hbar \bar\psi\Ddag\!_\sigma\!\rightarrow\!\bar\psi p_\sigma$, the energy-momentum tensors
(\ref{TKleinGordon}) for spin-0 and (\ref{TDirac}) for spin-1/2 both go to
the classical hydrodynamics case (\ref{Thydrodynamic}).

\section{\label{MaxwellEquations}Maxwell's Equations}
%Now let us derive the field equations resulting from the Lagrangian density (\ref{JSlag3}).
%Setting to zero the variational derivative of
%(\ref{JSlag3}) with respect to $\nGam^\alpha_{\nu\mu}$ will give the
%field equations (\ref{JSconnection},\ref{JScontractionsymmetric}),
%and Ampere's law can be derived from these. However,
%as discussed previously, the same field equations must result if we instead
%use $\tGam^\alpha_{\nu\mu}$ and $\Aphi_\tau$ as the independent variables,
%and since this is simpler we will follow this method.
Setting $\delta{\mathcal L}/\delta\Aphi_\tau\!=0$
and using the definition (\ref{fdef}) gives
\begin{eqnarray}
\fl~~~~~~~~~~~~~~0&=&\!\frac{4\pi}{\rmg}\!\left[\frac{\partial {\mathcal L}}{\partial \Aphi_\tau}
-\left(\frac{\partial {\mathcal L}}{\partial \Aphi_{\tau,\omega}}\right)\!{_{,\,\omega}}\right]\!\\
\label{Ampere0}
\fl &=&\frac{\rmt\Lambda_b^{\!1/2}}{2\rmg}\,(\rmN N^{\dashv[\omega\tau]})_{,\,\omega}-4\pi j^\tau
=\frac{(\rmg f^{\omega\tau})_{,\,\omega}}{\rmg}-4\pi j^\tau.
\end{eqnarray}
where
\begin{eqnarray}
\label{jdef}
\fl~~~~~~~~~~~~~~j^\tau&=&\!\frac{-1}{\rmg}\!\left[\frac{\partial {\mathcal L}_m}{\partial \Aphi_\tau}
-\left(\frac{\partial {\mathcal L}_m}{\partial \Aphi_{\tau,\omega}}\right)\!{_{,\,\omega}}\right].
\end{eqnarray}
From (\ref{Ampere0},\ref{Fdef}) we get Maxwell's equations,
\begin{eqnarray}
\label{Ampere}
{f^{\omega\tau}}_{;\,\omega}&=&4\pi j^\tau,\\
\label{Faraday}
F_{[\nu\mu,\alpha]}&=&0.
\end{eqnarray}
%where we define
%\begin{eqnarray}
%\label{fdef}
%f^{\rho\tau}=-\frac{\rmN}{\rmg} N^{\dashv[\rho\tau]}.
%\end{eqnarray}
Using $f_{\nu\mu}\!=\!F_{\nu\mu}\!+\!\ord(\Lambda_b^{\!-1})$ from (\ref{fapproxF}),
we see that equations (\ref{Ampere},\ref{Faraday}) differ from the ordinary Maxwell equations only by
terms with $\Lambda_b$ in the denominator, and these equations become identical to the ordinary Maxwell
equations in the limit as $\Lambda_b\!\rightarrow\!\infty$.
In \S\ref{Connections} we will examine how close the approximation is
for the very large value $\Lambda_b\!\sim\!10^{66}cm^{-2}$ from (\ref{Lambdab}).

From (\ref{jdef}) we see that $j^\tau$ is different for each ${\mathcal L}_m$ case.
For the electro-vac case, $j^\alpha$ may have a singularity at the position of a
particle, but otherwise we have,
\begin{eqnarray}
j^\alpha&=&0.
%Q\,u_\circ^\alpha\delta(x^\alpha\!-\!x_\circ^\alpha).
\end{eqnarray}
For the classical hydrodynamics case (\ref{Lhydrodynamic}),
\begin{eqnarray}
\label{uj}
j^\alpha&=&\frac{\boldsymbol\mu Q}{m\rmg} u^\alpha.
\end{eqnarray}
For the spin-0 case (\ref{LKleinGordon}) as in \cite{Birrell},
\begin{eqnarray}
\label{jKleinGordon}
j^\alpha
&=&\frac{i\hbar Q}{2m}(\bar\psi D^\alpha\psi-\bar\psi\Ddag{^\alpha}\psi).
\end{eqnarray}
For the spin-1/2 case (\ref{LDirac}) as in \cite{Birrell},
\begin{eqnarray}
\label{jDirac}
j^\alpha&=&Q\bar\psi\gamma^\alpha\psi.
\end{eqnarray}
A continuity equation follows from (\ref{Ampere}) regardless of the type of source,
\begin{eqnarray}
\label{continuity}
j^\rho{_{\!;\rho}}&=&\frac{1}{4\pi}f^{\tau\rho}{_{;[\tau;\rho]}}=0.
\end{eqnarray}
Note that the covariant derivative in (\ref{Ampere},\ref{continuity}) is done using the
Christoffel connection (\ref{Christoffel}) formed from the symmetric metric (\ref{gdef}).

\section{\label{Connections}The Connection Equations}
Setting $\delta{\mathcal L}/\delta\tGam^\alpha_{\nu\mu}\!=0$
requires some preliminary calculations. With the definition
\begin{eqnarray}
\frac{\Delta{\mathcal L}}{\Delta\tGam^\beta_{\tau\rho}}
=\frac{\partial{\mathcal L}}{\partial\tGam^\beta_{\tau\rho}}
-\left(\frac{\partial{\mathcal L}}
{\partial\tGam^\beta_{\tau\rho,\omega}}\right){_{\!,\,\omega}}~...
\end{eqnarray}
and (\ref{JSlag3},\ref{HermitianizedRiccit}) we can calculate,
\begin{eqnarray}
\fl-16\pi\frac{\Delta {\mathcal L}}
{\Delta\tGam^\beta_{\tau\rho}}&=&
\ifnum\ExpandDerivations=1
2\rmN N^{\dashv\mu\nu}
(\delta^\sigma_\beta\delta^\tau_\nu
\delta^\rho_{[\mu|}\tGam^\alpha_{\sigma|\alpha]}
+\tGam^\sigma_{\nu [\mu|}
\delta^\alpha_\beta\delta^\tau_\sigma\delta^\rho_{|\alpha]})\nonumber\\
\fl&&-2(\rmN N^{\dashv\mu\nu}\delta^\alpha_\beta\delta^\tau_\nu
\delta^\rho_{[\mu}\delta^\omega_{\alpha]}){_{\!,\,\omega}}
-(\rmN N^{\dashv\mu\nu}\delta^\alpha_\beta\delta^\tau_\alpha\delta^\rho_{[\nu}\delta^\omega_{\mu]})_{,\omega}\nonumber\\
\fl &=&
\fi
\label{semivarder}
-(\rmN N^{\dashv\rho\tau})_{\!,\,\beta}
-\tGam^\rho_{\beta\mu}\rmN N^{\dashv\mu\tau}
-\tGam^\tau_{\nu\beta}\rmN N^{\dashv\rho\nu}
+\tGam^\alpha_{\beta\alpha}\rmN N^{\dashv\rho\tau}\nonumber\\
\fl&&+\delta^\rho_\beta((\rmN N^{\dashv\,\omega\tau})_{\!,\,\omega}
+\tGam^\tau_{\nu\mu}\rmN N^{\dashv\mu\nu})
+\delta^\tau_\beta(\rmN N^{\dashv[\rho\omega]})_{,\omega},\\
\fl-16\pi\frac{\Delta{\mathcal L}}
{\Delta\tGam^\alpha_{\alpha\rho}}
&=&(n\!-\!2)(\rmN N^{\dashv[\rho\omega]})_{\!,\,\omega},\\
\fl-16\pi\frac{\Delta{\mathcal L}}
{\Delta\tGam^\alpha_{\tau\alpha}}
&=&(n\!-\!1)
((\rmN N^{\dashv\,\omega\tau})_{\!,\,\omega}
+\tGam^\tau_{\nu\mu}
\rmN N^{\dashv\mu\nu})
+(\rmN N^{\dashv[\tau\omega]})_{\!,\,\omega}.
\end{eqnarray}
In these last two equations, the index contractions occur after
the derivatives. At this point we must be careful.
Because ${\tGam^\alpha_{\nu\mu}}$ has the symmetry
(\ref{JScontractionsymmetric}), it has
only $n^3\!-n$ independent components, so there can only be $n^3\!-n$
independent field equations associated with it. It is shown in
%Appendix C of \cite{Shifflett}
\ref{VariationalDerivative}
that instead of just setting (\ref{semivarder}) to zero,
the field equations associated with such a field are given by the expression,
%Now, (\ref{semivarder}) is not the variational
%derivative of ${\mathcal L}$ because of the symmetry properties of
%${\tGam^\alpha_{\nu\mu}}$. In \cite{Shifflett}
%we show how to calculate the variational derivative with respect to a
%contraction-symmetric affinity. Setting the variational derivative
%to zero and using equation (C.7) of \cite{Shifflett} along with
%(\ref{Ampere}) gives the field equations,
\begin{eqnarray}
\fl0&=&16\pi\left[\frac{\Delta{\mathcal L}}{\Delta\!\tGam^\beta_{\tau\rho}}
\!-\!\frac{\delta^\tau_\beta}{(n\!-\!1)}
\frac{\Delta {\mathcal L}}{\Delta\!\tGam^\alpha_{\alpha\rho}}
\!-\!\frac{\delta^\rho_\beta}{(n\!-\!1)}\frac{\Delta
{\mathcal L}}{\Delta\!\tGam^\alpha_{\tau\alpha}}\right]\\
\ifnum\ExpandDerivations=1
\fl&=&(\rmN N^{\dashv\rho\tau})_{\!,\,\beta}
+\tGam^\rho_{\beta\mu}\rmN N^{\dashv\mu\tau}
+\tGam^\tau_{\nu\beta}
\rmN N^{\dashv\rho\nu}-\tGam^\alpha_{\beta\alpha}
\rmN N^{\dashv\rho\tau}\nonumber\\
\fl&&\!\!-\delta^\tau_\beta(\rmN N^{\dashv[\rho\omega]})_{,\omega}
\!+\!\frac{1}{(n\!-\!1)}((n\!-\!2)\delta^\tau_\beta(\rmN N^{\dashv[\rho\omega]})_{,\omega}
\!+\!\delta^\rho_\beta(\rmN N^{\dashv[\tau\omega]})_{,\omega})\nonumber\\
\fl&=&(\rmN N^{\dashv\rho\tau})_{\!,\,\beta}
+\tGam^\tau_{\nu\beta}\rmN N^{\dashv\rho\nu}
+\tGam^\rho_{\beta\mu}\rmN N^{\dashv\mu\tau}
-\tGam^\alpha_{\beta\alpha}\rmN N^{\dashv\rho\tau}\nonumber\\
\fl&&~~~~~~~~~~~~~~~~~~~~~~~~~~~~~~~
-\frac{1}{(n\!-\!1)}(\delta^\tau_\beta(\rmN N^{\dashv[\rho\omega]})_{,\omega}
-\delta^\rho_\beta(\rmN N^{\dashv[\tau\omega]})_{,\omega})\nonumber\\
\fi
\label{JSconnection}
\fl&=&(\rmN N^{\dashv\rho\tau})_{\!,\,\beta}
+\tGam^\tau_{\sigma\beta}\rmN N^{\dashv\rho\sigma}
+\tGam^\rho_{\beta\sigma}\rmN N^{\dashv\sigma\tau}
-\tGam^\alpha_{\beta\alpha}\rmN N^{\dashv\rho\tau}\nonumber\\
\fl&&~~~~~~~~~~~~~~~~~~~~~~~~~~~~~~~~~~~~~~~~~~~~~~~~~~~~~~~
-\frac{8\pi\rmt}{(n\!-\!1)\Lambda_b^{1/2}}\rmg j^{[\rho}\delta^{\tau]}_\beta.
\end{eqnarray}
These are the connection equations, like $(\rmg g^{\rho\tau}){_{;\beta}}\!=\!0$
in the symmetric case.

%which are equivalent to the uglier covariant form (\ref{JSconnection}).
From the definition of matrix inverse
$N^{\dashv\rho\tau}\!=(1/N)\partial N/\partial N_{\tau\rho}$,~
$N^{\dashv\rho\tau}N_{\tau\mu}\!=\nobreak\!\delta^\rho_\mu$ we get the identity
\begin{eqnarray}
\label{sqrtdetcomma}
\fl~~~~~(\!\rmN\,)_{,\sigma}
=\frac{\partial\rmN}
{\partial N_{\tau\rho}}N_{\tau\rho,\sigma}
=\frac{\rmN}{2}N^{\dashv\rho\tau}N_{\tau\rho,\sigma}
\label{sqrtdetcomma2}
=-\frac{\rmN}{2}
{N^{\dashv\rho\tau}}_{,\sigma} N_{\tau\rho}.
\end{eqnarray}
Contracting (\ref{JSconnection}) with $N_{\tau\rho}$
using (\ref{JScontractionsymmetric},\ref{sqrtdetcomma})
\ifnum\ExpandDerivations=1
\begin{eqnarray}
\fl~~~~~0=(n\!-\!2)((\rmN\,)_{,\,\beta}
-\tGam^\alpha_{\alpha\beta}\rmN\,)
+\frac{8\pi}{(n\!-\!1)}\rmg j^\rho N_{[\rho\beta]}\rmt\Lambda_b^{\!-1/2},\nonumber
\end{eqnarray}
\fi
and dividing this by $(n\!-\!2)$ gives,
\begin{eqnarray}
\label{der0}
\fl~~~~~(\rmN\,)_{,\,\beta}-\tGam^\alpha_{\alpha\beta}\rmN
=-\frac{8\pi\rmt}{(n\!-\!1)(n\!-\!2)\Lambda_b^{1/2}}\rmg j^\rho N_{[\rho\beta]}.
%\fl~~~~~\tGam^\alpha_{\alpha\nu}
%&=&\frac{(\rmN\,)_{,\nu}}{\rmN}
%+\frac{8\pi}{(n\!-\!1)(n\!-\!2)}
%\!\frac{\rmg}{\rmN}j^\rho N_{[\rho\nu]}.
\end{eqnarray}
%This will be important in showing that the Lagrangian density (\ref{JSlag3})
%transplants into itself when ${\mathcal L}_m=0,~\Lambda_z=0$.
From (\ref{der0}) we get
\begin{eqnarray}
\label{RisB}
\fl~~~~~\tGam^\alpha_{\alpha[\nu,\mu]}
-\frac{8\pi\rmt}{(n\!-\!1)(n\!-\!2)\Lambda_b^{1/2}}
\left(\!\frac{\rmg}{\rmN}j^\rho N_{[\rho[\nu]}\!\right)\!\!{_{,\mu]}}
=(ln\rmN\,)_{,[\nu,\mu]}=0.
\end{eqnarray}
%Because of (\ref{RisB}), the $\tGam^\alpha_{\nu[\alpha,\mu]}$ term in
%(\ref{HermitianizedRicci}) only affects the field equations
%if ${\mathcal L}_m$ contains $A_\mu$, that is if $j^\alpha\!\ne\!0$.
From (\ref{JSconnection},\ref{der0}) we get the contravariant connection equations,
\begin{eqnarray}
\label{contravariant}
\fl {N^{\dashv\rho\tau}}_{,\beta}
\!+\!\tGam^\tau_{\sigma\beta}N^{\dashv\rho\sigma}
\!+\!\tGam^\rho_{\beta\sigma}N^{\dashv\sigma\tau}
\!=\!\frac{8\pi\rmt}{(n\!-\!1)\Lambda_b^{\!1/2}}\frac{\rmg}{\rmN}\!\left(
j^{[\rho}\delta^{\tau]}_\beta
\!+\!\frac{1}{(n\!-\!2)}j^\alpha N_{[\alpha\beta]}N^{\dashv\rho\tau}\!\right)\!.
\end{eqnarray}
Multiplying this by
$-N_{\nu\rho}N_{\tau\mu}$ gives the covariant connection equations,
\begin{eqnarray}
\label{JSconnection0}
\fl N_{\nu\mu,\beta}\!-\!\tGam^\alpha_{\nu\beta}N_{\alpha\mu}
\!-\!\tGam^\alpha_{\beta\mu}N_{\nu\alpha}
\!=\!\frac{-8\pi\rmt}{(n\!-\!1)\Lambda_b^{\!1/2}}\frac{\rmg}{\rmN}\!\left(
N_{\nu[\alpha}N_{\beta]\mu}
\!+\!\frac{1}{(n\!-\!2)} N_{[\alpha\beta]}N_{\nu\mu}\!\right)\!j^\alpha.
\end{eqnarray}
Equation (\ref{JSconnection0}) together with (\ref{JSsymmetric},\ref{JScurl},\ref{JScontractionsymmetric})
are often used to define the Einstein-Schr\"{o}dinger theory,
particularly when $T_{\nu\mu}\!=\!0$, $j^\alpha\!=\!0$.

The connection equations (\ref{JSconnection}) can be solved
similar to the way that $g_{\rho\tau;\beta}\!=\nobreak\!0$ is solved to get the
Christoffel connection\cite{Tonnelat,Hlavaty,Shifflett2}.
An expansion in powers of $\Lambda_b^{\!-1}$ is derived in \ref{ApproximateGamma},
confirmed by tetrad methods in \cite{Shifflett2}, and is also stated without derivation in \cite{Antoci3},
\begin{eqnarray}
\label{gammadecomposition}
\fl\tGam^\alpha_{\nu\mu}&\!\!=\!&\Gamma^\alpha_{\nu\mu}
+\Upsilon^\alpha_{\nu\mu},\\
\fl \Upsilon^\alpha_{\!(\nu\mu)}
\label{upsilonsymmetric}
&\!\!=\!&\!-2\left[f^\tau{_{\!\!(\nu}}f_{\mu)}{\!^\alpha}{_{\!;\tau}}
\!+f^{\alpha\tau}\!f_{\tau(\nu;\mu)}
\!+\!\frac{1}{4(n\!-\!2)}((f^{\rho\sigma}\!f_{\sigma\rho})_,{^\alpha}g_{\nu\mu}
\!-2(f^{\rho\sigma}\!f_{\sigma\rho})_{,(\nu}\delta^\alpha_{\mu)})\right.\nonumber\\
\fl&&~~~~~~\left.+\frac{4\pi}{(n\!-\!2)}j^\rho\left(f^\alpha{_\rho}\,g_{\nu\mu}
+\frac{2}{(n\!-\!1)}f_{\rho(\nu}\delta^\alpha_{\mu)}\right)\right]\!\Lambda_b^{\!-1}
+(f^{4\prime})\Lambda_b^{\!-2}\dots,\\
\label{upsilonantisymmetric}
\fl \Upsilon^\alpha_{\![\nu\mu]}
&\!\!=\!&\!\left[\frac{1}{2}(f_{\nu\mu;}{^\alpha}+f^\alpha{_{\mu;\nu}}-f^\alpha{_{\nu;\mu}})
+\frac{8\pi}{(n\!-\!1)}j_{[\nu}\delta^\alpha_{\mu]}\right]\rmt\Lambda_b^{\!-1/2}
+(f^{3\prime})\Lambda_b^{\!-3/2}\dots\,,\\
\label{upsiloncontracted}
\fl \Upsilon^\alpha_{\alpha\nu}
&\!=&2\left[\frac{1}{2(n\!-\!2)}(f^{\rho\sigma}\!f_{\sigma\rho})_{,\nu}
-\frac{8\pi}{(n\!-\!1)(n\!-\!2)}j^\alpha f_{\alpha\nu}\right]\!\Lambda_b^{\!-1}
+(f^{4\prime})\Lambda_b^{\!-2}\dots\,.
\end{eqnarray}
In (\ref{gammadecomposition}), $\Gamma^\alpha_{\nu\mu}$ is the Christoffel
connection (\ref{Christoffel}).
The notation $(f^{3\prime})$ and $(f^{4\prime})$ refers to terms like
or $f^\alpha{_\tau}f^\tau{_\sigma}f^\sigma{_{[\nu;\mu]}}$ and
$f^\alpha{_\tau}f^\tau{_\sigma}f^\sigma{_\rho}f^\rho{_{(\nu;\mu)}}$.
%and $j^\tau$ can be considered as a substitute for $(c/4\pi){f^{\omega\tau}}_{;\,\omega}$
%because of Ampere's law (\ref{Ampere}).
As in (\ref{approximateNbar},\ref{approximateNhat}), we see from (\ref{highenergyskew}) that
the higher order terms in (\ref{upsilonsymmetric}-\ref{upsiloncontracted})
must be $<\!10^{-66}$ of the leading order terms, so they will be completely negligible for most purposes.

%Extracting $\Upsilon^\tau_{\sigma\beta}$ of (\ref{gammadecomposition}) from
%the Hermitianized Ricci tensor (\ref{HermitianizedRiccit}) gives from \ref{ExtractionofConnectionAddition},
%%\cite{Shifflett},
%\begin{eqnarray}
%%\label{Ricciaddition}
%%\fl~~~\tR_{\nu\mu}
%%&=&R_{\nu\mu}+\Upsilon^\alpha_{\nu\mu;\alpha}
%%\!-\Upsilon^\alpha_{\alpha(\nu;\mu)}
%%\!-\Upsilon^\sigma_{\nu\alpha}\Upsilon^\alpha_{\sigma\mu}
%%\!+\Upsilon^\sigma_{\nu\mu}\Upsilon^\alpha_{\sigma\alpha},\\
%\label{Ricciadditionsymmetric}
%\fl~~~\tR_{(\nu\mu)}
%&=&R_{\nu\mu}+\Upsilon^\alpha_{\!(\nu\mu);\alpha}
%\!-\Upsilon^\alpha_{\alpha(\nu;\mu)}
%\!-\Upsilon^\sigma_{\!(\nu\alpha)}\Upsilon^\alpha_{\!(\sigma\mu)}
%\!-\Upsilon^\sigma_{\![\nu\alpha]}\Upsilon^\alpha_{\![\sigma\mu]}
%\!+\Upsilon^\sigma_{\!(\nu\mu)}\Upsilon^\alpha_{\sigma\alpha},\\
%\label{Ricciadditionantisymmetric}
%\fl~~~\tR_{[\nu\mu]}
%&=&\Upsilon^\alpha_{\![\nu\mu];\alpha}
%\!-\Upsilon^\sigma_{\!(\nu\alpha)}\Upsilon^\alpha_{\![\sigma\mu]}
%\!-\Upsilon^\sigma_{\![\nu\alpha]}\Upsilon^\alpha_{\!(\sigma\mu)}
%\!+\Upsilon^\sigma_{\![\nu\mu]}\Upsilon^\alpha_{\sigma\alpha}.
%\end{eqnarray}
%Here the covariant derivative is done using $\Gamma^\alpha_{\nu\mu}$ from (\ref{Christoffel}), and
%$R_{\nu\mu}\!=\!R_{\nu\mu}(\Gamma)$.
Substituting (\ref{gammadecomposition}-\ref{upsiloncontracted},\ref{Ampere})
into (\ref{Ricciadditionsymmetric}),
\ifnum\ExpandDerivations=1
with $\ff=f^{\rho\sigma}\!f_{\sigma\rho}$,
\begin{eqnarray}
\fl \tR_{(\nu\mu)}\!
&=& R_{\nu\mu}+{\Upsilon}^\alpha_{\!(\nu\mu);\,\alpha}
\!-\Upsilon^\alpha_{\alpha(\nu;\mu)}
\!-{\Upsilon}^\sigma_{\![\nu\alpha]}{\Upsilon}^\alpha_{\![\sigma\mu]}\dots\nonumber\\
\fl &=&R_{\nu\mu}
-2\left[\left(f^\tau{_{(\nu}}f_{\mu)}{^\alpha}{_{;\,\tau}}
+f^{\alpha\tau}f_{\tau(\nu;\,\mu)}
+\frac{1}{4(n\!-\!2)}(\ff_,{^\alpha}g_{\nu\mu}
-2\ff_{,(\nu}\delta^\alpha_{\mu)})\right){_{;\,\alpha}}\right.\nonumber\\
\fl &&~~~~~~~~~~~~~~+\frac{4\pi}{(n\!-\!2)}j^\rho{_{;\alpha}}\!\left(f^\alpha{_\rho}g_{\nu\mu}
+\frac{2}{(n\!-\!1)}f_{\rho(\nu}\delta^\alpha_{\mu)}\right)\nonumber\\
\fl &&~~~~~~~~~~~~~~+\frac{4\pi}{(n\!-\!2)}j^\rho\!\left(f^\alpha{_\rho}g_{\nu\mu}
+\frac{2}{(n\!-\!1)}f_{\rho(\nu}\delta^\alpha_{\mu)}\right){_{\!;\alpha}}\nonumber\\
\fl &&~~~~~~~~~~~~~~+\frac{1}{2(n\!-\!2)}\ff_{,(\nu;\,\mu)}
-\frac{8\pi}{(n\!-\!1)(n\!-\!2)}(j^\alpha f_{\alpha(\nu})_{;\mu)}\nonumber\\
\fl &&~~~~~~~~~~~~~~-\frac{1}{4}\left(f_{\nu\alpha;}{^\sigma}\!+\!f^\sigma{_{\alpha;\nu}}
\!-\!f^\sigma{_{\nu;\,\alpha}}+\frac{16\pi}{(n\!-\!1)}j_{[\nu}\delta^\sigma_{\alpha]}\right)\nonumber\\
\fl &&~~~~~~~~~~~~~~~~~\left.\times\left(f_{\sigma\mu;}{^\alpha}\!+\!f^\alpha{_{\mu;\,\sigma}}
\!-\!f^\alpha{_{\sigma;\,\mu}}+\frac{16\pi}{(n\!-\!1)}j_{[\sigma}\delta^\alpha_{\mu]}\right)\right]\!\Lambda_b^{\!-1}\dots\nonumber\\
\label{Riccitensor}
\fl &=&R_{\nu\mu}
-\left[2f^\tau{_{(\nu}}f_{\mu)}{^\alpha}{_{;\tau;\alpha}}
+2f^{\alpha\tau}f_{\tau(\nu;\,\mu)}{_{;\,\alpha}}
+\frac{1}{2(n\!-\!2)}\ff_,{^\alpha}{_{;\alpha}}g_{\nu\mu}\right.\nonumber\\
\fl &&-f^\sigma{_{\nu;\alpha}}f^\alpha{_{\mu;\sigma}}
+f^\sigma{_{\nu;\,\alpha}}f_{\sigma\mu;}{^\alpha}
+\frac{1}{2}f^\sigma{_{\alpha;\nu}}f^\alpha{_{\sigma;\,\mu}}\nonumber\\
\fl &&\left.+8\pi j^\tau f_{\tau(\nu;\mu)}
\!-\!\frac{32\pi^2}{(n\!-\!1)}j_\nu j_\mu
\!+\!\frac{32\pi^2}{(n\!-\!2)}j^\rho j_\rho g_{\nu\mu}
\!+\!\frac{8\pi}{(n\!-\!2)}j^\rho{_{;\alpha}}f^\alpha{_\rho}g_{\nu\mu}\right]\!\Lambda_b^{\!-1}\dots,\nonumber\\
\fl~~\tR^\rho_\rho &=&R
-\left[2f^{\tau\beta}f_{\beta}{^\alpha}{_{;\tau;\alpha}}
+\frac{n}{2(n\!-\!2)}\ff_,{^\alpha}{_{;\alpha}}
-f{^{\sigma\beta}}_{;\alpha}f^\alpha{_{\beta;\sigma}}
+\frac{1}{2}f{^{\sigma\beta}}_{;\,\alpha}f_{\sigma\beta;}{^\alpha}\right.\nonumber\\
\fl &&\left.-8\pi f^{\alpha\tau}j_{\tau;\alpha}
\!-\!32\pi^2\left(1\!+\!\frac{1}{(n\!-\!1)}\!-\!\frac{n}{(n\!-\!2)}\right)j^\rho j_\rho
\!+\!\frac{8\pi n}{(n\!-\!2)}j^\rho{_{;\alpha}}f^\alpha{_\rho}\right]\!\Lambda_b^{\!-1}\dots\nonumber\\
\label{Ricciscalar}
\fl &=&R
+\left[-2f^{\tau\beta}f_{\beta}{^\alpha}{_{;\tau;\alpha}}
-\frac{n}{2(n\!-\!2)}\ff_,{^\alpha}{_{;\alpha}}
-\frac{3}{2}f_{[\sigma\beta;\alpha]}f^{[\sigma\beta}{_;}{^{\alpha]}}\right.\nonumber\\
\fl &&~~~~~~~~~\left.-\frac{32\pi^2n}{(n\!-\!1)(n\!-\!2)}j^\rho j_\rho
-\frac{16\pi}{(n\!-\!2)}f^{\alpha\tau}j_{\tau;\alpha}\right]\!\Lambda_b^{\!-1}\dots\,.\nonumber
\end{eqnarray}
\fi
and using (\ref{genEinstein}) gives
%\bigskip\\
%\bigskip
\begin{eqnarray}
\label{tGminusG}
\fl \tG_{\nu\mu}&=&G_{\nu\mu}
-\left(2f^\tau{_{(\nu}}f_{\mu)}{^\alpha}{_{;\tau;\alpha}}
+2f^{\alpha\tau}f_{\tau(\nu;\,\mu)}{_{;\alpha}}\phantom{\frac{1}{2}}\right.\nonumber\\
\fl &&~~~~~~~~~~~-f^\sigma{_{\nu;\alpha}}f^\alpha{_{\mu;\sigma}}
\!+f^\sigma{_{\nu;\alpha}}f_{\sigma\mu;}{^\alpha}
+\frac{1}{2}f^\sigma{_{\alpha;\nu}}f^\alpha{_{\sigma;\,\mu}}\nonumber\\
\fl &&~~~~~~~~~~~-g_{\nu\mu}f^{\tau\beta}f_{\beta}{^\alpha}{_{;\tau;\alpha}}
\!-\!\frac{1}{4}g_{\nu\mu}(f^{\rho\sigma}\!f_{\sigma\rho})_,{^\alpha}{_{;\alpha}}
-\frac{3}{4}g_{\nu\mu}f_{[\sigma\beta;\alpha]}f^{[\sigma\beta}{_;}{^{\alpha]}}\nonumber\\
\fl &&~~~~~~~~~~~\left.+8\pi j^\tau f_{\tau(\nu;\mu)}
-\frac{32\pi^2}{(n\!-\!1)}j_\nu j_\mu
+\frac{16\pi^2}{(n\!-\!1)}g_{\nu\mu}j^\rho j_\rho\right)\!\Lambda_b^{\!-1}\dots.
\end{eqnarray}
Substituting (\ref{upsilonantisymmetric},\ref{Ampere}) into (\ref{Ricciadditionantisymmetric}) gives
\begin{eqnarray}
\fl\tR_{[\nu\mu]}&=&
\ifnum\ExpandDerivations=1
{\Upsilon}^\alpha_{\![\nu\mu];\alpha}+\ord(\Lambda_b^{\!-3/2})\dots\nonumber\\
\fl &=&\!\left(\frac{1}{2}(f_{\nu\mu;}{^\alpha}
\!+\!f^\alpha{_{\mu;\nu}}\!-\!f^\alpha{_{\nu;\mu}}){_{;\alpha}}
+\frac{8\pi}{(n\!-\!1)}j_{[\nu,\mu]}\right)\!\rmt\Lambda_b^{\!-1/2}\dots\nonumber\\
\fl &=&\!\left(\!\frac{3}{2}f_{[\nu\mu,\alpha];}{^\alpha}
+f^\alpha{_{\mu;\nu;\alpha}}-f^\alpha{_{\nu;\mu;\alpha}}
+\frac{8\pi}{(n\!-\!1)}j_{[\nu,\mu]}\right)\!\rmt\Lambda_b^{\!-1/2}\dots\nonumber\\
\fl &=&
\fi
\label{antisymmetricpreliminary}
\!\left(\!\frac{3}{2}f_{[\nu\mu,\alpha];}{^\alpha}
\!+\!2f^\alpha{_{\mu;[\nu;\alpha]}}\!-\!2f^\alpha{_{\nu;[\mu;\alpha]}}
\!-\!\frac{8\pi(n\!-\!2)}{(n\!-\!1)}\,j_{[\nu,\mu]}\right)\!\rmt\Lambda_b^{\!-1/2}\dots.
\end{eqnarray}

As we have already noted in \S\ref{EinsteinEquations} and \S\ref{MaxwellEquations},
the $\Lambda_b$ in the denominator of (\ref{tGminusG},\ref{antisymmetricpreliminary})
causes our Einstein and Maxwell equations (\ref{Einstein2},\ref{Ampere},\ref{Faraday})
to become the ordinary Einstein and Maxwell equations in the limit as
$\omega_c\!\rightarrow\!\infty$, $|\Lambda_z|\!\rightarrow\!\infty$, $\Lambda_b\!\rightarrow\!\infty$,
and it also causes the relation $f_{\nu\mu}\!\approx\!F_{\nu\mu}$ from (\ref{fapproxF})
to become exact in this limit.
Let us examine how close these approximations are when $\Lambda_b\sim 10^{66}cm^{-2}$ as in (\ref{Lambdab}).

We will start with the Einstein equations (\ref{Einstein2}).
Let us consider worst-case values of $\tG_{\nu\mu}\!-G_{\nu\mu}$
accessible to measurement, and compare these
to the ordinary electromagnetic term in the Einstein equations (\ref{Einstein2}).
%For electric monopole fields, we see from
%(\ref{highenergyskew}-\ref{highenergyderiv2},\ref{Ampere}) that
%the extra terms of (\ref{Einstein4}) must be $<\!10^{-32}$ of the ordinary electromagnetic term.
If we assume that charged particles retain $f^1{_0}\!\sim\!Q/r^2$
down to the smallest radii probed by high energy particle physics
experiments ($10^{-17}{\rm cm}$) we have,
\begin{eqnarray}
\label{highenergyderiv1}
|f^1{_{0;1}}/f^1{_0}|^2/\Lambda_b&\sim& 4/\Lambda_b\,(10^{-17})^2\sim 10^{-32},\\
\label{highenergyderiv2}
|f^1{_{0;1;1}}/f^1{_0}|/\Lambda_b&\sim& 6/\Lambda_b\,(10^{-17})^2\sim 10^{-32}.
\end{eqnarray}
So for electric monopole fields, terms like $f^\sigma{_{\nu;\alpha}}f^\alpha{_{\mu;\sigma}}\Lambda_b^{\!-1}$
and $f^{\alpha\tau}f_{\tau(\nu;\,\mu)}{_{;\alpha}}\Lambda_b^{\!-1}$ in (\ref{tGminusG})
must be $<\!10^{-32}$ of the ordinary electromagnetic term in (\ref{Einstein2}).
And regarding $j^\tau$ as a substitute for $(1/4\pi){f^{\omega\tau}}_{;\,\omega}$ from (\ref{Ampere}),
the same is true for the $j_\nu$ terms.
For an electromagnetic plane-wave in a flat background space we have $j^\sigma=0$ and
\begin{eqnarray}
\label{planewaveA}
A_\mu&=&A\epsilon_\mu{\rm sin}(k_\alpha x^\alpha)
~~,~~\epsilon^\alpha\epsilon_\alpha=-1
~~,~~k^\alpha k_\alpha=k^\alpha\epsilon_\alpha=0,\\
\label{planewavef}
f_{\nu\mu}&=&2\Aphi_{[\mu,\nu]}
=2A\epsilon_{[\mu} k_{\nu]}{\rm cos}(k_\alpha x^\alpha).
\end{eqnarray}
Here $A$ is the magnitude, $k^\alpha$ is the frequency, and
$\epsilon^\alpha$ is the polarization.
Substituting (\ref{planewaveA},\ref{planewavef}) into (\ref{tGminusG}),
all of the terms vanish for a flat background space.
%In this case the additional terms of (\ref{Einstein4}) are all zero.
%We must also consider rates of change for the highest energy gamma rays ($10^{20}$eV) where
%$k\!=\!E/\hbar c\!=5\times 10^{24}\rm{rad/cm}$ so that,
%Assuming the field is localized to a volume $1/k^3$ and using (\ref{cgs},\ref{Lambdab}) gives
Also, for the highest energy gamma rays known in nature ($10^{20}$eV, $10^{34}$Hz)
we have from (\ref{Lambdab}),
\begin{eqnarray}
\label{gammaderiv1}
|f^1{_{0;1}}/f^1{_0}|^2/\Lambda_b&\sim& (E/\hbar c)^2/\Lambda_b\sim 10^{-16},\\
\label{gammaderiv2}
|f^1{_{0;1;1}}/f^1{_0}|/\Lambda_b&\sim& (E/\hbar c)^2/\Lambda_b\sim 10^{-16}.
\end{eqnarray}
So for electromagnetic plane-wave fields, even if some of the extra terms
in (\ref{Einstein2}) were non-zero because of spatial curvatures, they must still be
$<\!10^{-16}$ of the ordinary electromagnetic term.
%For the purpose of looking at worst-case curvatures, electrons could perhaps be
%considered as tiny Schwarzschild solutions with an effective Schwarzschild radius of
%$r_s=2Gm_e/c^2\!=1.4\nobreak\times\nobreak\!10^{-55}~{\rm cm}$.
%Existing particle accelerators can only create energies high
%enough to probe radii of around $r\!\sim\!10^{-17}~{\rm cm}$, where
%\begin{eqnarray}
%\label{electroncurvature}
%\frac{\tilde C_{trtr}}{\Lambda_b}
%&\sim&\frac{r_s}{\Lambda_br^3}
%=\frac{1}{\Lambda_br^3}\left(\frac{2G m_e}{c^2}\right)\!\sim\!10^{-70}.
%\end{eqnarray}
%Therefore (\ref{Einstein4}) is an extremely close approximation to the Einstein
%equations of ordinary general relativity and electromagnetism.
Therefore even for the most extreme worst-case fields accessible to measurement, the extra terms
in the Einstein equations (\ref{Einstein2}) must all be $<\!10^{-16}$ of the ordinary electromagnetic term.

Now let us look at the approximation $f_{\nu\mu}\!\approx\!F_{\nu\mu}$ from (\ref{fapproxF}).
From the covariant derivative commutation rule,
%the cyclic identity $2R_{\nu[\tau\alpha]\mu}=R_{\nu\mu\alpha\tau}$,
the definition of the Weyl tensor $C_{\nu\mu\alpha\tau}$, and
the Einstein equations $R_{\nu\mu}=-\Lambda g_{\nu\mu}+(f^2)\dots$ from
(\ref{Einstein2}) we get
\begin{eqnarray}
\fl~~~~~~2f^\alpha{_{\nu;[\mu;\alpha]}}&=&
\ifnum\ExpandDerivations=1
R^\tau{_{\nu\mu\alpha}}f^\alpha{_\tau}
+R_\tau{^\alpha}{_{\mu\alpha}}f^\tau{_\nu}
=\frac{1}{2}R_{\nu\mu\alpha\tau}f^{\alpha\tau}
+R^\tau{_\mu}f_{\tau\nu}\nonumber\\
%\fl\!&=&\frac{1}{2}\left(C_{\nu\mu\alpha\tau}+\!\frac{2}{(n\!-\!2)}
%(g_{\nu[\alpha}R_{\tau]\mu}-g_{\mu[\alpha}R_{\tau]\nu})
%-\frac{2}{(n\!-\!1)(n\!-\!2)}Rg_{\nu[\alpha}g_{\tau]\mu}\right)f^{\alpha\tau}\nonumber\\
%\fl &&+R^\tau{_\mu}f_{\tau\nu}\\
\fl\!&=&\frac{1}{2}\left(C_{\nu\mu}{^{\alpha\tau}}
+\!\frac{4}{(n\!-\!2)}\delta^{[\alpha}_{[\nu}R^{\tau]}_{\mu]}
-\frac{2}{(n\!-\!1)(n\!-\!2)}\delta^{[\alpha}_{[\nu}\delta^{\tau]}_{\mu]}R\right)f_{\alpha\tau}
-R^\tau{_\mu}f_{\nu\tau}\nonumber\\
%\fl\!&=&\frac{1}{2}C_{\nu\mu\alpha\tau}f^{\alpha\tau}
%+\frac{2}{(n\!-\!2)}R^\tau{_{[\mu}}f_{\nu]\tau}
%-\frac{1}{(n\!-\!1)(n\!-\!2)}Rf_{\nu\mu}
%+R^\tau{_\mu}f_{\tau\nu}.
\fl &=&
\fi
\label{Fterm0}
\frac{1}{2}f^{\alpha\tau}C_{\alpha\tau\nu\mu}
+\frac{(n\!-\!2)\Lambda}{(n\!-\!1)}f_{\nu\mu}+(f^3)\dots.
%\label{Rantisymmetric}
%\fl\tR_{[\nu\mu]}\!=\!-\!\left(\theta_{[\tau,\alpha]}\varepsilon_{\nu\mu}{^{\tau\alpha}}
%\!+\!f^{\alpha\tau}C_{\alpha\tau\nu\mu}
%\!+\!\frac{2(n\!-\!2)\Lambda}{(n\!-\!1)}f_{\nu\mu}
%\!+\!\frac{8\pi(n\!-\!2)}{(n\!-\!1)}j_{[\nu,\mu]}\!+\!(f^3)\right)\!\frac{\rmt}{\Lambda_b^{\!1/2}}\dots
\end{eqnarray}
Substituting (\ref{approximateNhat})
into the antisymmetric field equations (\ref{JSantisymmetric}) gives
\begin{eqnarray}
\label{antisymmetric0}
f_{\nu\mu}
\!&=&\!F_{\nu\mu}+\tR_{[\nu\mu]}\rmt\Lambda_b^{\!-1/2}/2+(f^3)\Lambda_b^{\!-1}\dots,
\end{eqnarray}
and using (\ref{antisymmetricpreliminary},\ref{Fterm0}) we get
\begin{eqnarray}
\fl f_{\nu\mu}
\label{threeparts}
\!&=&\!F_{\nu\mu}
\!+\!\!\left(\theta_{[\tau,\alpha]}\varepsilon_{\nu\mu}{^{\tau\alpha}}
\!+\!f^{\alpha\tau}C_{\alpha\tau\nu\mu}
\!+\!\frac{2(n\!-\!2)\Lambda}{(n\!-\!1)}f_{\nu\mu}
\!+\!\frac{8\pi(n\!-\!2)}{(n\!-\!1)}j_{[\nu,\mu]}\!+\!(f^3)\right)\!\Lambda_b^{\!-1}\dots
\end{eqnarray}
where
\begin{eqnarray}
\label{thdef}
~~~~~~\theta_\tau &=&\frac{\lower1pt\hbox{$1$}}{4}f_{[\nu\mu,\alpha]}\varepsilon_\tau{^{\nu\mu\alpha}},~~
f_{[\nu\mu,\alpha]}=-\frac{\lower1pt\hbox{$2$}}{3}
\,\theta_\tau\varepsilon^\tau{_{\nu\mu\alpha}},\\
~~\varepsilon_{\tau\nu\mu\alpha}
&=&{(\rm Levi\!-\!Civita~tensor)},\\
%&=&{[\,\nu\mu\tau\tau]}\rmg~~~~~~,~~
%\varepsilon^{\nu\mu\tau\tau}=-[\,\nu\mu\tau\tau]/\rmg~,\\
%{[\,\nu\mu\tau\tau]}
%&=&\left\{\matrix{
%+1~{\rm for~even~permutations~of~0123}\cr
%-1~{\rm for~odd~~permutations~of~0123}\cr
%~0~{\rm for~two~equal~indices~~~~~~~~~~~}
%}\right.\\
\label{Ctilde}
~C_{\alpha\tau\nu\mu}&=&({\rm Weyl~tensor}).
%\tC_{\nu\mu\alpha\tau}&=&R_{\nu\mu\alpha\tau}
%-g_{\nu[\alpha}R_{\tau]\mu}+g_{\mu[\alpha}R_{\tau]\nu},\\
%\label{Fdef}
%~~~~F_{\nu\mu}&=&2A_{[\mu,\nu]}.
\end{eqnarray}
The $\theta_{[\tau,\alpha]}\varepsilon_{\nu\mu}{^{\tau\alpha}}\Lambda_b^{\!-1}$
term in (\ref{threeparts}) is divergenceless so that it has no effect on Ampere's law (\ref{Ampere}).
The $f_{\nu\mu}\Lambda/\Lambda_b$ term is $\sim 10^{-122}$ of
$f_{\nu\mu}$ from (\ref{LambdaoverLambdab}).
The $(f^3)\Lambda_b^{\!-1}$ term is $<\!10^{-66}$ of $f_{\nu\mu}$ from (\ref{highenergyskew}).
The largest observable values of the Weyl tensor might be expected to occur
near the Schwarzschild radius, $r_s\!\nobreak=\nobreak\!2Gm/c^2$, of black
holes, where it takes on values around $r_s/r^3$. However, since
the lightest black holes have the smallest Schwarzschild radius,
they will create the largest value of
$r_s/r_s^3=1/r_s^2$. The lightest black hole that we might
observe would be of about one solar mass, where from (\ref{Lambdab}),
\begin{eqnarray}
\label{blackholecurvature}
\frac{C_{0101}}{\Lambda_b}
&\sim&\frac{1}{\Lambda_br^2_s}
=\frac{1}{\Lambda_b}\left(\frac{c^2}{2Gm_\odot}\right)^2\!\sim\!10^{-77}.
\end{eqnarray}
And regarding $j^\tau$ as a substitute for $(1/4\pi){f^{\omega\tau}}_{;\,\omega}$ from
(\ref{Ampere}), the $j_{[\nu,\mu]}\Lambda_b^{\!-1}$ term is
$<\!10^{-32}$ of $f_{\nu\mu}$ from (\ref{highenergyderiv2}).
Therefore even for the most extreme worst-case fields accessible to measurement,
the last four terms in (\ref{threeparts}) must all be $<\!10^{-32}$ of $f_{\nu\mu}$.
%\newpage

%Taking the divergence of (\ref{threeparts}) using (\ref{Ampere}),
%%and ignoring $C_{\nu\mu\alpha\tau}f^{\alpha\tau}/\Lambda_b$
%%and $f_{\nu\mu}\Lambda/\Lambda_b$
%%due to (\ref{blackholecurvature},\ref{LambdaoverLambdab}),
%the divergenceless term $\dual_{[\tau,\alpha]}\varepsilon_{\nu\mu}{^{\tau\alpha}}$
%falls out and we get an extremely close approximation to Maxwell's equations,
%\begin{eqnarray}
%\label{Maxwell}
%\fl F_{\nu\mu;}{^\nu}=4\pi j_\mu
%\!+\!\left[(f^{\tau\alpha}C_{\alpha\tau\nu\mu});{^\nu}
%\!-\!\frac{8\pi(n\!-\!2)\Lambda}{(n\!-\!1)}j_\mu
%\!-\!\frac{8\pi(n\!-\!2)}{(n\!-\!1)}\,j_{[\nu,\mu];}{^\nu}+(f^{3\prime})\right]\!\Lambda_b^{\!-1}\dots,\\
%\label{Faraday}
%\fl F_{[\nu\mu,\sigma]}=0.
%\end{eqnarray}
%As usual, Faraday's law (\ref{Faraday}) is just an identity which follows from
%the definition (\ref{Fdef}).
%The last four terms in Ampere's law (\ref{Ampere})
%are $<\!10^{-32}$ of the primary terms because this is true for (\ref{threeparts}).
From (\ref{threeparts},\ref{highenergyderiv2}), the extra terms in
Maxwell's equations (\ref{Ampere},\ref{Faraday}) must be
$<\nobreak\!10^{-32}$ of the ordinary terms.
In most so-called ``exact'' equations in physics, there are really many known
corrections due to quantum electrodynamics and other effects which are
ignored because they are too small to measure.
It should be emphasized that the extra terms in our Maxwell equations are at least 20 orders of magnitude
smaller than known corrections to these equations which are routinely ignored\cite{Jackson}.

The divergenceless term $\theta_{[\tau,\alpha]}\varepsilon_{\nu\mu}{^{\tau\alpha}}\Lambda_b^{\!-1}$ of
(\ref{threeparts}) should also be expected to be $<\!10^{-32}$ of $f_{\nu\mu}$ from
(\ref{highenergyderiv1},\ref{highenergyderiv2},\ref{thdef}).
However, we need to consider the possibility where $\theta_\tau$ changes extremely rapidly,
so let us consider the ``dual'' part of (\ref{threeparts}).
%but where there are no extreme spatial curvatures so that
%the Weyl tensor term in (\ref{threeparts}) is still negligible.
Taking the curl of (\ref{threeparts}),
the $F_{\nu\mu}$ and $j_{[\nu,\mu]}$ terms drop out,
\ifnum\ExpandDerivations=1
\begin{eqnarray}
\fl~~~ f_{[\nu\mu,\sigma]}
=\!\left(\theta_{\tau;\alpha;[\sigma}\,\varepsilon_{\nu\mu]}{^{\tau\alpha}}
+(f^{\alpha\tau}C_{\alpha\tau[\nu\mu}){_{,\sigma]}}
+\frac{2(n\!-\!2)\Lambda}{(n\!-\!1)}f_{[\nu\mu,\sigma]}
+(f^{3\prime})\right)\!\Lambda_b^{\!-1}\dots.\nonumber
\end{eqnarray}
%\bigskip\\
Contracting this with $\Lambda_b\varepsilon^{\rho\sigma\nu\mu}/2$ and using (\ref{thdef}) gives,
\begin{eqnarray}
\label{rawProca}
\fl~~~ 2\Lambda_b\theta^\rho
%\!&=&\!-2\,\dual^\tau{\!_;}{^\alpha}{_{;\,\sigma}}\delta^{[\rho}_{\,\tau}\delta^{\sigma]}_\alpha\\
\!&=&\!-2\,\theta^{[\rho}{_;}{^{\sigma]}}{_{\!;\,\sigma}}
+\frac{1}{2}\varepsilon^{\rho\sigma\nu\mu}(f^{\alpha\tau}C_{\alpha\tau[\nu\mu}){_{,\sigma]}}
+\frac{4(n\!-\!2)\Lambda}{(n\!-\!1)}\,\theta^\rho
+(f^{3\prime})\dots\nonumber
\end{eqnarray}
Using $\theta^\sigma{_{;\sigma}}\!=0$ from (\ref{thdef}), the covariant derivative commutation rule,
and the Einstein equations $R_{\nu\mu}\!=\!-\Lambda g_{\nu\mu}+(f^2)\dots$ from (\ref{Einstein2}),
gives $\theta^\sigma{_{;\rho;\sigma}}\!=\!R_{\sigma\rho}\theta^\sigma\!=\!-\theta_\rho\Lambda+(f^{3\prime})\dots$,
\fi
and we get something similar to the Proca equation,
%\begin{eqnarray}
%\Lambda_b\theta_\rho
%\label{rawProca}
%\approx-\theta_{[\rho,\sigma];}{^{\sigma}}.
%\end{eqnarray}
\begin{eqnarray}
\label{Proca}
\fl~~~~~~~2\Lambda_b\theta_\rho=-\theta_\rho{_{;\,\sigma;}}{^\sigma}
+\frac{1}{2}\varepsilon_\rho{^{\sigma\nu\mu}}(f^{\alpha\tau}C_{\alpha\tau[\nu\mu}){_{,\sigma]}}
+\frac{(3n\!-\!7)\Lambda}{(n\!-\!1)}\,\theta_\rho
+(f^{3\prime})\dots.
\end{eqnarray}
%From (\ref{Lambdab}), a $\dual_\rho$ particle would have an energy
%$mc^2\!=\!\sqrt{2\Lambda_b}\,\hbar c\!=\!\sqrt{60/\pi}\,\hbar c/l_P$
%which exceeds the zero-point cut-off energy
%$\hbar \omega_c c\!=\!\hbar c/l_P$, so it seems unlikely that
%$\dual_\rho$ could represent a real particle. The possibility of a $\dual_\rho$
%particle is discussed in \S6-\S7 of \cite{Shifflett}.
This equation suggests the possibility of $\theta_\rho$ waves,
%or a $\dual_\rho$ particle in a quantized theory,
and this is discussed in detail
in \cite{Shifflett}. There it is shown that if
$\theta_\rho$ waves do result from (\ref{Proca}), they would appear to
have a negative energy. However, this theory avoids ghosts in an unusual way.
Recall that this theory is the original Einstein-Schr\"{o}dinger theory,
but with a $\Lambda_z g_{\mu\nu}$ in the field equations
to account for zero-point fluctuations,
%The whole theory is based upon the assumption that zero-point fluctuations
%really do contribute to the cosmological constant,
and $\Lambda_z\!=\!-C_z\omega_c^4l_P^2$ from (\ref{Lambdab},\ref{Cz}) is finite only because
of a cutoff frequency $\omega_c\!\sim\nobreak\!1/l_P$ from (\ref{cutoff}). From these equations
and (\ref{Proca}), Proca waves would be cut off because they would have a minimum frequency
\begin{eqnarray}
\omega_{Proca}\!=\!\sqrt{2\Lambda_b}\!=\!\sqrt{-2\Lambda_z}\!=\!\!\sqrt{2C_z}\,\omega_c^2l_P>\omega_c.
\end{eqnarray}
Whether the cutoff of zero-point fluctuations is caused by a discreteness, uncertainty or foaminess of
spacetime near the Planck length\cite{Garay,Padmanabhan,Padmanabhan2,Ashtekar,Smolin}
or by some other effect, the same $\omega_c$ which cuts off $\Lambda_z$
should also cut off Proca waves in this theory. So we should expect to observe only the
trivial solution $\dual_\rho\!\approx\!0$ to (\ref{Proca}) and no ghosts.
Comparing $\omega_{Proca}$ and $\omega_c$ from above, we see that this argument only applies if
\begin{eqnarray}
\label{limit}
\omega_c>\frac{\lower1pt\hbox{$1$}}{l_P\sqrt{2C_z}}.
\end{eqnarray}
Here $C_z$ is defined by (\ref{Cz}), and the inequality
is satisfied for this theory when $\omega_c$ and $C_z$ are chosen as in
(\ref{cutoff},\ref{Lambdab}) to be consistent
with a cosmological constant caused by zero-point fluctuations.
Since the prediction of negative energy waves would probably be inconsistent with
reality, this theory should be approached cautiously when considering it with
values of $\omega_c$ and $C_z$ which do not satisfy (\ref{limit}).

Finally, if we fully renormalize with $\omega_c\!\rightarrow\!\infty$
as in quantum electrodynamics,
then $\Lambda_b\!\rightarrow\!\infty$ and $\omega_{Proca}\!\rightarrow\!\infty$,
so the potential ghost goes away completely.
%From this perspective, our theory does not contain ghosts any more than quantum electrodynamics
%contains ghosts when the renormalization is done using Pauli-Villars fields
%(which are negative energy fields).
In the limit $\omega_c\!\rightarrow\!\infty$ our theory becomes exactly Einstein-Maxwell theory.
Assuming such a full renormalization does not diminish the value of the theory in any way.
It still unifies gravitation with electromagnetism, it still suggests untried
approaches to a complete unified field theory, and it still
offers an untried approach to the quantization of gravity.
In any attempt to quantize this theory, the cutoff frequency $\omega_c$ would need to be the same
cutoff which is taken to infinity during renormalization.
For example, Pauli-Villars masses would probably go as $M\!=\!\hbar\omega_c$
if Pauli-Villars renormalization was used.
Since $\Lambda_b$ and $\Lambda_z$ in the Lagrangian density go as $\omega_c^4$,
quantization and renormalization would certainly need to be done a bit different than usual.
Also, because $\omega_{Proca}$ goes as $\omega_c^2$, Proca waves would not
represent a ghost from the standpoint of quantization.

\section{\label{LorentzForce}The Lorentz Force Equation}
%From a theorem of tensor calculus\cite{Adler}, (\ref{JScurl}) implies that
%$\tR_{[\nu\mu]}+\Lambda_b N_{[\nu\mu]}$ is a curl,
%so (\ref{JSsymmetric},\ref{JScurl}) are really equivalent to (\ref{para}).
A generalized contracted Bianchi identity is derived in \ref{Bianchi}
using only the connection equations (\ref{JSconnection}) and the symmetry
(\ref{JScontractionsymmetric}) of $\tGam^\alpha_{\nu\mu}$,
\begin{eqnarray}
\label{niceform}
\fl~~~~~~~(\rmN N^{\dashv\nu\sigma}\tR{_{\sigma\lambda}}
+\rmN N^{\dashv\sigma\nu}\tR_{\lambda\sigma}){_{,\nu}}
-\rmN N^{\dashv\nu\sigma}\tR{_{\sigma\nu,\lambda}}=0.
\end{eqnarray}
The identity can also be written in a manifestly covariant form
\begin{eqnarray}
\label{niceform2}
\fl~~~~~~~(\rmN N^{\dashv\nu\sigma}\tR{_{\sigma\lambda}}
+\rmN N^{\dashv\sigma\nu}\tR_{\lambda\sigma}){_{;\nu}}
-\rmN N^{\dashv\nu\sigma}\tR{_{\sigma\nu;\lambda}}=0,
\end{eqnarray}
%\bigskip\\
or in terms of $g^{\rho\tau},f^{\rho\tau}$ and $\tG_{\nu\mu}$ from (\ref{gdef},\ref{fdef},\ref{genEinstein}),
\begin{eqnarray}
\label{contractedBianchi}
\tG^\sigma_{\nu;\,\sigma}
=\left(\frac{\lower2pt\hbox{$3$}}{2}f^{\sigma\rho}\,\tR_{[\sigma\rho,\nu]}
+f^{\alpha\sigma}{_{\!;\alpha}}\tR_{[\sigma\nu]}\right)\!\rmt\Lambda_b^{\!-1/2}.
\end{eqnarray}
Clearly (\ref{niceform},\ref{contractedBianchi}) are simple generalizations of the ordinary
contracted Bianchi identity
$2(\rmg\,R^\nu{_\lambda})_{,\nu}\!-\!\rmg\,g^{\nu\sigma}\!R_{\sigma\nu,\lambda}\!=\!0$,
or $G^\sigma_{\nu;\sigma}\!=\!0$.
%and it applies even when $j^\tau\!\ne0$.
%It also does not depend on the choice of the matter term ${\mathcal L}_m$.
%although of course $j^\sigma$ represents something different for each case.
%which reduces to the usual contracted Bianchi identity $G^\sigma_{\nu;\,\sigma}\!=\!0$ for the symmetric case.
This identity was first derived in \cite{EinsteinBianchi,SchrodingerIII} without assuming charge currents,
and later expressed in terms of the metric (\ref{gdef}) by \cite{Hely,JohnsonI,Treder57}.
The derivation with charge currents was first done in \cite{Antoci3} by applying an infinitesimal
coordinate transformation to an invariant integral.
\ref{Bianchi} is included because \cite{Antoci3} does not include
the full derivation, and to confirm the earlier result by using a much different direct computation method.

Another useful identity is derived in \ref{UsefulIdentity}
using only the definitions (\ref{gdef},\ref{fdef}) of $g_{\mu\nu}$ and $f_{\mu\nu}$,
\begin{eqnarray}
\label{usefulidentity}
\fl~~~~~~~~~~\left(N^{(\mu}{_{\nu)}} \!-\!\frac{\lower2pt\hbox{$1$}}{2}\delta^\mu_\nu
N^\rho_\rho\right)\!{_{;\,\mu}}
=\left(\frac{\lower2pt\hbox{$3$}}{2}f^{\sigma\rho}N_{[\sigma\rho,\nu]}
+f^{\sigma\rho}{_{;\sigma}}N_{[\rho\nu]}\right)\!\rmt\Lambda_b^{\!-1/2}.
\end{eqnarray}

The ordinary Lorentz force equation of Einstein-Maxwell theory results from
taking the divergence of the Einstein equations (\ref{Einstein}) using
(\ref{contractedBianchi},\ref{Ampere},\ref{JSantisymmetric},\ref{usefulidentity},\ref{Fdef})
\begin{eqnarray}
\label{divergence}
\fl 8\pi T^\sigma_{\nu;\,\sigma}
&=&\left(\frac{\lower2pt\hbox{$3$}}{2}f^{\sigma\rho}\,\tR_{[\sigma\rho,\nu]}
+4\pi j^\sigma\tR_{[\sigma\nu]}\right)\!\rmt\Lambda_b^{\!-1/2}
+\Lambda_b\!\left(N^{(\mu}{_{\nu)}}
\!-\!\frac{\lower2pt\hbox{$1$}}{2}\delta^\mu_\nu N^\rho_\rho\right)\!{_{;\,\mu}}\\
\fl&=&\!\left(4\pi j^\sigma\tR_{[\sigma\nu]}
-\Lambda_b\frac{\lower2pt\hbox{$3$}}{2}f^{\sigma\rho}N_{[\sigma\rho,\nu]}\right)\!\rmt\Lambda_b^{\!-1/2}
+\Lambda_b\!\left(N^{(\mu}{_{\nu)}}
\!-\!\frac{\lower2pt\hbox{$1$}}{2}\delta^\mu_\nu N^\rho_\rho\right)\!{_{;\,\mu}}\\
\fl&=&(4\pi j^\sigma\tR_{[\sigma\nu]}
+\Lambda_b f^{\rho\sigma}{_{;\rho}}N_{[\sigma\nu]})\rmt\Lambda_b^{\!-1/2}\\
\fl&=&4\pi j^\sigma(\tR_{[\sigma\nu]}
+\Lambda_bN_{[\sigma\nu]})\rmt\Lambda_b^{\!-1/2}\\
\fl&=&16\pi j^\sigma A_{[\sigma,\nu]},\\
\label{Euler}
\fl ~~~T^\sigma_{\nu;\,\sigma}
\fl&=&F_{\nu\sigma}j^\sigma.
\end{eqnarray}
Note that the covariant derivatives in
(\ref{niceform2},\ref{contractedBianchi},\ref{usefulidentity},\ref{Euler}) are all done using the
Christoffel connection (\ref{Christoffel}) formed from the symmetric metric (\ref{gdef}).

%Equation (\ref{Euler}) applies for all of the ${\mathcal L}_m$ cases.
%For the classical hydrodynamics case, further simplification results by
%combining (\ref{Thydrodynamic},\ref{uj}) with the continuity equation
%(\ref{continuity}). Then (\ref{Euler}) is just the Lorentz force coupled to Newton's 2nd law,
%\begin{eqnarray}
%\label{Lorentz}
%\fl~~~~~F_{\nu\sigma}j^\sigma
%=T^\sigma_{\nu;\,\sigma}
%=(\mum u^\sigma u_{\nu})_{;\,\sigma}
%=\mum u^\sigma u_{\nu;\,\sigma}
%=\frac{\mum}{m}\left(m\frac{dx_\nu}{ds}\right){_{\!;\sigma}}
%\frac{dx^\sigma}{ds}.
%\end{eqnarray}

\section{\label{KleinGordonDirac}The Klein-Gordon and Dirac Equations}
For the spin-0 case, the Klein-Gordon equation
is obtained by setting $\delta{\mathcal L}/\delta\bar\psi\!=0$,
\begin{eqnarray}
0&=&\frac{-2}{\rmg}\left[\frac{\partial{\mathcal L}}{\partial\bar\psi}
-\left(\frac{\partial{\mathcal L}}{\partial\bar\psi_{,\lambda}}\right){_{\!,\lambda}}\right]\\
\fl &=&-\left[\frac{\hbar^2}{m}\left(-\frac{iQ}{\hbar}A_\mu\right)D^\mu\psi
-m\psi
-\frac{\hbar^2}{m\rmg}(\rmg D^\lambda\psi)_{,\lambda}\right]\\
\fl &=&\frac{-1}{m}\left[\frac{-\hbar^2}{\rmg}\left(\frac{\partial}{\partial x^\mu}+\frac{iQ}{\hbar}A_\mu\right)\rmg D^\mu \psi
-m^2\right]\psi\\
\label{KleinGordon1}
&=&\frac{1}{m}\left[\frac{\hbar^2}{\rmg}D_\mu \rmg D^\mu +m^2\right]\psi.
\end{eqnarray}
%\bigskip\\
%\bigskip\\
The conjugate Klein-Gordon equation is found by setting $\delta{\mathcal L}/\delta\psi\!=0$,
\begin{eqnarray}
0&=&\frac{-2}{\rmg}\left[\frac{\partial{\mathcal L}}{\partial\psi}
-\left(\frac{\partial{\mathcal L}}{\partial\psi_{,\lambda}}\right){_{\!,\lambda}}\right]\\
%\fl &=&-\left[\frac{\hbar^2}{m}\bar\psi\Ddag{^\mu}\left(-\frac{iQ\Lambda_b}{2l_P^2}A_\mu\right)-mc^2\bar\psi
%-\frac{\hbar^2}{m}(\bar\psi\rmg \Ddag{^\lambda})_{,\lambda}\frac{1}{\rmg}\right]\\
%\fl &=&\frac{-1}{m}\bar\psi\left[\Ddag{^\mu}\rmg\left(-\frac{\overleftarrow\partial}{\partial x^\mu}-\frac{iQ\Lambda_b}{2l_P^2}A_\mu\right)\frac{\hbar^2}{\rmg}-m^2c^2\right]\\
\label{conjKleinGordon}
&=&\frac{1}{m}\bar\psi\left[\Ddag{^\mu}\rmg \Ddag\!_\mu\frac{\hbar^2}{\rmg}+m^2\right].
%\fl &=&\frac{1}{m}\left(\left[\frac{\hbar^2}{\rmg}D_\mu \rmg D^\mu +m^2c^2\right]\bar\psi^*\right)^*.
\end{eqnarray}
This is just the complex conjugate of the Klein-Gordon equation (\ref{KleinGordon1})
if $\bar\psi=\psi^*$.

For the spin-1/2 case, the Dirac equation is found in a similar manner,
\begin{eqnarray}
0&=&\frac{1}{\rmg}\left[\frac{\partial{\mathcal L}}{\partial\bar\psi}
-\left(\frac{\partial{\mathcal L}}{\partial\bar\psi_{,\lambda}}\right){_{\!,\lambda}}\right]\\
\label{Dirac}
&=&i\hbar \gamma^\sigma D_\sigma\psi-m\psi.
\end{eqnarray}
The conjugate Dirac equation is,
\begin{eqnarray}
0&=&\frac{1}{\rmg}\left[\frac{\partial{\mathcal L}}{\partial\psi}
-\left(\frac{\partial{\mathcal L}}{\partial\psi_{,\lambda}}\right){_{\!,\lambda}}\right]\\
\label{conjDirac}
&=&-i\hbar\bar\psi\Ddag\!_\sigma\gamma^\sigma -m\bar\psi.
\end{eqnarray}
%This is the complex conjugate of the Dirac equation if $\bar\psi=\psi^\dag\gamma^0$.

%\bigbreak
Both the Klein-Gordon and Dirac equations match those of ordinary one-particle
quantum mechanics in curved space\cite{Birrell}.
%There is nothing analogous to these equations for the
%classical hydrodynamics case, where $\boldsymbol\mu$ and $u^\nu$ are unconstrained.
%There is also nothing analogous to these equations for the electro-vac case, where
%particles are represented by singularities.
Note that for the spin-0 case, instead of deriving the continuity equation
(\ref{continuity},\ref{jKleinGordon}) from the divergence
of Ampere's law, it can be derived from the Klein-Gordon equation,
\begin{eqnarray}
\fl 0 &=&\frac{i Q}{2\hbar}\left[\bar\psi
\left({ one~side~of}\atop {Klein\!-\!Gordon~equation}\right)
-\left({one~side~of~conjugate}\atop{Klein\!-\!Gordon~equation}\right)
\psi\right]\\
\fl &=&j^\mu{_{;\mu}}.
\end{eqnarray}
Similarly, instead of deriving the Lorentz force equation (\ref{Euler},\ref{TKleinGordon}) from
the divergence of the Einstein equations, it can be derived from the Klein-Gordon equation,
\begin{eqnarray}
\fl 0&=&\frac{\bar\psi\Ddag\!_\rho}{2}\left({one~side~of}\atop{Klein\!-\!Gordon~equation}\right)
+\left({one~side~of~conjugate}\atop {Klein\!-\!Gordon~equation}\right)\frac{D_\rho\psi}{2}\\
\fl&=&T^\lambda_{\rho;\,\lambda}+F_{\lambda\rho}j^\lambda.
\end{eqnarray}
The calculations can be found in a commented-out appendix of this paper's .tex file.
%It is not clear that this has been proven in the literature for curved space,
%so this is done in \ref{continuityEuler}.
Presumably, similar results occur for the spin-1/2 case, but this was not verified.

\section{\label{Discussion}Discussion}
%There are other ways of generalizing the symmetric field Lagrangian
%density (\ref{GR}) to nonsymmetric fields which are just as natural as (\ref{Palatini}).
%The literature on the original Einstein-Schr\"{o}dinger theory can be confusing because
%the theory results from many different Lagrangian densities.
The original Einstein-Schr\"{o}dinger theory results from many different Lagrangian densities.
In fact it results from any Lagrangian density of the form,
\begin{eqnarray}
\label{JSlag2_d}
\fl~~~{\mathcal L}(\nGam^{\lambda}_{\!\rho\tau},N_{\rho\tau})
&=&\!-\frac{\lower2pt\hbox{$1$}}{16\pi}\rmN\left[N^{\dashv\mu\nu}(R_{\nu\mu}(\tGam)
\!+\ca\tGam^\alpha_{\alpha[\nu,\mu]}
\!+2\Aphi_{[\nu,\mu]}\rmt\Lambda_b^{\!1/2})\right.\nonumber\\
\fl &&~~~~~~~~~~~~~~~~~~~~~~~~~~~~~~~~~~~~~~~~~~~~~~~~~~~~~~\left.\phantom{0_0^{|}}+(n\!-\!2)\Lambda_b\,\right]\!,
\end{eqnarray}
where $\ca,\cb,\cc$ are arbitrary constants and
\begin{eqnarray}
R_{\nu\mu}(\tGam)
&=&\tGam^\alpha_{\nu\mu,\alpha}
-\tGam^\alpha_{\nu\alpha,\mu}
+\tGam^\sigma_{\nu\mu}\tGam^\alpha_{\sigma\alpha}
-\tGam^\sigma_{\nu\alpha}\tGam^\alpha_{\sigma\mu},\\
\label{gamma_tilde_d}
~~~~~\tGam{^\alpha_{\nu\mu}}
&=&{\nGam}{^\alpha_{\nu\mu}}
\!+\frac{\lower2pt\hbox{$2$}}{(n\!-\!1)}\left[\,\cb\delta^\alpha_\mu{\nGam}{^\sigma_{\![\sigma\nu]}}
\!+\!(\cb-1)\,\delta^\alpha_\nu {\nGam}{^\sigma_{\![\sigma\mu]}}\right],\\
\label{A_d}
~~~~~~\Aphi_\nu&=&{\nGam}{^\sigma_{\![\sigma\nu]}}/\cc.
%\cc,\cb,zeta_3&=&({\rm arbitrary~constants).
\end{eqnarray}
Contracting (\ref{gamma_tilde_d}) on the right and left gives
\begin{eqnarray}
%\label{JScontractionsymmetric2}
\fl~~~~~~~~~\tGam^\alpha_{\beta\alpha}
=\frac{\lower2pt\hbox{$1$}}{(n\!-\!1)}\left[(\cb n+\cb-1)\nGam^\alpha_{\alpha\beta}
-(\cb n+\cb-n)\nGam^\alpha_{\beta\alpha}\right]=\tGam^\alpha_{\alpha\beta},
\end{eqnarray}
so $\tGam{^\alpha_{\nu\mu}}$
has only $n^3\!-n$ independent components.
%Also note that the term $\tGam^\alpha_{\alpha[\nu,\mu]}$ in (\ref{JSlag2}) is a tensor
%because $\tGam^\alpha_{\alpha[\nu,\mu]}\!=\!\tR^\alpha{_{\!\alpha\mu\nu}}/2$.
Also, from (\ref{gamma_tilde_d},\ref{A_d}) we have
\begin{eqnarray}
\label{gamma_natural_d}
{\nGam}{^\alpha_{\nu\mu}}\!&=&\!\tGam{^\alpha_{\nu\mu}}
\!-\frac{\lower2pt\hbox{$2\cc$}}{(n\!-\!1)}\left[\,\cb\delta^\alpha_\mu \Aphi_\nu
+(\cb-1)\delta^\alpha_\nu \Aphi_\mu\right],
\end{eqnarray}
so $\tGam^\alpha_{\nu\mu}$ and $A_\nu$ fully parameterize
$\nGam^\alpha_{\nu\mu}$ and can be treated as independent variables.
Therefore setting $\delta{\mathcal L}/\delta\tGam^\alpha_{\nu\mu}\!=0$
and $\delta{\mathcal L}/\delta A_\nu\!=0$
must give the same field equations as
$\delta{\mathcal L}/\delta\nGam^\alpha_{\nu\mu}\!=0$.
%Therefore when we set to zero the variational derivative of (\ref{JSlag2_d}) with respect to
%$\tGam^\alpha_{\nu\mu}$~and $A_\nu$, the same field equations must result as when
%$\nGam^\alpha_{\nu\mu}$ is used.
%In this theory $\nGam^\alpha_{\nu\mu}$ is really the
%``fundamental'' field, and $\tGam^\alpha_{\nu\mu}$ and $A_\nu$ are derived from it.
Because the field equations can be derived in this way,
the constants $\cb$ and $\cc$ are clearly arbitrary,
and because of (\ref{RisB}) with $j^\sigma\!=0$, the constant $\ca$ is also arbitrary.

For $\ca\!=\!1,\,\cb\!=\!1/2,\,\cc\!=\!-(n\!-\!1)\rmt\Lambda_b^{\!1/2}$,
(\ref{JSlag2_d}) reduces to (\ref{Palatini})
formed from the Hermitianized Ricci tensor (\ref{HermitianizedRicci}), where
we have the invariance properties from (\ref{transposition},\ref{gauge}),
\begin{eqnarray}
\label{transposition_d}
\fl~~A_\nu\!\rightarrow\!-A_\nu
,~\tGam^\alpha_{\nu\mu}\!\rightarrow\!\tGam^\alpha_{\mu\nu}
,~\nGam^\alpha_{\nu\mu}\!\rightarrow\!\nGam^\alpha_{\mu\nu}
,~N_{\nu\mu}\!\rightarrow\!N_{\mu\nu}
,~N^{\dashv\mu\nu}\!\!\rightarrow\!N^{\dashv\nu\mu}
~~\Rightarrow~{\mathcal L}\!\rightarrow\!{\mathcal L},\\
\label{gauge_d}
\fl~~A_\alpha\!\rightarrow\! A_\alpha\!-\!\frac{\hbar}{Q}\phi_{,\alpha},
~\tGam^\alpha_{\rho\tau}\!\rightarrow\!\tGam^\alpha_{\rho\tau}
,~\nGam^\alpha_{\rho\tau}\!\rightarrow\!\nGam^\alpha_{\rho\tau}\!+\frac{2\hbar}{Q}\delta^\alpha_{[\rho}\phi_{,\tau]}\rmt\Lambda_b^{\!1/2}
~\Rightarrow~{\mathcal L}\!\rightarrow\!{\mathcal L}.
\end{eqnarray}
For this case we have
$\tGam^\alpha_{\sigma\alpha}\!=\!{\tGam}{^\alpha_{\alpha\sigma}}\!=\!{\nGam}{^\alpha_{\!(\alpha\sigma)}}$,
and from (\ref{der0},\ref{para},\ref{JSlag2_d}) the field equations require
a generalization of the result ${\mathcal L}_{,\sigma}\!-\!{\Gamma}{^\alpha_{\alpha\sigma}}{\mathcal L}\!=\!0$
that occurs with the Lagrangian density (\ref{GR}) of ordinary vacuum general relativity,
that is
\begin{eqnarray}
\label{selftransplantation0}
{\mathcal L}_{,\sigma}\!-\!{\nGam}{^\alpha_{(\alpha\sigma)}}{\mathcal L}=0
~~~{\rm or}~~~
{\mathcal L}_{,\sigma}\!-\!Re({\nGam}{^\alpha_{\alpha\sigma}}){\mathcal L}=0.
\end{eqnarray}
For the alternative choice, $\ca\!=\!0,\,\cb\!=\!n/(n\!+\!1),\,\cc\!=\!-(n\!-\!1)\rmt\Lambda_b^{\!1/2}/2$,
we have $\tGam^\alpha_{\sigma\alpha}\!=\!{\tGam}{^\alpha_{\alpha\sigma}}\!=\!{\nGam}{^\alpha_{\alpha\sigma}}$ and
from (\ref{der0},\ref{para},\ref{JSlag2_d}) the field equations require
\begin{eqnarray}
\label{selftransplantation}
{\mathcal L}_{,\sigma}\!-\!{\nGam}{^\alpha_{\alpha\sigma}}{\mathcal L}=0.
\end{eqnarray}
%In \cite{Shifflett} it is shown that the
%theory appears to be unique in that it can be derived from a
%Lagrangian density with the property (\ref{selftransplantation}).
%at least when the Lagrangian density is
%assumed to depend only on a nonsymmetric ${^\natural\Gamma}{^\alpha_{\nu\mu}}$.
%This property is destroyed by including an extrinsic cosmological constant
%from zero-point fluctuations, and also by modelling matter with spin-0
%or spin-1/2 wave-functions rather than as singular solutions of the field equations.
%This suggests the hypothesis that nature requires a purely classical Lagrangian density
%to have the property (\ref{selftransplantation}), but that quantization modifies this requirement.
%A vague idea is presented in \cite{Shifflett} as to why the property (\ref{selftransplantation})
%might be expected.
%However, it is legitimate to propose a simple principle without knowing why it should apply.
%As an example which makes this obvious, consider that when Newton proposed
%that mass attracts itself with a $1/r^2$ force, he certainly could not explain why.
%After all, no one knows why the laws of physics are what they are,
%so it would be unreasonable to expect an exception in this case.
For the alternative choice $\ca\!=\!1,\,\cb\!=\!0,\,\cc\!=\!-(n\!-\!1)\rmt\Lambda_b^{\!1/2}$,
(\ref{JSlag2_d}) reduces to
\begin{eqnarray}
\label{Palatini_d}
\fl~~~~{\mathcal L}(\nGam^{\lambda}_{\!\rho\tau},N_{\rho\tau})
=-\frac{\lower2pt\hbox{$1$}}{16\pi}\rmN\left[N^{\dashv\mu\nu}
\Re_{\nu\mu}({\nGam})+(n\!-\!2)\Lambda_b\,\right],
\end{eqnarray}
where $\Re_{\nu\mu}({\nGam})$ is a fairly simple generalization of the ordinary Ricci tensor
\begin{eqnarray}
\label{HermitianizedRicciB}
\Re_{\nu\mu}(\nGam)
&=&\nGam^\alpha_{\nu\mu,\alpha}
-\nGam^\alpha_{\!(\nu|\alpha,|\mu)}
+\nGam^\sigma_{\nu\mu}\nGam^\alpha_{\sigma\alpha}
-\nGam^\sigma_{\nu\alpha}\nGam^\alpha_{\sigma\mu}.
\end{eqnarray}
For another alternative choice $\ca\!=\!0,\,\cb\!=\!0,\,\cc\!=\!-(n\!-\!1)\rmt\Lambda_b^{\!1/2}/2$,
(\ref{JSlag2_d}) reduces to
\begin{eqnarray}
\label{ordinary}
\fl~~~~{\mathcal L}(\nGam^{\lambda}_{\!\rho\tau},N_{\rho\tau})
=-\frac{\lower2pt\hbox{$1$}}{16\pi}\rmN\left[N^{\dashv\mu\nu}
R_{\nu\mu}({\nGam})+(n\!-\!2)\Lambda_b\,\right],
\end{eqnarray}
where $R_{\nu\mu}(\nGam)$ is the ordinary Ricci tensor
\begin{eqnarray}
R_{\nu\mu}(\nGam)
&=&\nGam^\alpha_{\nu\mu,\alpha}
-\nGam^\alpha_{\!\nu\alpha,\mu}
+\nGam^\sigma_{\nu\mu}\nGam^\alpha_{\sigma\alpha}
-\nGam^\sigma_{\nu\alpha}\nGam^\alpha_{\sigma\mu}.
\end{eqnarray}
The original Einstein-Schr\"{o}dinger theory (including the cosmological constant)
can even be derived from purely affine versions
of the Lagrangian densities described above, such as the Lagrangian density
used by Schr\"{o}dinger\cite{SchrodingerI},
\begin{eqnarray}
\label{Schrodingerslag}
{\mathcal L}(\nGam)=\sqrt{-det(R_{\nu\mu}(\nGam))}\,.
\end{eqnarray}

Whether one prefers the Lagrangian density (\ref{Palatini}) with the properties
(\ref{transposition_d},\ref{gauge_d},\ref{selftransplantation0})
or one of the alternatives, it is clear that the original
Einstein-Schr\"{o}dinger theory can be derived from rather simple principles.
The theory proposed in this paper is a natural extension of the original Einstein-Schr\"{o}dinger theory
to account for zero-point fluctuations and first quantization.
The search for simple principles has led to many advances in physics,
and is what led Einstein to general relativity and also
to the Einstein-Schr\"{o}dinger theory\cite{Schilpp,EinsteinBianchi}.
%He says ``I have learned something else from the theory of gravitation:
%No ever so inclusive collection of empirical facts can ever lead to the setting up of such
%complicated equations. A theory can be tested by experience, but there is no way from experience
%to the setting up of a theory. Equations of such complexity as are the equations of the
%gravitational field can be found only through the discovery of a logically simple mathematical
%condition which determines them completely, or [at least] almost completely. Once one
%has those sufficiently strong formal conditions, one requires only little knowledge of
%facts for the setting up of a theory; in the case of the equations of gravitation it is
%the four-dimensionality and the symmetric tensor as expression for the structure of
%space which, together with the invariance concerning the continuous transformation-group,
%determine the equations almost completely.''
Einstein disliked the term
$\rmg F^{\nu\mu}\!F_{\mu\nu}/16\pi$ in the Einstein-Maxwell Lagrangian density. Referring to the equation
$G_{\nu\mu}\!\nobreak=\nobreak\!8\pi T_{\nu\mu}$ he states\cite{Schilpp} ``The right side is a
formal condensation of all things whose comprehension in the sense of a field-theory is still
problematic. Not for a moment, of course, did I doubt that this formulation was merely a
makeshift in order to give the general principle of relativity a preliminary closed
expression. For it was essentially not anything more than a theory of the gravitational field,
which was somewhat artificially isolated from a total field of as yet unknown structure.''
In modern times the term $\rmg F^{\nu\mu}\!F_{\mu\nu}/16\pi$ has become standard and
is rarely questioned.
The theory presented here suggests that this term should be questioned, and offers an
alternative which is based on simple principles
and which genuinely unifies gravitation and electromagnetism.
%and suggests an alternative and more natural explanation for electromagnetism.
%The author believes that this term should be questioned, and that the theory presented here
%offers a more natural explanation for electromagnetism.

%But why should it be there? Also, the Lagrangian density of classical general relativity
%and electromagnetism is evidently rather important, yet what does it's value actually represent?
%It is not (kinetic energy)-(potential energy) as in classical mechanics, so what is it?
%Why should its integral over all space be extremized? Of what significance is the covariant
%derivative of a Lagrangian density, and what happens if it is required to vanish?
%Such questions can guide the search for the fundamental physical laws,
%and as such they are legitimate questions to investigate.

\section{\label{Conclusions}Conclusions}

The Einstein-Schr\"{o}dinger theory is extended to include spin-0 and spin-1/2 sources.
The theory is also modified by including a cosmological constant
caused by zero-point fluctuations. This cosmological constant
which multiplies the symmetric metric
is assumed to be nearly cancelled by
Schr\"{o}dinger's ``bare'' cosmological constant
which multiplies the nonsymmetric fundamental tensor,
such that the total ``physical'' cosmological constant matches measurement.
%Other fields could be included in a similar manner.
The resulting
$\Lambda$-renormalized Einstein-Schr\"{o}dinger
theory closely
approximates ordinary Einstein-Maxwell theory and one-particle quantum mechanics.
In particular, the field equations match the ordinary Einstein and Maxwell
equations except for additional terms which are $<\!10^{-16}$ of the usual terms for
worst-case field strengths and rates-of-change accessible to measurement.
The theory also predicts the ordinary Lorentz force equation and the ordinary Klein-Gordon and Dirac equations.
And the theory becomes exactly Einstein-Maxwell theory and one-particle quantum mechanics
in the limit as the cosmological constant from zero-point fluctuations goes to infinity.
%We show that the theory closely approximates ordinary
%Einstein-Maxwell theory and one-particle quantum mechanics,
%except for some extra terms in the field equations, all of which have a $\Lambda_b$ in the denominator.
%When $\Lambda_b\!\approx\!-\Lambda_z$ and $\Lambda_z$ is caused by zero-point fluctuations,
%the extra terms in the field equations are $<\!10^{-16}$ of the ordinary terms
%for worst-case field strengths and rates-of-change accessible to measurement,
%and these extra terms go to zero in the limit as $\Lambda_b\!\rightarrow\!\infty$.

\section*{Acknowledgements}
%I am grateful to Clifford Will for discussions
%and for helpful comments on drafts of this manuscript.
%Thanks also to Claude Bernard for his help.
This work was supported in part by the National Science Foundation under grant PHY~03-53180.
%\bigskip\\

%\newpage
\appendix

\section{\label{UsefulIdentity}A divergence identity}
Using only the definitions (\ref{gdef},\ref{fdef}) of $g_{\nu\mu}$
and $f_{\nu\mu}$, and the identity (\ref{sqrtdetcomma}) gives,
\begin{eqnarray}
\fl&&\left(N^{(\mu}{_{\nu)}} \!-\!\frac{1}{2}\delta^\mu_\nu
N^\rho_\rho\right)\!{_{;\,\mu}}
-\frac{3}{2}f^{\sigma\rho}N_{[\sigma\rho,\nu]}\rmt\Lambda_b^{\!-1/2}\\
\fl &&=\frac{1}{2}g^{\sigma\rho}
(N_{(\rho\nu);\sigma}\!+\!N_{(\nu\sigma);\rho}\!-\!N_{(\rho\sigma);\nu})
%-\frac{3}{2}f^{\sigma\rho}N_{[\sigma\rho,\nu]}\\
\!-\!\frac{1}{2}f^{\sigma\rho}(N_{[\sigma\rho];\nu}\!+\!N_{[\rho\nu];\sigma}
\!+\!N_{[\nu\sigma];\rho})\rmt\Lambda_b^{\!-1/2}\\
\fl &&=\frac{1}{2}\frac{\rmN}{\rmg} \!\left[N^{\dashv(\sigma\rho)}
(N_{(\rho\nu);\sigma} \!+\!N_{(\nu\sigma);\rho}
\!-\!N_{(\rho\sigma);\nu}) \!+\!N^{\dashv[\sigma\rho]}
(N_{[\sigma\rho];\nu} \!+\!N_{[\rho\nu];\sigma}
\!+\!N_{[\nu\sigma];\rho})\right]\\
\fl &&=\frac{1}{2}\frac{\rmN}{\rmg} \!\left[\,N^{\dashv\sigma\rho}
(N_{(\rho\nu);\sigma} \!+\!N_{(\nu\sigma);\rho}
\!-\!N_{(\rho\sigma);\nu}) \!+\!N^{\dashv\sigma\rho}
(N_{[\rho\nu];\sigma} \!+\!N_{[\nu\sigma];\rho}
\!-\!N_{[\rho\sigma];\nu})\right]\\
\fl &&=\frac{1}{2}\frac{\rmN}{\rmg}N^{\dashv\sigma\rho}
(N_{\rho\nu;\sigma}+N_{\nu\sigma;\rho}-N_{\rho\sigma;\nu})\\
\fl &&=\frac{1}{2}\frac{\rmN}{\rmg}\left[N^{\dashv\sigma\rho}
(N_{\rho\nu;\sigma}+N_{\nu\sigma;\rho})
-N^{\dashv\sigma\rho}(N_{\rho\sigma,\nu}
-\Gamma^\alpha_{\rho\nu}N_{\alpha\sigma}
-\Gamma^\alpha_{\sigma\nu}N_{\rho\alpha})\right]\\
\fl &&=-\frac{1}{2}\frac{\rmN}{\rmg}
(N^{\dashv\sigma\rho}{_{;\sigma}}N_{\rho\nu}
+N^{\dashv\sigma\rho}{_{;\rho}}N_{\nu\sigma})
-\frac{1}{\rmg}(\rmN\,)_{;\nu}\\
\fl &&=-\frac{1}{2}\left[
\left(\frac{\rmN}{\rmg}N^{\dashv\sigma\rho}\right)
{_{\!\!;\sigma}}N_{\rho\nu}
+\left(\frac{\rmN}{\rmg}N^{\dashv\sigma\rho}\right)
{_{\!\!;\rho}}N_{\nu\sigma}\right]\\
\fl &&=-\frac{1}{2}\left[
(g^{\rho\sigma}+f^{\rho\sigma}\rmt\Lambda_b^{\!-1/2}){_{;\sigma}}N_{\rho\nu}
+(g^{\rho\sigma}+f^{\rho\sigma}\rmt\Lambda_b^{\!-1/2}){_{;\rho}}N_{\nu\sigma}\right]\\
\fl&&=f^{\sigma\rho}{_{;\sigma}}N_{[\rho\nu]}\rmt\Lambda_b^{\!-1/2}.
\end{eqnarray}

%\newpage
\section{\label{Bianchi}Derivation of the generalized contracted Bianchi identity}
Here we derive the generalized contracted Bianchi identity
(\ref{contractedBianchi}) from the connection equations
%in their contravariant density form
(\ref{JSconnection}), and from the
symmetry (\ref{JScontractionsymmetric}) of $\tGam^\alpha_{\nu\mu}$.
Whereas \cite{Antoci3} derived the identity
by performing an infinitesimal coordinate transformation on an invariant integral,
%although the full derivation was not shown.
we will instead use a direct method similar to \cite{EinsteinBianchi},
but generalized to include charge currents.  First we make the
following definitions,
\begin{eqnarray}
\label{Mdef}
\fl~~~~~~~~~~~~~~~ \mathbf{W}^{\tau\rho}&=&\rmg\,W^{\tau\rho}=\rmN N^{\dashv\rho\tau}
=\rmg\,(g^{\tau\rho}+\hf^{\tau\rho}),\\
\label{notation0}
\fl~~~~~~~~~~~~~~~~\hf^{\nu\mu}&=&f^{\nu\mu}\rmt\Lambda_b^{\!-1/2},~~~~\hbj^\alpha=\rmg j^\alpha\rmt\Lambda_b^{\!-1/2},\\
\label{HermitianizedRiemann}
\fl~~~~~~~~~~~~~\tR^\tau{_{\nu\alpha\mu}}&=&\tGam^\tau_{\nu\mu,\alpha}-\tGam^\tau_{\nu\alpha,\mu}
+\tGam^\sigma_{\nu\mu}\tGam^\tau_{\sigma\alpha}
-\tGam^\sigma_{\nu\alpha}\tGam^\tau_{\sigma\mu}
+\delta^\tau_\nu\tGam^\sigma_{\sigma[\alpha,\mu]},\\
\fl~~~\tR_{\nu\mu}=\tR^\alpha{_{\nu\alpha\mu}}
\label{HermitianizedRicci2}
&=&\tGam^\alpha_{\nu\mu,\alpha}-\tGam^\alpha_{\nu\alpha,\mu}
+\tGam^\sigma_{\nu\mu}\tGam^\alpha_{\sigma\alpha}
-\tGam^\sigma_{\nu\alpha}\tGam^\alpha_{\sigma\mu}
+\tGam^\sigma_{\sigma[\nu,\mu]}.
\end{eqnarray}
Here $\tR_{\nu\mu}$ is the Hermitianized Ricci tensor
(\ref{HermitianizedRiccit}), which has the property from (\ref{transpositionsymmetric}),
\begin{eqnarray}
\label{Hermiticity}
{\mathcal R}_{\nu\mu}(\tGam^T)=\tR_{\mu\nu}.
\end{eqnarray}
The tensors $\tR_{\nu\mu}$ and $\tR^\tau{_{\nu\alpha\mu}}$ reduce to the
ordinary Ricci and Riemann tensors for symmetric fields where
$\Gamma^\sigma_{\sigma[\nu,\mu]}\!=\!R^\sigma_{~\sigma\mu\nu}/2\!=\!0$.

Rewriting the
%contravariant density
connection equations (\ref{JSconnection})
in terms of the definitions above gives,
\begin{eqnarray}
\label{Lconnections}
\fl 0&=&\mathbf{W}^{\tau\rho}{_{,\lambda}}
+\tGam^\tau_{\sigma\lambda}\mathbf{W}^{\sigma\rho}
+\tGam^\rho_{\lambda\sigma}\mathbf{W}^{\tau\sigma}
-\tGam^\sigma_{\sigma\lambda}\mathbf{W}^{\tau\rho}
-\frac{4\pi}{(n\!-\!1)}(\hbj^{\rho}\delta^{\tau}_\lambda-\hbj^{\tau}\delta^{\rho}_\lambda).
\end{eqnarray}
Differentiating (\ref{Lconnections}), antisymmetrizing, and substituting
(\ref{Lconnections}) for $\mathbf{W}^{\tau\rho}{_{,\lambda}}$ gives,
\begin{eqnarray}
\fl 0&=&\!\left(\mathbf{W}^{\tau\rho}{_{,[\lambda}}
+\tGam^\tau_{\sigma[\lambda}\mathbf{W}^{\sigma\rho}
+\tGam^\rho_{[\lambda\vert\sigma}\mathbf{W}^{\tau\sigma}
-\tGam^\sigma_{\sigma[\lambda}\mathbf{W}^{\tau\rho}
-\frac{4\pi}{(n\!-\!1)}(\hbj^\rho\delta^{\tau}_{[\lambda}
-\hbj^\tau\delta^{\rho}_{[\lambda})\right){_{\!\!,\,\nu]}}\\
\fl &=&\tGam^\tau_{\sigma[\lambda,\nu]}\mathbf{W}^{\sigma\rho}
+\tGam^\rho_{[\lambda\vert\sigma,\vert\nu]}\mathbf{W}^{\tau\sigma}
-\tGam^\sigma_{\sigma[\lambda,\nu]}\mathbf{W}^{\tau\rho}
-\frac{4\pi}{(n\!-\!1)}(\hbj^\rho{_{,[\nu}}\delta^\tau_{\lambda]}
-\hbj^\tau{_{,[\nu}}\delta^\rho_{\lambda]})\nonumber\\
\fl &&+\tGam^\tau_{\sigma[\lambda}\mathbf{W}^{\sigma\rho}{_{,\nu]}}
+\tGam^\rho_{[\lambda\vert\sigma}\mathbf{W}^{\tau\sigma}{_{,\nu]}}
-\tGam^\sigma_{\sigma[\lambda}\mathbf{W}^{\tau\rho}{_{,\nu]}}\\
%\end{eqnarray}
%%\bigskip
%\begin{eqnarray}
\label{mess}
\fl &=&\tGam^\tau_{\sigma[\lambda,\nu]}\mathbf{W}^{\sigma\rho}
+\tGam^\rho_{[\lambda\vert\sigma,\vert\nu]}\mathbf{W}^{\tau\sigma}
-\tGam^\sigma_{\sigma[\lambda,\nu]}\mathbf{W}^{\tau\rho}
-\frac{4\pi}{(n\!-\!1)}(\hbj^\rho{_{,[\nu}}\delta^\tau_{\lambda]}
-\hbj^\tau{_{,[\nu}}\delta^\rho_{\lambda]})\nonumber\\
\fl &&-\tGam^\tau_{\sigma[\lambda}\left(
\tGam^\sigma_{\alpha\vert\nu]}\mathbf{W}^{\alpha\rho}
+\tGam^\rho_{\nu]\alpha}\mathbf{W}^{\sigma\alpha}
-\tGam^\alpha_{\nu]\alpha}\mathbf{W}^{\sigma\rho}
-\frac{4\pi}{(n\!-\!1)}(\hbj^\rho\delta^\sigma_{\nu]}
-\hbj^\sigma\delta^\rho_{\nu]})\right)\nonumber\\
\fl &&-\tGam^\rho_{[\lambda\vert\sigma}\!\left(
\tGam^\tau_{\alpha\vert\nu]}\mathbf{W}^{\alpha\sigma}
+\tGam^\sigma_{\nu]\alpha}\mathbf{W}^{\tau\alpha}
-\tGam^\alpha_{\nu]\alpha}\mathbf{W}^{\tau\sigma}
-\frac{4\pi}{(n\!-\!1)}(\hbj^\sigma\delta^\tau_{\nu]}
-\hbj^\tau\delta^\sigma_{\nu]})\right)\nonumber\\
\fl &&+\tGam^\sigma_{\sigma[\lambda}\left(
\tGam^\tau_{\alpha\vert\nu]}\mathbf{W}^{\alpha\rho}
+\tGam^\rho_{\nu]\alpha}\mathbf{W}^{\tau\alpha}
-\tGam^\alpha_{\nu]\alpha}\mathbf{W}^{\tau\rho}
-\frac{4\pi}{(n\!-\!1)}(\hbj^\rho\delta^\tau_{\nu]}
-\hbj^\tau\delta^\rho_{\nu]})\right).
\end{eqnarray}
Cancelling the terms 2B-3A, 2C-4A, 3C-4B and using (\ref{HermitianizedRiemann}) gives,
\begin{eqnarray}
\label{result}
\fl 0&=&\frac{1}{2}\left[\mathbf{W}^{\sigma\rho}\tR^\tau{_{\sigma\nu\lambda}}
+\mathbf{W}^{\tau\sigma}{\mathcal R}^\rho{_{\sigma\nu\lambda}}(\tGam^T)\right]
+\frac{4\pi}{(n\!-\!1)}\left[\tGam{^\tau_{\![\nu\lambda]}}\hbj^\rho
-\tGam{^\rho_{\![\lambda\nu]}}\hbj^\tau\right]\nonumber\\
\fl &&+\frac{4\pi}{(n\!-\!1)}\left[(\hbj^\tau{_{,[\nu}}
\!+\!\tGam^\tau_{\sigma[\nu}\hbj^\sigma
\!-\!\tGam^\sigma_{\sigma[\nu}\hbj^\tau)\delta^\rho_{\lambda]}
-(\hbj^\rho{_{,[\nu}}
\!+\!\tGam^\rho_{\![\nu\vert\sigma}\hbj^\sigma
\!-\!\tGam^\sigma_{\sigma[\nu}\hbj^\rho)\delta^\tau_{\lambda]}\right].
\end{eqnarray}
%\bigskip\\
Multiplying by 2, contracting over $^\rho_\nu$, and using (\ref{Hermiticity})
and $\hbj^\nu_{,\nu}\!=\!0$ from (\ref{continuity}) gives,
\begin{eqnarray}
\fl 0&=&\mathbf{W}^{\sigma\nu}\tR^\tau{_{\sigma\nu\lambda}}
+\mathbf{W}^{\tau\sigma}{\mathcal R}^\nu{_{\sigma\nu\lambda}}(\tGam^T)
+\frac{8\pi}{(n\!-\!1)}\left[\tGam{^\tau_{\![\nu\lambda]}}\hbj^\nu
-\tGam{^\nu_{\![\lambda\nu]}}\hbj^\tau\right]\nonumber\\
\fl &&+\frac{8\pi}{(n\!-\!1)}\left[(\hbj^\tau{_{,[\nu}}
\!+\!\tGam^\tau_{\sigma[\nu}\hbj^\sigma
\!-\!\tGam^\sigma_{\sigma[\nu}\hbj^\tau)\delta^\nu_{\lambda]}
-(\hbj^\nu{_{,[\nu}}
\!+\!\tGam^\nu_{\![\nu\vert\sigma}\hbj^\sigma
\!-\!\tGam^\sigma_{\sigma[\nu}\hbj^\nu)\delta^\tau_{\lambda]}\right]\\
\label{intermediate}
\fl &=&\mathbf{W}^{\sigma\nu}\tR^\tau{_{\sigma\nu\lambda}}
+\mathbf{W}^{\tau\sigma}\tR_{\lambda\sigma}
-\frac{4\pi(n\!-\!2)}{(n\!-\!1)}(\hbj^\tau{_{,\lambda}}
\!+\!\tGam^\tau_{\sigma\lambda}\hbj^\sigma
\!-\!\tGam^\sigma_{\sigma\lambda}\hbj^\tau).
\end{eqnarray}
This is a generalization of the symmetry $R^\tau{_\lambda}=R_\lambda{^\tau}$
%that occurs with
of the ordinary Ricci tensor.

%\newpage
Next we will use the generalized uncontracted Bianchi identity\cite{EinsteinBianchi},
which can be verified by direct computation,
\begin{eqnarray}
\label{PBianchi}
  \tR{^{\Stacksymbols{+}{\tau}{0}{1}}}{_{\Stacksymbols{\sigma}{+}{6}{1}{\Stacksymbols{\nu}{-}{6}{1}}{\Stacksymbols{\alpha}{+}{6}{1}};\lambda}}
 +\tR{^{\Stacksymbols{+}{\tau}{0}{1}}}{_{\Stacksymbols{\sigma}{+}{6}{1}{\Stacksymbols{\alpha}{+}{6}{1}}{\Stacksymbols{\lambda}{+}{6}{1}};\nu}}
 +\tR{^{\Stacksymbols{+}{\tau}{0}{1}}}{_{\Stacksymbols{\sigma}{+}{6}{1}{\Stacksymbols{\lambda}{-}{6}{1}}{\Stacksymbols{\nu}{-}{6}{1}};\alpha}}
=0.
\end{eqnarray}
The $+/-$ notation is from \cite{EinsteinBianchi} and indicates that covariant
derivative is being done with $\tGam^\alpha_{\nu\mu}$ instead of the usual
$\Gamma^\alpha_{\nu\mu}$. A plus by
an index means that the associated derivative index is to be placed on the right side of the
connection, and a minus means that it is to be placed on the left side.
Note that the identity (\ref{PBianchi}) is true for either the ordinary Riemann tensor or
for our definition (\ref{HermitianizedRiemann}).
This is because the two tensors differ by the term $\delta^\tau_\nu\tGam^\sigma_{\sigma[\alpha,\mu]}$,
so that the expression (\ref{PBianchi}) would differ by the term
$\delta^\tau_\sigma(
 \tGam^\rho_\rho{_{[{\Stacksymbols{\nu}{-}{6}{1}},{\Stacksymbols{\alpha}{+}{6}{1}]};\lambda}}
+\tGam^\rho_\rho{_{[{\Stacksymbols{\alpha}{+}{6}{1}},{\Stacksymbols{\lambda}{+}{6}{1}]};\nu}}
+\tGam^\rho_\rho{_{[{\Stacksymbols{\lambda}{-}{6}{1}},{\Stacksymbols{\nu}{-}{6}{1}]};\alpha}})$.
But this difference vanishes because for an arbitrary curl $Y_{[\alpha,\lambda]}$ we have
\begin{eqnarray}
\fl Y{_{[{\Stacksymbols{\nu}{-}{6}{1}},{\Stacksymbols{\alpha}{+}{6}{1}]};\lambda}}
   +Y{_{[{\Stacksymbols{\alpha}{+}{6}{1}},{\Stacksymbols{\lambda}{+}{6}{1}]};\nu}}
   +Y{_{[{\Stacksymbols{\lambda}{-}{6}{1}},{\Stacksymbols{\nu}{-}{6}{1}]};\alpha}}
&=&Y_{[\nu,\alpha],\lambda}
-\tGam^\sigma_{\lambda\nu}Y_{[\sigma,\alpha]}
-\tGam^\sigma_{\alpha\lambda}Y_{[\nu,\sigma]}\nonumber\\
&+&Y_{[\alpha,\lambda],\nu}
-\tGam^\sigma_{\alpha\nu}Y_{[\sigma,\lambda]}
-\tGam^\sigma_{\lambda\nu}Y_{[\alpha,\sigma]}\nonumber\\
&+&Y_{[\lambda,\nu],\alpha}
-\tGam^\sigma_{\alpha\lambda}Y_{[\sigma,\nu]}
-\tGam^\sigma_{\alpha\nu}Y_{[\lambda,\sigma]}=0.
\end{eqnarray}

A simple form of the generalized contracted Bianchi identity results if we contract (\ref{PBianchi})
over $\mathbf{W}^{\sigma\nu}$ and $^\tau_\alpha$, then substitute (\ref{intermediate}) for
$\mathbf{W}^{\sigma\nu}\tR^\tau{_{\sigma\nu\lambda}}$ and (\ref{Lconnections}) for
$\mathbf{W}^{\Stacksymbols{+}{\sigma}{0}{1}\Stacksymbols{-}{\nu}{0}{1}}{_{;\tau}}$,
\begin{eqnarray}
\fl 0&=&\mathbf{W}^{\sigma\nu}( \tR{^{\Stacksymbols{+}{\tau}{0}{1}}}{_{\Stacksymbols{\sigma}{+}{6}{1}{\Stacksymbols{\nu}{-}{6}{1}}{\Stacksymbols{\tau}{+}{6}{1}};\lambda}}
   +\tR{^{\Stacksymbols{+}{\tau}{0}{1}}}{_{\Stacksymbols{\sigma}{+}{6}{1}{\Stacksymbols{\tau}{+}{6}{1}}{\Stacksymbols{\lambda}{+}{6}{1}};\nu}}
   +\tR{^{\Stacksymbols{+}{\tau}{0}{1}}}{_{\Stacksymbols{\sigma}{+}{6}{1}{\Stacksymbols{\lambda}{-}{6}{1}}{\Stacksymbols{\nu}{-}{6}{1}};\tau}})\\
\fl &=&-\mathbf{W}^{\sigma\nu}\tR{_{\Stacksymbols{\sigma}{+}{6}{1}{\Stacksymbols{\nu}{-}{6}{1}};\lambda}}
   +\mathbf{W}^{\sigma\nu}\tR{_{\Stacksymbols{\sigma}{+}{6}{1}{\Stacksymbols{\lambda}{+}{6}{1}};\nu}}
   -\mathbf{W}^{\sigma\nu}\tR{^{\Stacksymbols{+}{\tau}{0}{1}}}{_{\Stacksymbols{\sigma}{+}{6}{1}{\Stacksymbols{\nu}{-}{6}{1}}{\Stacksymbols{\lambda}{-}{6}{1}};\tau}}\\
\fl &=&-\mathbf{W}^{\sigma\nu}\tR{_{\Stacksymbols{\sigma}{+}{6}{1}{\Stacksymbols{\nu}{-}{6}{1}};\lambda}}
   +(\mathbf{W}^{\sigma\Stacksymbols{-}{\nu}{0}{1}}\tR{_{\sigma{\Stacksymbols{\lambda}{+}{6}{1}}}}){_{;\nu}}
   -(\mathbf{W}^{\sigma\nu}\tR{^{\Stacksymbols{+}{\tau}{0}{1}}}{_{\sigma\nu{\Stacksymbols{\lambda}{-}{6}{1}}}}){_{;\tau}}\nonumber\\
\fl &&~~~~~~~~~~~~~~~~~~~~~~
 -\mathbf{W}^{\Stacksymbols{+}{\sigma}{0}{1}\Stacksymbols{-}{\nu}{0}{1}}{_{;\nu}}\tR_{\sigma\lambda}
 +\mathbf{W}^{\Stacksymbols{+}{\sigma}{0}{1}\Stacksymbols{-}{\nu}{0}{1}}{_{;\tau}}\tR^\tau{_{\sigma\nu\lambda}}\\
\fl &=&-\mathbf{W}^{\sigma\nu}\tR{_{\Stacksymbols{\sigma}{+}{6}{1}{\Stacksymbols{\nu}{-}{6}{1}};\lambda}}
 +(\mathbf{W}^{\sigma\Stacksymbols{-}{\nu}{0}{1}}\tR{_{\sigma{\Stacksymbols{\lambda}{+}{6}{1}}}}){_{;\nu}}\nonumber\\
\fl &&+\left(\mathbf{W}^{\Stacksymbols{+}{\tau}{0}{1}\sigma}\tR_{\Stacksymbols{\lambda}{-}{6}{1}\sigma}
-\frac{4\pi(n\!-\!2)}{(n\!-\!1)}(\hbj^{\Stacksymbols{+}{\tau}{0}{1}}{_{,\Stacksymbols{\lambda}{-}{6}{1}}}
\!+\!\tGam^{\Stacksymbols{+}{\tau}{0}{1}}_{\sigma\Stacksymbols{\lambda}{-}{6}{1}}\hbj^\sigma
\!-\!\tGam^\sigma_{\sigma\Stacksymbols{\lambda}{-}{6}{1}}\hbj^{\Stacksymbols{+}{\tau}{0}{1}})\right){_{\!;\,\tau}}\nonumber\\
\fl &&-\frac{4\pi}{(n\!-\!1)}(\hbj^{\nu}\delta^{\sigma}_\nu-\hbj^{\sigma}\delta^{\nu}_\nu)\tR_{\sigma\lambda}
 +\frac{4\pi}{(n\!-\!1)}(\hbj^{\nu}\delta^{\sigma}_\tau-\hbj^{\sigma}\delta^{\nu}_\tau)\tR^\tau{_{\sigma\nu\lambda}}\\
%\end{eqnarray}
%\begin{eqnarray}
\fl~~&=&-\mathbf{W}^{\sigma\nu}\tR{_{\Stacksymbols{\sigma}{+}{6}{1}{\Stacksymbols{\nu}{-}{6}{1}};\lambda}}
 +(\mathbf{W}^{\sigma\Stacksymbols{-}{\nu}{0}{1}}\tR{_{\sigma{\Stacksymbols{\lambda}{+}{6}{1}}}}){_{;\nu}}
+(\mathbf{W}^{\Stacksymbols{+}{\nu}{0}{1}\sigma}\tR_{\Stacksymbols{\lambda}{-}{6}{1}\sigma}){_{;\nu}}\nonumber\\
\fl &&-\frac{4\pi(n\!-\!2)}{(n\!-\!1)}(\hbj^{\Stacksymbols{+}{\tau}{0}{1}}{_{,\Stacksymbols{\lambda}{-}{6}{1}}}
\!+\!\tGam^{\Stacksymbols{+}{\tau}{0}{1}}_{\sigma\Stacksymbols{\lambda}{-}{6}{1}}\hbj^\sigma
\!-\!\tGam^\sigma_{\sigma\Stacksymbols{\lambda}{-}{6}{1}}\hbj^{\Stacksymbols{+}{\tau}{0}{1}}){_{;\tau}}\nonumber\\
\fl &&+\frac{4\pi(n\!-\!2)}{(n\!-\!1)}\hbj^{\sigma}\tR_{\sigma\lambda}
 +\frac{4\pi}{(n\!-\!1)}\hbj^{\nu}\tR^\sigma{_{\sigma\nu\lambda}}
\end{eqnarray}
\begin{eqnarray}
\label{manyterms}
\fl~~&=&-\mathbf{W}^{\sigma\nu}(\tR{_{\sigma\nu,\lambda}}
-\tGam^\alpha_{\sigma\lambda}\tR{_{\alpha\nu}}
-\tGam^\alpha_{\lambda\nu}\tR{_{\sigma\alpha}})\nonumber\\
\fl &&+(\mathbf{W}^{\sigma\nu}\tR{_{\sigma\lambda}}){_{,\nu}}
+\tGam^\nu_{\nu\alpha}\mathbf{W}^{\sigma\alpha}\tR{_{\sigma\lambda}}
-\tGam^\alpha_{\lambda\nu}\mathbf{W}^{\sigma\nu}\tR{_{\sigma\alpha}}
-\tGam^\alpha_{\alpha\nu}\mathbf{W}^{\sigma\nu}\tR{_{\sigma\lambda}}\nonumber\\
\fl &&+(\mathbf{W}^{\nu\sigma}\tR_{\lambda\sigma}){_{,\nu}}
+\tGam^\nu_{\alpha\nu}\mathbf{W}^{\alpha\sigma}\tR_{\lambda\sigma}
-\tGam^\alpha_{\nu\lambda}\mathbf{W}^{\nu\sigma}\tR_{\alpha\sigma}
-\tGam^\alpha_{\alpha\nu}\mathbf{W}^{\nu\sigma}\tR_{\lambda\sigma}\nonumber\\
\fl  && -\frac{4\pi(n\!-\!2)}{(n\!-\!1)}[\hbj^\tau{_{,\lambda,\tau}}
\!+\!\tGam^\tau_{\sigma\lambda,\tau}\hbj^\sigma
\!+\!\tGam^\tau_{\sigma\lambda}\hbj^\sigma_{,\tau}
\!-\!\tGam^\sigma_{\sigma\lambda,\tau}\hbj^\tau
\!-\!\tGam^\sigma_{\sigma\lambda}\hbj^\tau_{,\tau}\nonumber\\
\fl&&~~~~~~~~~~~~~~~~~
+\tGam^\tau_{\alpha\tau}(\hbj^\alpha{_{,\lambda}}
\!+\!\tGam^\alpha_{\sigma\lambda}\hbj^\sigma
\!-\!\tGam^\sigma_{\sigma\lambda}\hbj^\alpha)\nonumber\\
\fl&&~~~~~~~~~~~~~~~~~
-\tGam^\alpha_{\tau\lambda}(\hbj^\tau{_{,\alpha}}
\!+\!\tGam^\tau_{\sigma\alpha}\hbj^\sigma
\!-\!\tGam^\sigma_{\sigma\alpha}\hbj^\tau)\nonumber\\
\fl&&~~~~~~~~~~~~~~~~~
-\tGam^\alpha_{\alpha\tau}(\hbj^\tau{_{,\lambda}}
\!+\!\tGam^\tau_{\sigma\lambda}\hbj^\sigma
\!-\!\tGam^\sigma_{\sigma\lambda}\hbj^\tau)\nonumber\\
\fl&&~~~~~~~~~~~~~~~~~
-\hbj^{\sigma}(\tR_{\sigma\lambda}-\tGam^\alpha_{\alpha[\sigma,\lambda]})
 -\hbj^{\sigma}(\tGam^\alpha_{\alpha\sigma,\lambda}\!-\!\tGam^\alpha_{\alpha\lambda,\sigma})]
\end{eqnarray}
With the $\hbj^\sigma$ terms of (\ref{manyterms}), 4C-6A,4D-8D,5A-7A,5B-7B,5C-7C all cancel,
4A and 4E are zero because $\hbj^\nu_{,\nu}\!=\!0$ from (\ref{continuity}),
and 4B,6B,6C,8C cancel the Ricci tensor term 8A,8B.
With the $\mathbf{W}^{\tau\sigma}$ terms of (\ref{manyterms}), all those
with a $\tGam^\alpha_{\nu\mu}$ factor cancel, which are the terms
1C-2C,1B-3C,2B-2D,3B-3D. Doing the cancellations and using (\ref{Mdef}) we get
\begin{eqnarray}
\label{simpleBianchi}
\fl~~~~~~~0&=&(\rmN N^{\dashv\nu\sigma}\tR{_{\sigma\lambda}}
+\rmN N^{\dashv\sigma\nu}\tR_{\lambda\sigma}){_{,\nu}}
-\rmN N^{\dashv\nu\sigma}\tR{_{\sigma\nu,\lambda}}.
\end{eqnarray}
Equation (\ref{simpleBianchi}) is a simple generalization of the ordinary contracted Bianchi
identity $2(\rmg\,R^\nu{_\lambda})_{,\nu}\!-\!\rmg\,g^{\nu\sigma}\!R_{\sigma\nu,\lambda}\!=\!0$,
and it applies even when $j^\tau\!\ne0$.
Because $\tGam^\alpha_{\nu\mu}$ has cancelled out of (\ref{simpleBianchi}),
the Christoffel connection $\Gamma^\alpha_{\nu\mu}$ would also cancel, so
a manifestly tensor
relation can be obtained by replacing the ordinary derivatives with covariant
derivatives done with $\Gamma^\alpha_{\nu\mu}$,
\begin{eqnarray}
\label{simpleBianchi2}
\fl~~~~~~~0&=&(\rmN N^{\dashv\nu\sigma}\tR{_{\sigma\lambda}}
+\rmN N^{\dashv\sigma\nu}\tR_{\lambda\sigma}){_{;\nu}}
-\rmN N^{\dashv\nu\sigma}\tR{_{\sigma\nu;\lambda}}.
\end{eqnarray}
%\bigskip\\
Rewriting the identity in terms of $g^{\rho\tau}$ and $\hf^{\rho\tau}$
as defined by (\ref{Mdef},\ref{notation0}) gives,
\begin{eqnarray}
\fl 0&=&(\rmg\,(g^{\sigma\nu}\!+\!\hf^{\sigma\nu})\tR{_{\sigma\lambda}}
\!+\rmg\,(g^{\nu\sigma}\!+\!\hf^{\nu\sigma})\tR_{\lambda\sigma}){_{;\nu}}
\!-\!\rmg\,(g^{\sigma\nu}\!+\!\hf^{\sigma\nu})\tR{_{\sigma\nu;\lambda}}\\
%\fl&+&(\rmg\,f^{\sigma\nu}\tR{_{\sigma\lambda}}
%\!+\rmg\,f^{\nu\sigma}\tR_{\lambda\sigma}){_{;\nu}}\rmt\Lambda_b^{\!-1/2}
%\!-\!\rmg\,f^{\sigma\nu}\tR{_{\sigma\nu;\lambda}}\rmt\Lambda_b^{\!-1/2}\\
\fl &=&\rmg\,[2\tR^{(\nu}{_{\lambda);\nu}}
-\tR^\sigma_{\sigma;\lambda}]
+\rmg\,[2(\hf^{\nu\sigma}\tR{_{[\lambda\sigma]}}){_{;\nu}}
\!+\!\hf^{\nu\sigma}\tR{_{[\sigma\nu];\lambda}}]\\
\fl &=&\rmg\,[2\tR^{(\nu}{_{\lambda)}}{_{;\nu}}
-\tR^\sigma_{\sigma;\lambda}]
+\rmg\,[3\hf^{\nu\sigma}\tR{_{[\sigma\nu,\lambda]}}
\!+2\hf^{\nu\sigma}{_{;\nu}}\tR{_{[\lambda\sigma]}}].
\end{eqnarray}
Dividing by $2\rmg$ gives another form of the generalized contracted Bianchi identity
\begin{eqnarray}
\fl~~~~~ \left(\tR^{(\nu}{_{\lambda)}}
-\frac{1}{2}\delta^\nu_\lambda\tR^\sigma_\sigma\right){_{\!;\,\nu}}
=\frac{3}{2}\hf^{\nu\sigma}\tR{_{[\nu\sigma,\lambda]}}
+\hf^{\nu\sigma}{_{\!;\nu}}\tR{_{[\sigma\lambda]}}.
\end{eqnarray}
From (\ref{notation0},\ref{genEinstein}) we get the final result (\ref{contractedBianchi}).
%\bigskip\\

\section{\label{ExtractionofConnectionAddition}Extraction of a connection
addition from the Hermitianized Ricci tensor}
Substituting $\tGam^\alpha_{\nu\mu}\!=\!\Gamma^\alpha_{\nu\mu}
\!+\!\Upsilon^\alpha_{\nu\mu}$ from (\ref{gammadecomposition},\ref{Christoffel})
into (\ref{HermitianizedRiccit})
%and using the notation $\bUps^\alpha_{\nu\mu}
%=\Upsilon^\alpha_{(\nu\mu)}$,
%$\cUps^\alpha_{\nu\mu}=\Upsilon^\alpha_{[\nu\mu]}$
gives
\begin{eqnarray}
%\label{Ricciaddition}
\fl \hR_{\nu\mu}(\tGam)
\!&=&2[(\Gamma^\alpha_{\nu[\mu}
\!+\!\Upsilon^\alpha_{\nu[\mu}){_{,\alpha]}}
+(\Gamma^\sigma_{\nu[\mu}
\!+\!\Upsilon^\sigma_{\nu[\mu})(\Gamma^\alpha_{\sigma|\alpha]}
\!+\!\Upsilon^\alpha_{\sigma|\alpha]})]
+\!(\Gamma^\alpha_{\alpha[\nu}
\!+\!\Upsilon^\alpha_{\alpha[\nu})_{,\mu]}\\
\iftrue
\fl &=&R_{\nu\mu}(\Gamma)+\Upsilon^\alpha_{\nu\mu,\alpha}
-\Gamma^\sigma_{\nu\alpha}\Upsilon^\alpha_{\sigma\mu}
+\Gamma^\alpha_{\sigma\alpha}\Upsilon^\sigma_{\nu\mu}
-\Gamma^\alpha_{\sigma\mu}\Upsilon^\sigma_{\nu\alpha}\nonumber\\
\nopagebreak
\fl &&~~~~~~~~~-\Upsilon^\alpha_{\alpha(\nu,\mu)}
+\Gamma^\sigma_{\nu\mu}\Upsilon^\alpha_{\sigma\alpha}
-\Upsilon^\sigma_{\nu\alpha}\Upsilon^\alpha_{\sigma\mu}
+\Upsilon^\sigma_{\nu\mu}\Upsilon^\alpha_{\sigma\alpha}\\
\fi
%\label{Ricciaddition2}
\fl \!&=&R_{\nu\mu}(\Gamma)+\Upsilon^\alpha_{\nu\mu;\alpha}
-\Upsilon^\alpha_{\alpha(\nu;\mu)}
-\Upsilon^\sigma_{\nu\alpha}\Upsilon^\alpha_{\sigma\mu}
+\Upsilon^\sigma_{\nu\mu}\Upsilon^\alpha_{\sigma\alpha},\\
\fl\!R_{(\nu\mu)}(\tGam)
\label{Ricciadditionsymmetric}
%\fl
\!&=& R_{\nu\mu}(\Gamma)+{\Upsilon}^\alpha_{\!(\nu\mu);\alpha}
\!-\!\Upsilon^\alpha_{\alpha(\nu;\mu)}
\!-\!{\Upsilon}^\sigma_{\!(\nu\alpha)}{\Upsilon}^\alpha_{\!(\sigma\mu)}
\!-\!{\Upsilon}^\sigma_{\![\nu\alpha]}{\Upsilon}^\alpha_{\![\sigma\mu]}
\!+\!{\Upsilon}^\sigma_{\!(\nu\mu)}\Upsilon^\alpha_{\sigma\alpha},\\
\fl\!R_{[\nu\mu]}(\tGam)
\label{Ricciadditionantisymmetric}
\!&=&{\Upsilon}^\alpha_{\![\nu\mu];\alpha}
%\!-\!\Upsilon^\alpha_{\alpha[\nu;\mu]}
\!-\!{\Upsilon}^\sigma_{\!(\nu\alpha)}{\Upsilon}^\alpha_{\![\sigma\mu]}
\!-\!{\Upsilon}^\sigma_{\![\nu\alpha]}{\Upsilon}^\alpha_{\!(\sigma\mu)}
\!+\!{\Upsilon}^\sigma_{\![\nu\mu]}\Upsilon^\alpha_{\sigma\alpha}.
\end{eqnarray}
Also, substituting $\nGam^\alpha_{\!\nu\mu}\!=\!\tGam^\alpha_{\!\nu\mu}
\!+[\,\delta^\alpha_\mu \Aphi_\nu\!-\delta^\alpha_\nu \Aphi_\mu]\!\rmt\Lambda_b^{\!1/2}$
from (\ref{gamma_natural}) into (\ref{HermitianizedRicci}) and using
$\tGam^\alpha_{\nu\alpha}\!=\!\nGam^\alpha_{\!(\nu\alpha)}\!=\!\tGam^\alpha_{\alpha\nu}$ gives
\begin{eqnarray}
\fl \hR_{\nu\mu}(\nGam)
\!&=&\tGam^\alpha_{\!\nu\mu,\alpha}
\!+[\,\delta^\alpha_\mu \Aphi_\nu\!-\delta^\alpha_\nu \Aphi_\mu]_{,\alpha}\!\rmt\Lambda_b^{\!1/2}
\!-\tGam^\alpha_{\!\alpha(\nu,\mu)}\nonumber\\
\fl &&+\left(\tGam^\sigma_{\!\nu\mu}+[\,\delta^\sigma_\mu \Aphi_\nu
\!-\delta^\sigma_\nu \Aphi_\mu]\rmt\Lambda_b^{\!1/2}\right)\tGam^\alpha_{\!\sigma\alpha}\nonumber\\
\fl &&-\left(\tGam^\sigma_{\!\nu\alpha}\!+[\,\delta^\sigma_\alpha \Aphi_\nu
-\delta^\sigma_\nu \Aphi_\alpha]\rmt\Lambda_b^{\!1/2}\right)\!
\left(\tGam^\alpha_{\!\sigma\mu}\!+[\,\delta^\alpha_\mu \Aphi_\sigma-\delta^\alpha_\sigma \Aphi_\mu]\rmt\Lambda_b^{\!1/2}\right)\nonumber\\
\fl &&+2(n\!-\!1)A_\nu A_\mu\Lambda_b\\
\fl\!&=&\tGam^\alpha_{\!\nu\mu,\alpha}
\!+2\Aphi_{[\nu,\mu]}\rmt\Lambda_b^{\!1/2}
-\tGam^\alpha_{\!\alpha(\nu,\mu)}+\tGam^\sigma_{\!\nu\mu}\tGam^\alpha_{\!\sigma\alpha}
+[\,\Aphi_\nu\tGam^\alpha_{\!\mu\alpha}
-\Aphi_\mu\tGam^\alpha_{\!\nu\alpha}]\rmt\Lambda_b^{\!1/2}\nonumber\\
\fl &&-\tGam^\sigma_{\!\nu\alpha}\tGam^\alpha_{\!\sigma\mu}
-[\,\tGam^\sigma_{\!\nu\mu}\Aphi_\sigma-\tGam^\sigma_{\!\nu\sigma}\Aphi_\mu]\rmt\Lambda_b^{\!1/2}
-[\,\Aphi_\nu\tGam^\alpha_{\!\alpha\mu}
-\Aphi_\alpha\tGam^\alpha_{\!\nu\mu}]\rmt\Lambda_b^{\!1/2}\nonumber\\
\fl &&+2\Aphi_\nu\Aphi_\mu(1-n-1+1)\Lambda_b
+2(n\!-\!1)A_\nu A_\mu\Lambda_b\\
\label{RnGam}
\fl &=&\tGam^\alpha_{\!\nu\mu,\alpha}
-\tGam^\alpha_{\alpha(\nu,\mu)}+\tGam^\sigma_{\!\nu\mu}\tGam^\alpha_{\!\sigma\alpha}
-\tGam^\sigma_{\!\nu\alpha}\tGam^\alpha_{\!\sigma\mu}
\!+2\Aphi_{[\nu,\mu]}\rmt\Lambda_b^{\!1/2}\\
\fl &=&\hR_{\nu\mu}(\tGam)\!+2\Aphi_{[\nu,\mu]}\rmt\Lambda_b^{\!1/2}.
\end{eqnarray}

\section{\label{VariationalDerivative}Variational derivatives for fields with
the symmetry $\tGam^\sigma_{\![\mu\sigma]}\!=\!0$}
The field equations associated with a field with symmetry properties must have the
same number of independent components as the field. For a field with the symmetry
$\tGam^\sigma_{\![\mu\sigma]}=0$, the field equations can be found by introducing
a Lagrange multiplier $\Omega^\mu$,
\begin{eqnarray}
0=\delta\int({\mathcal L}+\Omega^\mu\tGam^\sigma_{\![\mu\sigma]})d^n x.
\end{eqnarray}
Minimizing the integral with respect to $\Omega^\mu$ shows that the symmetry is enforced.
Using the definition,
\begin{eqnarray}
\frac{\Delta{\mathcal L}}{\Delta\tGam^\beta_{\tau\rho}}
=\frac{\partial{\mathcal L}}{\partial\tGam^\beta_{\tau\rho}}
-\left(\frac{\partial{\mathcal L}}
{\partial\tGam^\beta_{\tau\rho,\omega}}\right)\!{_{,\,\omega}}~...~,
\end{eqnarray}
and minimizing the integral with respect to $\tGam^\beta_{\!\tau\rho}$ gives
\begin{eqnarray}
\label{minimization}
0=\frac{\Delta{\mathcal L}}{\Delta\tGam^\beta_{\tau\rho}}
+\Omega^\mu\delta^\sigma_\beta\delta^\tau_{[\mu}\delta^\rho_{\sigma]}
=\frac{\Delta{\mathcal L}}{\Delta\tGam^\beta_{\tau\rho}}
+\frac{1}{2}(\Omega^\tau\delta^\rho_\beta-\delta^\tau_\beta\Omega^\rho).
\end{eqnarray}
Contracting this on the left and right gives
\begin{eqnarray}
\label{lagrangemultiplier}
\Omega^\rho=\frac{2}{(n\!-\!1)}\frac{\Delta{\mathcal L}}{\Delta\tGam^\alpha_{\alpha\rho}}
=-\frac{2}{(n\!-\!1)}\frac{\Delta{\mathcal L}}{\Delta\tGam^\alpha_{\rho\alpha}}.
\end{eqnarray}
Substituting (\ref{lagrangemultiplier}) back into (\ref{minimization}) gives
\begin{eqnarray}
\label{usefulresult}
0&=&\frac{\Delta{\mathcal L}}
{\Delta\tGam^\beta_{\tau\rho}}
-\frac{\delta^\tau_\beta}{(n\!-\!1)}
\frac{\Delta{\mathcal L}}{\Delta\tGam^\alpha_{\alpha\rho}}
-\frac{\delta^\rho_\beta}{(n\!-\!1)} \frac{\Delta{\mathcal L}}
{\Delta\tGam^\alpha_{\tau\alpha}}.
\end{eqnarray}
In (\ref{lagrangemultiplier},\ref{usefulresult}) the index contractions occur after
the derivatives. Contracting (\ref{usefulresult}) on the right and left gives the
same result, so it has the same number of independent components as
$\tGam^\alpha_{\mu\nu}$. This is a general expression for the field equations
associated with a field having the symmetry $\tGam^\sigma_{\![\mu\sigma]}=0$.

\section{\label{ApproximateGamma}Solution for
${\tGam^\alpha_{\nu\mu}}$ in terms of $g_{\nu\mu}$ and
$f_{\nu\mu}$, with sources}
Here we derive the approximate solution
(\ref{upsilonsymmetric},\ref{upsilonantisymmetric})
to the connection equations (\ref{JSconnection}).
First let us define the notation
\begin{eqnarray}
\label{notation}
\fl~~~\hf^{\nu\mu}\!=\!f^{\nu\mu}\rmt\,\Lambda_b^{\!-1/2},~~~
\hj^\sigma\!=\!j^\sigma\rmt\,\Lambda_b^{\!-1/2},~~~
\bUps^\alpha_{\nu\mu}\!=\!\Upsilon^\alpha_{(\nu\mu)},~~~
\cUps^\alpha_{\nu\mu}\!=\!\Upsilon^\alpha_{[\nu\mu]}.
\end{eqnarray}
We assume that $|\hf^\nu{_\mu}|\!\ll\!1$ for all components of the
unitless field $\hf^\nu{_\mu}$, and find a solution
in the form of a power series expansion in $\hf^\nu{_\mu}$.
%It is easier to work with the contravariant connection equations
%(\ref{JSconnection}) than the covariant connection equations (\ref{JSconnection0}).
Using (\ref{contravariant}) and
\begin{eqnarray}
\fl\tGam^\sigma_{ \sigma\alpha}
&=&\frac{(\rmN)_{,\alpha}}{\rmN}+\frac{8\pi}{(n\!-\!2)(n\!-\!1)}\hj^\sigma N_{[\sigma\alpha]}
\end{eqnarray}
from (\ref{der0}) and
%$(\!\rmN)_{,\alpha}\!=\!\rmN\,\tGam^\sigma_{ \sigma\alpha}$,
$\tGam^\alpha_{\nu\mu}\!=\!\Gamma^\alpha_{\nu\mu}\!+\!\Upsilon^\alpha_{\nu\mu}$,
$(\rmN/\rmg\,)N^{\dashv\mu\nu}\!=\!g^{\nu\mu}\!+\!\hf^{\nu\mu}$
from (\ref{gammadecomposition},\ref{gdef},\ref{fdef}) we get
\begin{eqnarray}
\label{contravariant2}
\fl 0&=&\frac{\rmN}{\rmg}(N^{\dashv \mu\nu}{_{,\alpha}}
+\tGam^\nu_{\tau\alpha}N^{\dashv\mu\tau}
+\tGam^\mu_{\alpha\tau}N^{\dashv\tau\nu})\nonumber\\
\fl && -\frac{8\pi}{(n\!-\!1)}\!\left(
\hj^{[\mu}\delta^{\nu]}_\alpha
+\frac{1}{(n\!-\!2)}\hj^\tau N_{[\tau\alpha]}N^{\dashv\mu\nu}\!\right)\\
\fl &=&\left(\!\frac{\rmN N^{\dashv \mu\nu}}{\rmg}\right)\!{_{,\alpha}}
+\frac{\rmN}{\rmg}(\tGam^\nu_{\tau\alpha}N^{\dashv\mu\tau}
+\tGam^\mu_{\alpha\tau}N^{\dashv\tau\nu}
-(\tGam^\sigma_{ \sigma\alpha}
\!-\!\Gamma^\sigma_{\sigma\alpha})N^{\dashv\mu\nu})\nonumber\\
\fl && -\frac{8\pi}{(n\!-\!1)}\hj^{[\mu}\delta^{\nu]}_\alpha\\
\fl &=&(g^{\nu\mu}+\hf^{\nu\mu}){_{;\alpha}}
+\Upsilon^\nu_{\tau\alpha}(g^{\tau\mu}   +\hf^{\tau\mu})
+\Upsilon^\mu_{   \alpha\tau}(g^{\nu\tau}+\hf^{\nu\tau})
-\Upsilon^\sigma_{ \sigma\alpha}(g^{\nu\mu}  +\hf^{\nu\mu})\nonumber\\
\fl && -\frac{8\pi}{(n\!-\!1)}\hj^{[\mu}\delta^{\nu]}_\alpha\\
\label{contravariantreduced}
\fl &=&\hf^{\nu\mu}{_{;\alpha}}
+\Upsilon^\nu_{\tau\alpha}g^{\tau\mu}
+\Upsilon^\nu_{\tau\alpha}\hf^{\tau\mu}
+\Upsilon^\mu_{\alpha\tau}g^{\nu\tau}
+\Upsilon^\mu_{\alpha\tau}\hf^{\nu\tau}
-\Upsilon^\sigma_{\sigma\alpha}g^{\nu\mu}
-\Upsilon^\sigma_{\sigma\alpha}\hf^{\nu\mu}\nonumber\\
\fl && +\frac{4\pi}{(n\!-\!1)}(\hj^\nu\delta^\mu_\alpha-\hj^\mu\delta^\nu_\alpha).
\end{eqnarray}
Contracting this with $g_{\nu\mu}$ gives
\begin{eqnarray}
\label{contractedupsilon}
\fl~~~~~~~~ 0=(2-n)\Upsilon^\sigma_{\sigma\alpha}
-2{\cUps}^\sigma_{\alpha\tau}\hf^\tau{_\sigma}
~~\Rightarrow ~~\Upsilon^\sigma_{\sigma\alpha}
=\frac{2}{(n\!-\!2)}{\cUps}_{\sigma\tau\alpha}\hf^{\tau\sigma}.
\end{eqnarray}
Lowering the indices of (\ref{contravariantreduced}) and making linear
combinations of its permutations gives
\begin{eqnarray}
\fl\Upsilon_{\alpha\nu\mu}
=\Upsilon_{\alpha\nu\mu}
&+&\frac{1}{2}\left(\hf_{\nu\mu;\alpha}
+\!\Upsilon_{\nu\mu\alpha}+\!\Upsilon_{\nu\tau\alpha}\hf^\tau{_\mu}
+\!\Upsilon_{\mu\alpha\nu}+\!\Upsilon_{\mu\alpha\tau}\hf_\nu{^\tau}
-\!\Upsilon^\sigma_{\sigma\alpha}g_{\nu\mu}
-\!\Upsilon^\sigma_{\sigma\alpha}\hf_{\nu\mu}\phantom{\frac{1}{2}}\right.\nonumber\\
\fl && ~~~~+\left.\frac{4\pi}{(n\!-\!1)}(\hj_\nu g_{\alpha\mu}-\hj_\mu g_{\nu\alpha})\right)\nonumber\\
\fl &-&\frac{1}{2}\left(\hf_{\mu\alpha;\nu}
+\!\Upsilon_{\mu\alpha\nu}+\!\Upsilon_{\mu\tau\nu}\hf^\tau{_\alpha}
+\!\Upsilon_{\alpha\nu\mu}+\!\Upsilon_{\alpha\nu\tau}\hf_\mu{^\tau}
-\!\Upsilon^\sigma_{\sigma\nu}g_{\mu\alpha}
-\!\Upsilon^\sigma_{\sigma\nu}\hf_{\mu\alpha}\phantom{\frac{1}{2}}\right.\nonumber\\
\fl && ~~~~+\left.\frac{4\pi}{(n\!-\!1)}(\hj_\mu g_{\nu\alpha}-\hj_\alpha g_{\mu\nu})\right)\nonumber\\
\fl &-&\frac{1}{2}\left(\hf_{\alpha\nu;\mu}
+\!\Upsilon_{\alpha\nu\mu}+\!\Upsilon_{\alpha\tau\mu}\hf^\tau{_\nu}
+\!\Upsilon_{\nu\mu\alpha}+\!\Upsilon_{\nu\mu\tau}\hf_\alpha{^\tau}
-\!\Upsilon^\sigma_{\sigma\mu}g_{\alpha\nu}
-\!\Upsilon^\sigma_{\sigma\mu}\hf_{\alpha\nu}\phantom{\frac{1}{2}}\right.\nonumber\\
\fl && ~~~~+\left.\frac{4\pi}{(n\!-\!1)}(\hj_\alpha g_{\mu\nu}-\hj_\nu g_{\alpha\mu})\right).
\end{eqnarray}
Cancelling out the $\Upsilon_{\alpha\nu\mu}$ terms on the right-hand side, collecting terms,
and separating out the symmetric and antisymmetric parts gives,
\begin{eqnarray}
\label{barupsilonofN}
\fl\bUps_{\alpha\nu\mu}&=&
{\cUps}_{[\alpha\mu]\tau}\hf^\tau{_\nu}
+{\cUps}_{[\alpha\nu ]\tau}\hf^\tau{_\mu}
+{\cUps}_{(\nu\mu)\tau}\hf^\tau{_\alpha}
-\frac{1}{2}\Upsilon^\sigma_{\sigma\alpha}g_{\nu\mu}
+\Upsilon^\sigma_{\sigma(\nu}g_{\mu)\alpha}\\
\label{hatupsilonofN}
\fl\cUps_{\alpha\nu\mu}&=&
-{\bUps}_{(\alpha\mu)\tau}\hf^\tau{_\nu}
+{\bUps}_{(\alpha\nu )\tau}\hf^\tau{_\mu}
+{\bUps}_{[\nu\mu]\tau}\hf^\tau{_\alpha}
-\frac{1}{2}\Upsilon^\sigma_{\sigma\alpha}\hf_{\nu\mu}
+\Upsilon^\sigma_{\sigma[\nu}\hf_{\mu]\alpha}\nonumber\\
\fl &&+\frac{1}{2}
(\hf_{\nu\mu;\alpha}+\hf_{\alpha\mu;\nu}-\hf_{\alpha\nu;\mu})
+\frac{8\pi}{(n\!-\!1)}\hj_{[\nu} g_{\mu]\alpha}.
\end{eqnarray}
Substituting (\ref{barupsilonofN}) into (\ref{hatupsilonofN})
\ifnum\ExpandDerivations=1
%\pagebreak
\begin{eqnarray}
\fl\cUps_{\alpha\nu\mu}
&=&-\frac{1}{2}\left({\cUps}_{[\alpha\tau]\sigma}\hf^\sigma{_\mu}
+{\cUps}_{[\alpha\mu ]\sigma}\hf^\sigma{_\tau}
+{\cUps}_{(\mu\tau)\sigma}\hf^\sigma{_\alpha}
-\frac{1}{2}\Upsilon^\sigma_{\sigma\alpha}g_{\mu\tau}
+\Upsilon^\sigma_{\sigma(\mu}g_{\tau)\alpha}\right)\hf^\tau{_\nu}\nonumber\\
\fl &&-\frac{1}{2}\left({\cUps}_{[\mu\tau]\sigma}\hf^\sigma{_\alpha}
+{\cUps}_{[\mu\alpha ]\sigma}\hf^\sigma{_\tau}
+{\cUps}_{(\alpha\tau)\sigma}\hf^\sigma{_\mu}
-\frac{1}{2}\Upsilon^\sigma_{\sigma\mu}g_{\alpha\tau}
+\Upsilon^\sigma_{\sigma(\alpha}g_{\tau)\mu}\right)\hf^\tau{_\nu}\nonumber\\
\fl &&+\frac{1}{2}\left({\cUps}_{[\alpha\tau]\sigma}\hf^\sigma{_\nu}
+{\cUps}_{[\alpha\nu ]\sigma}\hf^\sigma{_\tau}
+{\cUps}_{(\nu\tau)\sigma}\hf^\sigma{_\alpha}
-\frac{1}{2}\Upsilon^\sigma_{\sigma\alpha}g_{\nu\tau}
+\Upsilon^\sigma_{\sigma(\nu}g_{\tau)\alpha}\right)\hf^\tau{_\mu}\nonumber\\
\fl &&+\frac{1}{2}\left({\cUps}_{[\nu\tau]\sigma}\hf^\sigma{_\alpha}
+{\cUps}_{[\nu\alpha ]\sigma}\hf^\sigma{_\tau}
+{\cUps}_{(\alpha\tau)\sigma}\hf^\sigma{_\nu}
-\frac{1}{2}\Upsilon^\sigma_{\sigma\nu}g_{\alpha\tau}
+\Upsilon^\sigma_{\sigma(\alpha}g_{\tau)\nu}\right)\hf^\tau{_\mu}\nonumber\\
\fl &&+\frac{1}{2}\left({\cUps}_{[\nu\tau]\sigma}\hf^\sigma{_\mu}
+{\cUps}_{[\nu\mu ]\sigma}\hf^\sigma{_\tau}
+{\cUps}_{(\mu\tau)\sigma}\hf^\sigma{_\nu}
-\frac{1}{2}\Upsilon^\sigma_{\sigma\nu}g_{\mu\tau}
+\Upsilon^\sigma_{\sigma(\mu}g_{\tau)\nu}\right)\hf^\tau{_\alpha}\nonumber\\
\fl &&-\frac{1}{2}\left({\cUps}_{[\mu\tau]\sigma}\hf^\sigma{_\nu}
+{\cUps}_{[\mu\nu ]\sigma}\hf^\sigma{_\tau}
+{\cUps}_{(\nu\tau)\sigma}\hf^\sigma{_\mu}
-\frac{1}{2}\Upsilon^\sigma_{\sigma\mu}g_{\nu\tau}
+\Upsilon^\sigma_{\sigma(\nu}g_{\tau)\mu}\right)\hf^\tau{_\alpha}\nonumber\\
\fl &&-\frac{1}{2}\Upsilon^\sigma_{\sigma\alpha}\hf_{\nu\mu}
+\Upsilon^\sigma_{\sigma[\nu}\hf_{\mu]\alpha}\nonumber\\
\fl &&+\frac{1}{2}
(\hf_{\nu\mu;\alpha}+\hf_{\alpha\mu;\nu}-\hf_{\alpha\nu;\mu})
+\frac{8\pi}{(n\!-\!1)}\hj_{[\nu} g_{\mu]\alpha}\nonumber\\
%\fl\nonumber\\
%\fl\nonumber\\
\fl &=&-\frac{1}{2}\left({\cUps}_{\alpha\tau\sigma}\hf^\sigma{_\mu}
+{\cUps}_{\mu\tau\sigma}\hf^\sigma{_\alpha}
\right)\hf^\tau{_\nu}\nonumber\\
\fl &&+\frac{1}{2}\left({\cUps}_{\alpha\tau\sigma}\hf^\sigma{_\nu}
+{\cUps}_{\nu\tau\sigma}\hf^\sigma{_\alpha}
\right)\hf^\tau{_\mu}\nonumber\\
\fl &&+\frac{1}{2}\left({\cUps}_{\tau\mu\sigma}\hf^\sigma{_\nu}
+2{\cUps}_{[\nu\mu]\sigma}\hf^\sigma{_\tau}
-{\cUps}_{\tau\nu\sigma}\hf^\sigma{_\mu}
\right)\hf^\tau{_\alpha}\nonumber\\
\fl &&+\frac{1}{2}\Upsilon^\sigma_{\sigma\alpha}\hf_{\mu\nu}
+\Upsilon^\sigma_{\sigma\tau}\hf^\tau{_{[\mu}}g_{\nu]\alpha}\nonumber\\
\fl &&+\frac{1}{2}
(\hf_{\nu\mu;\alpha}+\hf_{\alpha\mu;\nu}-\hf_{\alpha\nu;\mu})
+\frac{8\pi}{(n\!-\!1)}\hj_{[\nu} g_{\mu]\alpha}\nonumber,
\end{eqnarray}
\fi
and using (\ref{contractedupsilon}) gives,
\begin{eqnarray}
\label{upsilonhateq}
\fl\cUps_{\alpha\nu\mu}
&=&{\cUps}_{\alpha\sigma\tau}\hf^\sigma{_\mu}\hf^\tau{_\nu}
+{\cUps}_{(\mu\sigma)\tau}\hf^\sigma{_\alpha}\hf^\tau{_\nu}
-{\cUps}_{(\nu\sigma)\tau}\hf^\sigma{_\alpha}\hf^\tau{_\mu}
+{\cUps}_{[\nu\mu]\sigma}\hf^\sigma{_\tau}
\hf^\tau{_\alpha}\nonumber\\
\fl &&+\frac{1}{(n\!-\!2)}{\cUps}_{\sigma\tau\alpha}\hf^{\tau\sigma}\hf_{\mu\nu}
+\frac{2}{(n\!-\!2)}{\cUps}_{\sigma\rho\tau}\hf^{\rho\sigma}
\hf^\tau{_{[\mu}}g_{\nu]\alpha}\nonumber\\
\fl &&+\frac{1}{2}
(\hf_{\nu\mu;\alpha}+\hf_{\alpha\mu;\nu}-\hf_{\alpha\nu;\mu})
+\frac{8\pi}{(n\!-\!1)}\hj_{[\nu} g_{\mu]\alpha}.
\end{eqnarray}
Equation (\ref{upsilonhateq}) is useful for finding exact solutions
to the connection equations because it
consists of only $n^2(n-1)/2$ equations in the $n^2(n-1)/2$ unknowns
${\cUps}_{\alpha\nu\mu}$.
Also, from (\ref{upsilonhateq}) we can immediately see that
\begin{eqnarray}
\label{upsilonantisymmetriclowered}
\fl\cUps_{\alpha\nu\mu}
=\frac{1}{2}(\hf_{\nu\mu;\alpha}+\hf_{\alpha\mu;\nu}-\hf_{\alpha\nu;\mu})
+\frac{4\pi}{(n\!-\!1)}(\hj_{\nu} g_{\mu\alpha}-\hj_{\mu} g_{\nu\alpha})+(\hf^{3\prime})\dots.
\end{eqnarray}
Here the notation $(\hf^{3\prime})$ refers to terms like
$\hf_{\alpha\tau}\hf^\tau{_\sigma}\hf^\sigma{_{[\nu;\mu]}}$.
With (\ref{upsilonantisymmetriclowered}) as a starting point, one can calculate more
accurate $\cUps_{\alpha\nu\mu}$ by recursively substituting
the current $\cUps_{\alpha\nu\mu}$ into (\ref{upsilonhateq}).
%When the desired accuracy is achieved,
Then this
$\cUps_{\alpha\nu\mu}$ can be substituted into
(\ref{contractedupsilon},\ref{barupsilonofN}) to get $\bUps_{\alpha\nu\mu}$.
For our purposes (\ref{upsilonantisymmetriclowered}) will be accurate enough.
Substituting (\ref{upsilonantisymmetriclowered}) into (\ref{contractedupsilon}) we get
\begin{eqnarray}
%\label{upsilonantisymmetriclowered}
%\fl {\cUps}_{\alpha\nu\mu}
%&\approx&\frac{1}{2}(\hf_{\nu\mu;\alpha}
%+\hf_{\alpha\mu;\nu}-\hf_{\alpha\nu;\mu})
%+\frac{8\pi}{(n\!-\!1)}\hj_{[\nu} g_{\mu]\alpha},\\
\fl {\cUps}_{(\alpha\nu)\mu}
&=&-\hf_{\mu(\nu;\alpha)}
+\frac{4\pi}{(n\!-\!1)}(\hj_{(\nu} g_{\alpha)\mu}-\hj_\mu g_{\nu\alpha})+(\hf^{3\prime})\dots,\\
\fl {\cUps}_{[\alpha\nu]\mu}&=&\frac{1}{2}\hf_{\nu\alpha;\mu}
+\frac{4\pi}{(n\!-\!1)}\hj_{[\nu} g_{\alpha]\mu}+(\hf^{3\prime})\dots,\\
\fl\Upsilon^\sigma_{\sigma\alpha}&=&
\ifnum\ExpandDerivations=1
\frac{2}{(n\!-\!2)}\left(\frac{1}{2}\hf_{\tau\sigma;\alpha}
+\frac{4\pi}{(n\!-\!1)}\hj_{[\tau} g_{\sigma]\alpha}\right)\hf^{\tau\sigma}+(\hf^{4\prime})\dots\nonumber\\
\fl &=&
\fi
\label{upsiloncontractedappendix}
\frac{-1}{2(n\!-\!2)}(\hf^{\rho\sigma}\!\hf_{\sigma\rho})_{,\alpha}
+\frac{8\pi}{(n\!-\!1)(n\!-\!2)}\hj^\tau \hf_{\tau\alpha}+(\hf^{4\prime})\dots.
\end{eqnarray}
%where
%\begin{eqnarray}
%\ff=\hf^{\tau\sigma}\hf_{\sigma\tau}.
%\end{eqnarray}
Substituting these equations into (\ref{barupsilonofN}) gives
\begin{eqnarray}
\fl {\bUps}_{\alpha\nu\mu}&=&
\ifnum\ExpandDerivations=1
-\left(\frac{1}{2}\hf_{\alpha\mu;\tau}
+\frac{2\pi}{(n\!-\!1)}(\hj_{\alpha}g_{\mu\tau}-\hj_{\mu}g_{\alpha\tau})\right)\hf^\tau{_\nu}\nonumber\\
\fl&&+\left(\frac{1}{2}\hf_{\nu\alpha;\tau}
+\frac{2\pi}{(n\!-\!1)}(\hj_{\nu}g_{\alpha\tau}-\hj_{\alpha}g_{\nu\tau})\right)\hf^\tau{_\mu}\nonumber\\
\fl&&+\left(-\hf_{\tau(\mu;\nu)}
+\frac{2\pi}{(n\!-\!1)}(\hj_\mu g_{\nu\tau}+\hj_\nu g_{\mu\tau}-2 \hj_\tau g_{\mu\nu})\right)\hf^\tau{_\alpha}\nonumber\\
\fl&&-\frac{1}{2}\left(\frac{-1}{2(n\!-\!2)}(\hf^{\rho\sigma}\!\hf_{\sigma\rho})_{,\alpha}
+\frac{8\pi}{(n\!-\!1)(n\!-\!2)}\hj^\tau \hf_{\tau\alpha}\right)g_{\nu\mu}\nonumber\\
\fl&&+\left(\frac{-1}{2(n\!-\!2)}(\hf^{\rho\sigma}\!\hf_{\sigma\rho})_{,(\nu}
+\frac{8\pi}{(n\!-\!1)(n\!-\!2)}\hj^\tau \hf_{\tau(\nu}\right)g_{\mu)\alpha}+(\hf^{4\prime})\dots\nonumber\\
\fl &=&
\fi
\label{upsilonsymmetriclowered2}
\hf^\tau{_{(\nu}}\hf_{\mu)}{_{\alpha;\tau}}
+\hf_\alpha{^\tau}\hf_{\tau(\nu;\mu)}
+\frac{1}{4(n\!-\!2)}\left((\hf^{\rho\sigma}\!\hf_{\sigma\rho})_{,\alpha}g_{\nu\mu}
\!-2(\hf^{\rho\sigma}\!\hf_{\sigma\rho})_{,(\nu}g_{\mu)\alpha}\right)\nonumber\\
\fl &&+\frac{4\pi}{(n\!-\!2)}\hj^\tau\left(\hf_{\alpha\tau}g_{\nu\mu}
+\frac{2}{(n\!-\!1)}\hf_{\tau(\nu}g_{\mu)\alpha}\right)+(\hf^{4\prime})\dots.
\end{eqnarray}
Here the notation $(\hf^{4\prime})$ refers to terms like
$\hf_{\alpha\tau}\hf^\tau{_\sigma}\hf^\sigma{_\rho}\hf^\rho{_{(\nu;\mu)}}$.
Raising the indices on
(\ref{upsilonsymmetriclowered2},\ref{upsilonantisymmetriclowered},\ref{upsiloncontractedappendix})
and using (\ref{notation}) gives the final result
(\ref{upsilonsymmetric},\ref{upsilonantisymmetric},\ref{upsiloncontracted}).

\section{\label{ApproximateFandg}Solution for $N_{\nu\mu}$ in terms
of $g_{\nu\mu}$ and $f_{\nu\mu}$}
Here we invert the definitions (\ref{gdef},\ref{fdef}) of
$g_{\nu\mu}$ and $f_{\nu\mu}$ to obtain
(\ref{approximateNbar},\ref{approximateNhat}), the approximation of
$N_{\nu\mu}$ in terms of $g_{\nu\mu}$ and $f_{\nu\mu}$.
First let us define the notation
\begin{eqnarray}
\label{hfdef}
\hf^{\nu\mu}\!=\!f^{\nu\mu}\rmt\,\Lambda_b^{\!-1/2}.
\end{eqnarray}
We assume that $|\hf^\nu{_\mu}|\!\ll\!1$ for all components of the
unitless field $\hf^\nu{_\mu}$, and find a solution
in the form of a power series expansion in $\hf^\nu{_\mu}$.
Lowering an index on the equation
$(\rmN/\rmg\,)N^{\dashv\mu\nu}\!=g^{\nu\mu}\!+\hf^{\nu\mu}$
from (\ref{gdef},\ref{fdef}) gives
\begin{eqnarray}
\label{gminusF2}
\frac{\lower2pt\hbox{$\rmN$}}{\rmg}N^{\dashv\mu}{_\alpha}
=\delta^\mu_\alpha-\hf^\mu{_\alpha}.
\end{eqnarray}
Let us consider the tensor $\hf^\mu{_\alpha}\!=\!\hf^{\mu\nu}g_{\nu\alpha}$.
Because $g_{\nu\alpha}$ is symmetric and $\hf^{\mu\nu}$ is antisymmetric, it is
clear that $\hf^\alpha{_\alpha}\!=\!0$. Also because $\hf_{\nu\sigma}\hf^\sigma{_\mu}$ is
symmetric it is clear that $\hf^\nu{_\sigma}\hf^\sigma{_\mu}\hf^\mu{_\nu}=0$.
In matrix language therefore $tr(\hf)\!=\!0,~tr(\hf^3)\!=\!0$,
and in fact $tr(\hf^p)\!=\!0$ for any odd p.
Using the well known formula $det(e^M)=exp\,(tr(M))$ and
the power series $ln(1\!-\!x)=-x-x^2/2-x^3/3-x^4/4\dots$
we then get\cite{Deif},
\begin{eqnarray}
\label{lndetspecial}
ln(det(I\!-\!\hf))&=&tr(ln(I\!-\!\hf))
=-\frac{1}{2}\hf^\rho{_\sigma}\hf^\sigma{_\rho}+(\hf^4)\dots
\end{eqnarray}
Here the notation $(\hf^4)$ refers to terms like
$\hf^\tau{_\alpha}\hf^\alpha{_\sigma}\hf^\sigma{_\rho}\hf^\rho{_\tau}$.
Taking $ln(det())$ on both sides of (\ref{gminusF2}) using the result (\ref{lndetspecial})
and the identities $det(sM^{})\!=s^n det(M^{})$ and $det(M^{-1}_{})\!=1/det(M^{})$
gives
\begin{eqnarray}
\fl~~~~~~~~~ln\!\left(\!\frac{\lower2pt\hbox{$\rmN$}}{\rmg}\right)
\!&=&\!\frac{1}{(n\!-\!2)}\,ln\!\left(\!\frac{\lower2pt\hbox{$N^{(n/2-1)}$}}{g^{(n/2-1)}}\right)
\label{lnapproxdetN}
=-\frac{1}{2(n\!-\!2)}\,\hf^\rho{_\sigma}\hf^\sigma{_\rho}
+(\hf^4)\dots
\end{eqnarray}
Taking $e^x$ on both sides of (\ref{lnapproxdetN}) and using $e^x=1+x+x^2/2\dots$ gives
\begin{eqnarray}
\label{approxdetN}
\frac{\lower2pt\hbox{$\rmN$}}{\rmg}
=1\!-\!\frac{1}{2(n\!-\!2)}\,\hf^{\rho\sigma}\!\hf_{\sigma\rho}
+(\hf^4)\dots
\end{eqnarray}
Using the power series $(1\!-\!x)^{-1}\!=\!1+x+x^2+x^3\dots$,
or multiplying (\ref{gminusF2}) term by term,
we can calculate the inverse of (\ref{gminusF2}) to get\cite{Deif}
\begin{eqnarray}
\frac{\rmg}{\lower2pt\hbox{$\rmN$}} N^\nu{_\mu}
=\delta^\nu_\mu+\hf^\nu{_\mu}+\hf^\nu{_\sigma}\hf^\sigma{_\mu}
+\hf^\nu{_\rho}\hf^\rho{_\sigma}\hf^\sigma{_\mu}+(\hf^4)\dots\\
%\end{eqnarray}
%and therefore,
%\begin{eqnarray}
\label{approxN}
N_{\nu\mu}
=\frac{\lower2pt\hbox{$\rmN$}}{\rmg}(g_{\nu\mu}+\hf_{\nu\mu}+\hf_{\nu\sigma}\hf^\sigma{_\mu}
+\hf_{\nu\rho}\hf^\rho{_\sigma}\hf^\sigma{_\mu}+(\hf^4)\dots).
\end{eqnarray}
Here the notation $(\hf^4)$ refers to terms like
$\hf_{\nu\alpha}\hf^\alpha{_\sigma}\hf^\sigma{_\rho}\hf^\rho{_\mu}$.
Since $\hf_{\nu\sigma}\hf^\sigma{_\mu}$ is symmetric and
$\hf_{\nu\rho}\hf^\rho{_\sigma}\hf^\sigma{_\mu}$ is antisymmetric, we obtain from
(\ref{approxN},\ref{approxdetN},\ref{hfdef}) the final result
(\ref{approximateNbar},\ref{approximateNhat}).
\iffalse
\begin{eqnarray}
N_{(\nu\mu)}&=&g_{\nu\mu}+\hf_{\nu\sigma}\h\hf^\sigma{_\mu}
-\frac{\lower2pt\hbox{$1$}}{2(n\!-\!2)}g_{\nu\mu}\h\hf^{\rho\sigma}\!\hf_{\sigma\rho}+(\hf^4),\\
\label{approxNhat}
N_{[\nu\mu]}&=&\hf_{\nu\mu}+(\hf^3).
\end{eqnarray}
\fi

\ifnum\ExpandDerivations=1
\section{\label{continuityEuler}Derivation of the continuity equation and Lorentz force equation
from the Klein-Gordon equation in curved space}
For the spin-0 case, instead of deriving the continuity equation (\ref{continuity},\ref{jKleinGordon})
from the divergence of Ampere's law, it can also be derived from the Klein-Gordon equation.
Using (\ref{KleinGordon1},\ref{conjKleinGordon},\ref{derivKleinGordon},\ref{jKleinGordon}) we get,
\begin{eqnarray}
\fl 0 &=&\frac{i Q}{2\hbar}\left[\bar\psi
\left({ one~side~of}\atop {Klein\!-\!Gordon~equation}\right)
-\left({one~side~of~conjugate}\atop{Klein\!-\!Gordon~equation}\right)
\psi\right]\\
\fl &=&\frac{iQ}{2m\hbar }\bar\psi\left(\frac{\hbar^2}{\rmg}D_\mu\rmg D^\mu
+m^2-\Ddag{^\mu}\rmg\Ddag\!_\mu\frac{\hbar^2}{\rmg}-m^2\right)\psi\\
\fl &=&\frac{i\hbar Q}{2m}\bar\psi\left(\left(\Ddag\!_\mu\!+\!\frac{1}{\rmg}D_\mu\rmg\right)D^\mu
\!-\!\Ddag{^\mu}\left(\rmg\Ddag\!_\mu\frac{1}{\rmg}\!+\!D_\mu\right)\right)\psi\\
\fl &=&\frac{i\hbar Q}{2m}\bar\psi\left(\left(\frac{\overleftarrow\partial}{\partial x^\mu}
+\frac{1}{\rmg}\frac{\partial}{\partial x^\mu}\rmg\right)D^\mu
-\Ddag{^\mu}\left(\rmg\frac{\overleftarrow\partial}{\partial x^\mu}\frac{1}{\rmg}
+\frac{\partial}{\partial x^\mu}\right)\right)\psi\\
\fl &=&\frac{i\hbar Q}{2m}\left(\frac{1}{\rmg}\frac{\partial}{\partial x^\mu}(\bar\psi\rmg D^\mu \psi)
-(\bar\psi \Ddag{^\mu}\rmg\psi)\frac{\overleftarrow\partial}{\partial x^\mu}\frac{1}{\rmg}\right)\\
\fl &=&\frac{1}{\rmg}\frac{\partial}{\partial x^\mu}\left(\rmg\frac{i\hbar Q}{2m}(\bar\psi D^\mu\psi-\bar\psi\Ddag{^\mu}\psi)\right)\\
\label{continuity1}
\fl &=&j^\mu{_{;\mu}}.
\end{eqnarray}
Similarly, instead of deriving the Lorentz-force equation (\ref{Euler},\ref{TKleinGordon}) from the
divergence of the Einstein equations, it can also be derived from the Klein-Gordon equation.
Using (\ref{KleinGordon1},\ref{conjKleinGordon},\ref{derivKleinGordon},\ref{TKleinGordon}) we get,
\begin{eqnarray}
\fl 0&=&\left({one~side~of~conjugate}\atop {Klein\!-\!Gordon~equation}\right)\frac{D_\rho\psi}{2}
 +\frac{\bar\psi\Ddag\!_\rho}{2}\left({one~side~of}\atop{Klein\!-\!Gordon~equation}\right)\\
\fl &=&\bar\psi\left(\Ddag{^\lambda}\rmg\Ddag\!_\lambda\frac{\hbar^2}{\rmg}+m^2\right)\frac{D_\rho\psi}{2m}
 +\frac{\bar\psi\Ddag\!_\rho}{2m}\left(\frac{\hbar^2}{\rmg}D_\lambda\rmg D^\lambda+m^2\right)\psi\\
\fl &=&\frac{\hbar^2}{2m}\left[\frac{\partial(\bar\psi\Ddag{^\lambda}\rmg)}{\partial x^\lambda}\frac{1}{\rmg}D_\rho\psi
+\bar\psi\Ddag\!_\rho\frac{1}{\rmg}\frac{\partial(\rmg D^\lambda\psi)}{\partial x^\lambda}\right.\nonumber\\
\fl &&~~~~~~\left.+\bar\psi\Ddag{^\lambda}\left(-\frac{iQ}{\hbar}A_\lambda\right)D_\rho\psi
+\bar\psi\Ddag\!_\rho\left(\frac{iQ}{\hbar}A_\lambda\right)D^\lambda\psi\right]\nonumber\\
\fl &&~~~~~~+  \frac{m}{2}(\bar\psi D_\rho\psi+\bar\psi\Ddag\!_\rho\psi)\\
\fl &=&\frac{\hbar^2}{2m}\left[\frac{\partial(\bar\psi\Ddag{^\lambda}\rmg)}{\partial x^\lambda}\frac{1}{\rmg}D_\rho\psi
+\bar\psi\Ddag\!_\rho\frac{1}{\rmg}\frac{\partial(\rmg D^\lambda\psi)}{\partial x^\lambda}\right.\nonumber\\
\fl &&~~~~~~+\bar\psi\Ddag{^\lambda}\left(-\frac{iQ}{\hbar}A_\lambda\right)\frac{\partial\psi}{\partial x^\rho}
+\bar\psi\Ddag{^\lambda}\left(-\frac{iQ}{\hbar}A_\lambda\right)\left(\frac{iQ}{\hbar}A_\rho\right)\psi\nonumber\\
\fl &&~~~~~~\left. +\frac{\partial\bar\psi}{\partial x^\rho}\left(\frac{iQ}{\hbar}A_\lambda\right)D^\lambda\psi
+\left(-\frac{iQ}{\hbar}A_\rho\right)\bar\psi\left(\frac{iQ}{\hbar}A_\lambda\right)D^\lambda\psi\right]\nonumber\\
\fl&&~~~~~~+\frac{m}{2}\left(\bar\psi\frac{\partial\psi}{\partial x^\rho}+\frac{\partial\bar\psi}{\partial x^\rho}\psi\right)\\
\fl&=&\frac{\hbar^2}{2m}\left[\frac{\partial(\bar\psi\Ddag{^\lambda}\rmg)}{\partial x^\lambda}\frac{1}{\rmg}D_\rho\psi
+\bar\psi\Ddag\!_\rho\frac{1}{\rmg}\frac{\partial(\rmg D^\lambda\psi)}{\partial x^\lambda}\right.\nonumber\\
\fl&&~~~~~~-\bar\psi\Ddag{^\lambda}\left(\frac{iQ}{\hbar}A_\lambda\right)\frac{\partial\psi}{\partial x^\rho}
-\bar\psi\Ddag{^\lambda}\left(-\frac{iQ}{\hbar}A_\rho\right)\frac{\partial\psi}{\partial x^\lambda}
+\bar\psi\Ddag{^\lambda}\left(-\frac{iQ}{\hbar}A_\rho\right)D^\lambda\psi\nonumber\\
\fl&&~~~~~~\left.-\frac{\partial\bar\psi}{\partial x^\rho}\left(-\frac{iQ}{\hbar} A_\lambda\right) D^\lambda \psi
+\left(-\frac{iQ}{\hbar}A_\rho\right)\frac{\partial\bar\psi}{\partial x^\lambda}D^\lambda\psi
-\bar\psi\Ddag{^\lambda}\left(-\frac{iQ}{\hbar}A_\rho\right)D^\lambda\psi\right]\nonumber\\
\fl&&~~~~~~+\frac{m}{2}\frac{\partial(\bar\psi\psi)}{\partial x^\rho}\\
\fl&=&\frac{\hbar^2}{2m}\left[\frac{\partial(\bar\psi\Ddag{^\lambda}\rmg)}{\partial x^\lambda}\frac{1}{\rmg}D_\rho\psi
+\bar\psi\Ddag\!_\rho\frac{1}{\rmg}\frac{\partial(\rmg D^\lambda\psi)}{\partial x^\lambda}\right.\nonumber\\
\fl&&~~~~~~+\bar\psi\Ddag{^\lambda}\left(\frac{\partial}{\partial x^\lambda}\left(\frac{iQ}{\hbar}A_\rho\psi\right)-\frac{\partial}{\partial x^\rho}\left(\frac{iQ}{\hbar}A_\lambda\psi\right)-\frac{2iQ}{\hbar}A_{[\rho,\lambda]}\psi\right)\nonumber\\
\fl&&~~~~~~~~~~~~~\left.+\left(\frac{\partial}{\partial x^\lambda}\left(-\frac{iQ}{\hbar}A_\rho\bar\psi\right)-\frac{\partial}{\partial x^\rho}\left(-\frac{iQ}{\hbar}A_\lambda\bar\psi\right)+\frac{2iQ}{\hbar}A_{[\rho,\lambda]}\bar\psi\right)D^\lambda\psi\right]\nonumber\\
\fl&&~~~~~~+\frac{m}{2}\frac{\partial(\bar\psi\psi)}{\partial x^\rho}\\
\fl&=&\frac{\hbar^2}{2m}\left[\frac{\partial(\bar\psi\Ddag{^\lambda}\rmg)}{\partial x^\lambda}\frac{1}{\rmg}D_\rho\psi
+\bar\psi\Ddag\!_\rho\frac{1}{\rmg}\frac{\partial(\rmg D^\lambda\psi)}{\partial x^\lambda}\right.\nonumber\\
\fl&&~~~~~~+\bar\psi\Ddag\!_\nu\frac{\partial g^{\nu\lambda}}{\partial x^\rho}D_\lambda\psi
-\bar\psi\Ddag\!_\mu\frac{\partial g^{\mu\lambda}}{\partial x^\rho} D_\lambda\psi\nonumber\\
\fl&&~~~~~~+\bar\psi\Ddag{^\lambda}\left(\frac{\partial(D_\rho\psi)}{\partial x^\lambda}-\frac{\partial(D_\lambda\psi)}{\partial x^\rho}\right)
+\left(\frac{\partial(\bar\psi\Ddag\!_\rho)}{\partial x^\lambda}-\frac{\partial(\bar\psi\Ddag\!_\lambda)}{\partial x^\rho}\right)D^\lambda\psi\nonumber\\
\fl&&~~~~~~\left. +\frac{2iQ}{\hbar}(\bar\psi D^\lambda\psi-\bar\psi\Ddag{^\lambda}\psi)A_{[\rho,\lambda]}\right]
+\frac{m}{2}\frac{\partial(\bar\psi\psi)}{\partial x^\rho}\\
\fl&=&\frac{\hbar^2}{2m}\left[\frac{1}{\rmg}\frac{\partial}{\partial x^\lambda}\!\left(\rmg(\bar\psi\Ddag{^\lambda}D_\rho\psi\!+\bar\psi\Ddag\!_\rho D^\lambda\psi)\right)
\!+\bar\psi\Ddag\!_\lambda\frac{\partial g^{\lambda\nu}}{\partial x^\rho}D_\nu\psi
-\frac{\partial}{\partial x^\rho}\bar\psi\Ddag\!_\lambda D^\lambda\psi\right]\nonumber\\
\fl&&~~~~~~+\frac{m}{2}\frac{\partial(\bar\psi\psi)}{\partial x^\rho}
+\frac{iQ\hbar}{m}(\bar\psi D^\lambda\psi-\bar\psi\Ddag{^\lambda}\psi)A_{[\rho,\lambda]}\\
\label{Euler1}
\fl&=&\frac{1}{2m}\left[\hbar^2(\bar\psi\Ddag{^\lambda}D_\rho\psi+\bar\psi\Ddag\!_\rho D^\lambda\psi)_{;\lambda}
-(\hbar^2\bar\psi\Ddag\!_\lambda D^\lambda\psi-m^2\bar\psi\psi)_{,\rho}\right]\nonumber\\
\fl&&~~~~~~+\frac{iQ\hbar}{m}(\bar\psi D^\lambda\psi-\bar\psi\Ddag{^\lambda}\psi)A_{[\rho,\lambda]}\\
\fl&=&T^\lambda_{\rho;\,\lambda}+2j^\lambda A_{[\rho,\lambda]}.
\end{eqnarray}

\fi
%\newpage
%\bibliography{npshifflett}% Produces the bibliography via BibTeX.
%\section*{References}
%\begin{thebibliography}{68}
\Bibliography{68}
\expandafter\ifx\csname natexlab\endcsname\relax\def\natexlab#1{#1}\fi
\expandafter\ifx\csname bibnamefont\endcsname\relax
  \def\bibnamefont#1{#1}\fi
\expandafter\ifx\csname bibfnamefont\endcsname\relax
  \def\bibfnamefont#1{#1}\fi
\expandafter\ifx\csname citenamefont\endcsname\relax
  \def\citenamefont#1{#1}\fi
\expandafter\ifx\csname url\endcsname\relax
  \def\url#1{\texttt{#1}}\fi
\expandafter\ifx\csname urlprefix\endcsname\relax\def\urlprefix{URL }\fi
\providecommand{\bibinfo}[2]{#2}
\providecommand{\eprint}[2][]{#2}
\providecommand{\bitem}[2][]{\bibitem {#2}}

\bitem[{\citenamefont{Hlavaty}(1957)}]{Hlavaty}
\bibinfo{author}{\bibfnamefont{V.}~\bibnamefont{Hlavaty}},
  \emph{\bibinfo{title}{Geometry of Einstein's Unified Field Theory}}
  (\bibinfo{publisher}{P. Noordhoff Ltd.},
  \bibinfo{address}{Groningen},
  \bibinfo{year}{1957}).

\bitem[{\citenamefont{Shifflett}(2003)}]{Shifflett}
\bibinfo{author}{\bibfnamefont{J.~A.} \bibnamefont{Shifflett}},
 \eprint{arXiv:gr-qc/0310124}.

\bitem[{\citenamefont{Shifflett}(2004)}]{Shifflett2}
\bibinfo{author}{\bibfnamefont{J.~A.} \bibnamefont{Shifflett}},
 \eprint{arXiv:gr-qc/0403052}.

\bitem[{\citenamefont{Zeldovich}(1968)}]{Zeldovich}
\bibinfo{author}{\bibfnamefont{Ya.B.}~\bibnamefont{Zeldovich}},
  \bibinfo{journal}{Sov.\ Phys. - Uspekhi} \textbf{\bibinfo{volume}{11}}
  (\bibinfo{year}{1968}) \bibinfo{pages}{381}.

\bitem[{\citenamefont{Sahni and Starobinsky}(2000)}]{Sahni}
\bibinfo{author}{\bibfnamefont{V.}~\bibnamefont{Sahni}} \bibnamefont{and}
  \bibinfo{author}{\bibfnamefont{A.}~\bibnamefont{Starobinsky}},
  \bibinfo{journal}{Int.\ J.\ Mod.\ Phys.} \textbf{\bibinfo{volume}{D9}}
  (\bibinfo{year}{2000}) \bibinfo{pages}{373} [arXiv:astro-ph/9904398].

%\bitem[{\citenamefont{Griffiths}(1995)}]{Griffiths}
%\bibinfo{author}{\bibfnamefont{D.}~\bibnamefont{Griffiths}},
%  \emph{\bibinfo{title}{Introduction to Quantum Mechanics}}
%  (\bibinfo{publisher}{Prentice Hall}, \bibinfo{address}{Upper Saddle River, NJ},
%  \bibinfo{year}{1956}).

\bitem[{\citenamefont{Peskin}(1964)}]{Peskin}
\bibinfo{author}{\bibfnamefont{M.E.}~\bibnamefont{Peskin}},
\bibinfo{author}{\bibfnamefont{D.V.}~\bibnamefont{Schroeder}},
  \emph{\bibinfo{title}{An Introduction to Quantum Field Theory}}
  (\bibinfo{publisher}{Westview Press},
  \bibinfo{year}{1995}) p 790-791.

\bitem[{\citenamefont{Einstein and Infeld}(1949)}]{EinsteinInfeld}
\bibinfo{author}{\bibfnamefont{A.}~\bibnamefont{Einstein}} \bibnamefont{and}
  \bibinfo{author}{\bibfnamefont{L.}~\bibnamefont{Infeld}},
  \bibinfo{journal}{Canad.\ J.\ Math.} \textbf{\bibinfo{volume}{1}}
  (\bibinfo{year}{1949}) \bibinfo{pages}{209}.

\bitem[{\citenamefont{Einstein and Straus}(1946)}]{EinsteinStraus}
\bibinfo{author}{\bibfnamefont{A.}~\bibnamefont{Einstein}} \bibnamefont{and}
  \bibinfo{author}{\bibfnamefont{E.~G.} \bibnamefont{Straus}},
  \bibinfo{journal}{Ann.\ Math.} \textbf{\bibinfo{volume}{47}}
  (\bibinfo{year}{1946}) \bibinfo{pages}{731}.

\bitem[{\citenamefont{Einstein}(1948)}]{Einstein3}
\bibinfo{author}{\bibfnamefont{A.}~\bibnamefont{Einstein}},
  \bibinfo{journal}{Rev.\ Mod.\ Phys.} \textbf{\bibinfo{volume}{20}}
  (\bibinfo{year}{1948}) \bibinfo{pages}{35}.

\bitem[{\citenamefont{Einstein}(1950)}]{EinsteinBianchi}
\bibinfo{author}{\bibfnamefont{A.}~\bibnamefont{Einstein}},
  \bibinfo{journal}{Can.\ J.\ Math.} \textbf{\bibinfo{volume}{2}}
  (\bibinfo{year}{1949}) \bibinfo{pages}{120}.

\bitem[{\citenamefont{Einstein and Kaufman}(1955)}]{EinsteinKaufman}
\bibinfo{author}{\bibfnamefont{A.}~\bibnamefont{Einstein}} \bibnamefont{and}
  \bibinfo{author}{\bibfnamefont{B.}~\bibnamefont{Kaufman}},
  \bibinfo{journal}{Ann.\ Math.} \textbf{\bibinfo{volume}{62}}
  (\bibinfo{year}{1955}) \bibinfo{pages}{128}.

\bitem[{\citenamefont{Einstein}(1956)}]{EinsteinMOR}
\bibinfo{author}{\bibfnamefont{A.}~\bibnamefont{Einstein}},
  \emph{\bibinfo{title}{The Meaning of Relativity, 5th ed. revised}}
  (\bibinfo{publisher}{Princeton U. Press}, \bibinfo{address}{Princeton
  NJ}, \bibinfo{year}{1956}).

\bitem[{\citenamefont{Schilpp}(1949)}]{Schilpp}
\bibinfo{author}{\bibfnamefont{P.A.}~\bibnamefont{Schilpp}},
  \emph{\bibinfo{title}{Albert Einstein: Philosopher-Scientist}}
  (\bibinfo{publisher}{Open Court Publishing}, \bibinfo{address}{Peru, IL},
  \bibinfo{year}{1949}) p 75,89-95.

%\bitem[{\citenamefont{Callaway}(1953)}]{Callaway}
%\bibinfo{author}{\bibfnamefont{J.}~\bibnamefont{Callaway}},
%  \bibinfo{journal}{Phys.\ Rev.} \textbf{\bibinfo{volume}{92}}
%  (\bibinfo{year}{1953}) \bibinfo{pages}{1567}.

%\bitem[{\citenamefont{Infeld}(1950)}]{Infeld}
%\bibinfo{author}{\bibfnamefont{L.}~\bibnamefont{Infeld}},
%  \bibinfo{journal}{Acta Phys. Pol.} \textbf{\bibinfo{volume}{X}}
%  (\bibinfo{year}{1950}) \bibinfo{pages}{284}.

%\bitem[{\citenamefont{Adler et~al.}(1975)\citenamefont{Adler, Bazin, and
%  Schiffer}}]{Adler}
%\bibinfo{author}{\bibfnamefont{R.}~\bibnamefont{Adler}},
%  \bibinfo{author}{\bibfnamefont{M.}~\bibnamefont{Bazin}}, \bibnamefont{and}
%  \bibinfo{author}{\bibfnamefont{M.}~\bibnamefont{Schiffer}},
%  \emph{\bibinfo{title}{Introduction to General Relativity }}
%  (\bibinfo{publisher}{McGraw-Hill}, \bibinfo{address}{New York},
%  \bibinfo{year}{1975}) p 87.

\bitem[{\citenamefont{Schr\"{o}dinger}(1947)}]{SchrodingerI}
\bibinfo{author}{\bibfnamefont{E.}~\bibnamefont{Schr\"{o}dinger}},
  \bibinfo{journal}{Proc.\ Royal Irish Acad.} \textbf{\bibinfo{volume}{51A}}
  (\bibinfo{year}{1947}) \bibinfo{pages}{163}.

%\bitem[{\citenamefont{Schr\"{o}dinger}(1948)}]{SchrodingerII}
%\bibinfo{author}{\bibfnamefont{E.}~\bibnamefont{Schr\"{o}dinger}},
%  \bibinfo{journal}{Proc.\ Royal Irish Acad.} \textbf{\bibinfo{volume}{51A}}
%  (\bibinfo{year}{1948}) \bibinfo{pages}{205}.

\bitem[{\citenamefont{Schr\"{o}dinger}(1948)}]{SchrodingerIII}
\bibinfo{author}{\bibfnamefont{E.}~\bibnamefont{Schr\"{o}dinger}},
  \bibinfo{journal}{Proc.\ Royal Irish Acad.} \textbf{\bibinfo{volume}{52A}}
  (\bibinfo{year}{1948}) \bibinfo{pages}{1}.

\bitem[{\citenamefont{Schr\"{o}dinger}(1950)}]{SchrodingerSTS}
\bibinfo{author}{\bibfnamefont{E.}~\bibnamefont{Schr\"{o}dinger}},
  \emph{\bibinfo{title}{Space-Time Structure}}
  (\bibinfo{publisher}{Cambridge Press}, \bibinfo{address}{London},
  \bibinfo{year}{1950}) p 93,108,112.

\bitem[{\citenamefont{H\'{e}ly}(1954)}]{Hely}
\bibinfo{author}{\bibfnamefont{J.}~\bibnamefont{H\'{e}ly}},
  \bibinfo{journal}{C.\ R.\ Acad.\ Sci.\ (Paris)}
  \textbf{\bibinfo{volume}{239}}
  (\bibinfo{year}{1954}) \bibinfo{pages}{385}.

\bitem[{\citenamefont{Treder}(1957)}]{Treder57}
\bibinfo{author}{\bibfnamefont{H.-J.}~\bibnamefont{Treder}},
  \bibinfo{journal}{Annalen der Physik} \textbf{\bibinfo{volume}{19}}
  (\bibinfo{year}{1957}) \bibinfo{pages}{369}.

\bitem[{\citenamefont{Johnson}(1985)}]{JohnsonI}
\bibinfo{author}{\bibfnamefont{C.~R.} \bibnamefont{Johnson}},
  \bibinfo{journal}{Phys.\ Rev.\ D} \textbf{\bibinfo{volume}{31}}
  (\bibinfo{year}{1985}) \bibinfo{pages}{1236}.

\bitem[{\citenamefont{Antoci}(1991)}]{Antoci3}
\bibinfo{author}{\bibfnamefont{S.}~\bibnamefont{Antoci}},
  \bibinfo{journal}{Gen.\ Relativ.\ Gravit.} \textbf{\bibinfo{volume}{23}}
  (\bibinfo{year}{1991}) \bibinfo{pages}{47}  [arXiv:gr-qc/0108052].

\bitem[{\citenamefont{Borchsenius}(1978)}]{Borchsenius}
\bibinfo{author}{\bibfnamefont{K.}~\bibnamefont{Borchsenius}},
  \bibinfo{journal}{Nuovo Cimento} \textbf{\bibinfo{volume}{46A}}
  (\bibinfo{year}{1978}) \bibinfo{pages}{403}.

\bitem[{\citenamefont{Papapetrou}(1948)}]{Papapetrou}
\bibinfo{author}{\bibfnamefont{A.}~\bibnamefont{Papapetrou}},
  \bibinfo{journal}{Proc.\ Royal Irish Acad.} \textbf{\bibinfo{volume}{52A}}
  (\bibinfo{year}{1948}) \bibinfo{pages}{69}.

\bitem[{\citenamefont{Takeno et~al.}(1951)\citenamefont{Takeno, Ikeda, and
  Abe}}]{Takeno}
\bibinfo{author}{\bibfnamefont{H.}~\bibnamefont{Takeno}},
  \bibinfo{author}{\bibfnamefont{M.}~\bibnamefont{Ikeda}}, \bibnamefont{and}
  \bibinfo{author}{\bibfnamefont{S.}~\bibnamefont{Abe}},
  \bibinfo{journal}{Prog.\ Theor.\ Phys.} \textbf{\bibinfo{volume}{VI}}
  (\bibinfo{year}{1951}) \bibinfo{pages}{837}.

\bitem[{\citenamefont{Birrell}(1964)}]{Birrell}
\bibinfo{author}{\bibfnamefont{N.D.}~\bibnamefont{Birrell}},
\bibinfo{author}{\bibfnamefont{P.C.}~\bibnamefont{Davies}},
  \emph{\bibinfo{title}{Quantum Fields in Curved Space}}
  (\bibinfo{publisher}{Cambridge University Press}, \bibinfo{address}{Cambridge},
  \bibinfo{year}{1982}) p 81-88.

\bitem[{\citenamefont{Tonnelat}(1954)}]{Tonnelat}
\bibinfo{author}{\bibfnamefont{M.~A.} \bibnamefont{Tonnelat}},
  \bibinfo{journal}{C.\ R.\ Acad.\ Sci.} \textbf{\bibinfo{volume}{239}}
  (\bibinfo{year}{1954}) \bibinfo{pages}{231}.

\bitem[{\citenamefont{Wallace}(1941)}]{Wallace}
\bibinfo{author}{\bibfnamefont{P.~R.} \bibnamefont{Wallace}},
  \bibinfo{journal}{Am.\ J.\ Math.} \textbf{\bibinfo{volume}{63}}
  (\bibinfo{year}{1941}) \bibinfo{pages}{729}.

\bitem[{\citenamefont{Padmanabhan2}(2002)}]{Padmanabhan2}
\bibinfo{author}{\bibfnamefont{T.} \bibnamefont{Padmanabhan}},
  \bibinfo{journal}{Class.\ Quantum\ Grav.} \textbf{\bibinfo{volume}{19}}
  (\bibinfo{year}{2002}) \bibinfo{pages}{3551}.

\bitem[{\citenamefont{Padmanabhan}(1985)}]{Padmanabhan}
\bibinfo{author}{\bibfnamefont{T.} \bibnamefont{Padmanabhan}},
  \bibinfo{journal}{Gen.\ Rel.\ Grav.} \textbf{\bibinfo{volume}{17}}
  (\bibinfo{year}{1985}) \bibinfo{pages}{215}.

\bitem[{\citenamefont{Sakharov}(1968)}]{Sakharov}
\bibinfo{author}{\bibfnamefont{A.~D.} \bibnamefont{Sakharov}},
  \bibinfo{journal}{Sov.\ Phys.\ Doklady} \textbf{\bibinfo{volume}{12}}
  (\bibinfo{year}{1968}) \bibinfo{pages}{1040}.

\bitem[{\citenamefont{Ashtekar, Rovelli and Smolin}(2000)}]{Ashtekar}
\bibinfo{author}{\bibfnamefont{A.}~\bibnamefont{Ashtekar}} \bibnamefont{and}
\bibinfo{author}{\bibfnamefont{C.}~\bibnamefont{Rovelli}} \bibnamefont{and}
  \bibinfo{author}{\bibfnamefont{L.}~\bibnamefont{Smolin}},
  \bibinfo{journal}{Phys.\ Rev.\ Let.} \textbf{\bibinfo{volume}{69}}
  (\bibinfo{year}{1992}) \bibinfo{pages}{237}.

\bitem[{\citenamefont{Garay}(1995)}]{Garay}
\bibinfo{author}{\bibfnamefont{Luis.\ L.} \bibnamefont{Garay}},
  \bibinfo{journal}{Int. J. Mod. Phys.} \textbf{\bibinfo{volume}{A10}}
  (\bibinfo{year}{1995}) \bibinfo{pages}{145}.

\bitem[{\citenamefont{Smolin}(2004)}]{Smolin}
\bibinfo{author}{\bibfnamefont{L.} \bibnamefont{Smolin}},
  \bibinfo{journal}{Sci. Am.} \textbf{\bibinfo{volume}{290}}
  (\bibinfo{year}{2004}) \bibinfo{pages}{1}.

\bitem[{\citenamefont{Jackson}(1999)}]{Jackson}
\bibinfo{author}{\bibfnamefont{J.~D.}~\bibnamefont{Jackson}},
  \emph{\bibinfo{title}{Classical Electrodynamics}}
  (\bibinfo{publisher}{John Wiley and Sons},
  \bibinfo{year}{1999}) p 2-22.

\bitem[{\citenamefont{Deif}(1982)}]{Deif}
\bibinfo{author}{\bibfnamefont{A.~S.}~\bibnamefont{Deif}},
  \emph{\bibinfo{title}{Advanced Matrix Theory For Scientists and Engineers}}
  (\bibinfo{publisher}{John Wiley and Sons},
  \bibinfo{address}{NY}, (\bibinfo{year}{1982}) p~20,153,183,171.

%\bitem[{\citenamefont{Infeld}(1950)}]{Infeld}
%\bibinfo{author}{\bibfnamefont{L.}~\bibnamefont{Infeld}},
%  \bibinfo{journal}{Acta Phys. Pol.} \textbf{\bibinfo{volume}{X}}
%  (\bibinfo{year}{1950}) \bibinfo{pages}{284}.

%\bitem[{\citenamefont{Infeld and Wallace}(1940)}]{InfeldWallace}
%\bibinfo{author}{\bibfnamefont{L.}~\bibnamefont{Infeld}} \bibnamefont{and}
%  \bibinfo{author}{\bibfnamefont{P.~R.} \bibnamefont{Wallace}},
%  \bibinfo{journal}{Phys.\ Rev.} \textbf{\bibinfo{volume}{57}}
%  (\bibinfo{year}{1940}) \bibinfo{pages}{797}.
%
%\bitem[{\citenamefont{Fock}(1929)}]{Fock}
%\bibinfo{author}{\bibfnamefont{V.~A.}~\bibnamefont{Fock}},
%  \bibinfo{journal}{Z.\ Phys.} \textbf{\bibinfo{volume}{57}},
%   (\bibinfo{year}{1929}) \bibinfo{pages}{261}.

%\bitem[{\citenamefont{Misner}(1973)}]{Misner}
%\bibinfo{author}{\bibfnamefont{C.W.}~\bibnamefont{Misner}},
%\bibinfo{author}{\bibfnamefont{K.S.}~\bibnamefont{Thorne}},
%\bibinfo{author}{\bibfnamefont{J.A.}~\bibnamefont{Wheeler}},
%  \emph{\bibinfo{title}{Gravitation}}
%  (\bibinfo{publisher}{W. H. Freeman},
%  \bibinfo{year}{1973}) p 480.
\endbib
%\end{thebibliography}

\end{document}